\documentclass{emulateapj}
\usepackage{natbib}
\def\gsim{\;\rlap{\lower 2.5pt
 \hbox{$\sim$}}\raise 1.5pt\hbox{$>$}\;}
\def\lsim{\;\rlap{\lower 2.5pt
   \hbox{$\sim$}}\raise 1.5pt\hbox{$<$}\;}
\newcommand{\Reff}{$R_{\rm eff}$}
\def\dML{\mbox{$\nabla_{\ell} \Upsilon$}}

\newcommand{\Chandra}{\emph{Chandra}}
\newcommand{\ciao}{\textsc{ciao}}
\newcommand{\caldb}{\textsc{caldb}}
\newcommand{\asca}{\textsc{asca}}
\newcommand{\xspec}{\textsc{xspec}}
\newcommand{\apec}{\textsc{apec}}

\slugcomment{submitted to The Astronomical Journal, 26 June 2008; revised 28 February 2009; accepted 16 March 2009}

\shorttitle{Mapping the Dark Side with DEIMOS}
\shortauthors{Romanowsky, Strader, Spitler, Johnson, Brodie, Forbes, \& Ponman}

\begin{document}

\title{Mapping the Dark Side with DEIMOS:\\ Globular Clusters, X-ray Gas, and Dark Matter in the NGC 1407 Group}
\author{
Aaron J. Romanowsky\altaffilmark{1,2},
Jay Strader\altaffilmark{1,3,4},
Lee R. Spitler\altaffilmark{5},
Ria Johnson\altaffilmark{6},\\
Jean P. Brodie\altaffilmark{1},
Duncan A. Forbes\altaffilmark{5},
Trevor Ponman\altaffilmark{6}
}
\altaffiltext{1}{UCO/Lick Observatory, University of California, Santa Cruz, CA 95064, USA}
\altaffiltext{2}{Departamento de F\'{\i}sica, Universidad de Concepci\'on, Casilla 160-C, Concepci\'on, Chile}
\altaffiltext{3}{Harvard-Smithsonian Center for Astrophysics, 60 Garden St., Cambridge, MA 02138, USA}
\altaffiltext{4}{Hubble Fellow}
\altaffiltext{5}{Centre for Astrophysics \& Supercomputing, Swinburne University, Hawthorn, VIC 3122, Australia}
\altaffiltext{6}{School of Physics and Astronomy, University of Birmingham, Edgbaston, Birmingham B15 2TT, UK}
\email{romanow@ucolick.org}


\begin{abstract}
NGC~1407 is the central elliptical in a nearby evolved group of galaxies
apparently destined to become a galaxy cluster core.
We use the kinematics of globular clusters (GCs) to probe
the dynamics and mass profile of the group's center,
out to a radius of 60 kpc ($\sim$~10 galaxy effective radii)---the most
extended data set to date around an early-type galaxy.
This sample consists of 172 GC line-of-sight velocities,
most of them newly obtained using Keck/DEIMOS, with
a few additional objects identified as dwarf-globular transition objects or as intra-group GCs.
We find weak rotation for the outer parts of the GC system ($v/\sigma \sim 0.2$),
with a rotational misalignment between the metal-poor and metal-rich GCs.
The velocity dispersion profile declines rapidly to a radius of $\sim$~20~kpc,
and then becomes flat or rising to $\sim$~60~kpc.
There is evidence that the GC orbits have a tangential bias that is
strongest for the metal-poor GCs---in possible contradiction to theoretical expectations.
We construct cosmologically-motivated galaxy+dark halo dynamical models 
and infer a total mass within 60~kpc of $\sim 3\times10^{12} M_{\odot}$,
which extrapolates to a virial mass of 
$\sim 6\times 10^{13} M_{\odot}$ for a typical $\Lambda$CDM halo---in
agreement with results from kinematics of the group galaxies.
We present an independent {\it Chandra}-based analysis,
whose relatively high mass at $\sim$~20~kpc disagrees strongly with the GC-based result
unless the GCs are assumed
to have a peculiar orbit distribution, and we therefore discuss more generally some
comparisons between X-ray and optical results.
The group's $B$-band mass-to-light ratio of 
$\sim$~800 (uncertain by a factor of $\sim$~2)
in Solar units is extreme even for a rich galaxy cluster, much less a poor group---placing
it among the most dark matter dominated systems in the universe,
and also suggesting a massive reservoir of baryons lurking in an unseen phase,
in addition to the nonbaryonic dark matter.
We compare the kinematical and mass properties of the NGC~1407 group to
other nearby groups and clusters, and
discuss some implications of this system for structure formation.
\end{abstract}

\section{Introduction}\label{sec:intro}

In our current understanding of the universe, stellar systems larger than
a globular cluster (GC; baryon mass $\sim 10^6 M_{\odot}$) are generally embedded in
massive halos of dark matter (DM).
Their systemic properties versus increasing DM halo mass show that
there must be shifts between the various regimes of physical processes
shaping the baryonic properties
(e.g., \citealt{1977MNRAS.179..541R};
\citealt{2003ApJ...599...38B};
\citealt{2004MNRAS.355..694M};
\citealt{2006MNRAS.368....2D,2007MNRAS.381..389K,2007MNRAS.382.1481N,2008MNRAS.385L.116G}).
In particular, somewhere between the scales of individual galaxies and clusters
of galaxies, the mass of a halo's primary galaxy stops scaling with
the halo mass, and reaches a universal upper limit 
(see e.g., 
\citealt{2004A&A...415..931R};
\citealt{2005ApJ...627L..85C};
\citealt{2006AJ....131.2018B};
\citealt{2007arXiv0709.2192K};
\citealt{2007arXiv0710.3780H}),
while the bulk of the baryons become spread out as a hot intergalactic medium (IGM).
The high-mass systems behave less like galaxies with their own enveloping halos
and unique histories, and more like extended halos shaping the destinies of
their captive galaxian members
(e.g., \citealt{2006PASA...23...38F}).

In collapsed galaxy clusters, the dominant processes affecting the larger galaxies
involve cooling, heating, and stripping by the IGM, while
some level of harassment and stripping also occurs by galaxy fly-bys within the cluster.
However, it is thought that processes such as gas strangulation and major galaxy mergers
occurred earlier, within the groups that went on to conglomerate into the cluster
(e.g.,
\citealt{1985MNRAS.215..517B,1998ApJ...496...39Z,2001MNRAS.326..637K,2001ApJ...563..736C,2004PASJ...56...29F,2006MNRAS.370.1223B}, hereafter B+06a; 
\citealt{2007ApJ...671.1503M,2008ApJ...672L.103K,2008ApJ...680.1009W};
\citealt{2008ApJ...688L...5K,2008arXiv0812.2021P}).
In this picture, the characteristic properties of cluster galaxies (dominated by
early-types) would primarily arise by ``pre-processing'' in groups.
Thus by studying galaxy groups, one can learn not only about the ecosystems 
of most of the galaxies in the universe (\citealt{2004MNRAS.348..866E}),
but also about the pre-history of the rarer but interesting galaxy clusters.

The nearby group surrounding the elliptical galaxy NGC~1407,
sometimes called the ``Eridanus A'' group 
(\citealt{1989AJ.....98.1531W};
\citealt{1990AJ....100....1F};
\citealt{2006MNRAS.369.1351B}, hereafter B+06b), 
is a potentially informative transitional system.
It has the X-ray and optical luminosities of a group 
(with 1--3 galaxies of at least $L^*$ luminosity,
and highly dominated by early-types), but possibly the mass of a cluster.
An early study of its constituent galaxy redshifts suggested a group 
mass-to-light ratio of 
$\Upsilon \sim$~2500~$\Upsilon_{\odot}$  ($B$-band; 
\citealt{1993ApJ...403...37G}), and
while subsequent studies have brought this estimate down
to $\sim$~300--1100 
(partially through longer distance estimates;
\citealt{1994A&A...283..722Q,2005ApJ...618..214T}; B+06b;
\citealt{2006MNRAS.369.1375T,2006MNRAS.372.1856F};
\citealt{2007ApJ...656..805Z}, hereafter Z+07),
these are still high values more reminiscent of clusters than of groups.

The NGC~1407 group may provide insight into the origins of the
properties of galaxy cluster cores and of brightest cluster galaxies.
The group is part of a larger structure or ``supergroup'' that should eventually
collapse to form a cluster with NGC~1407---and possibly its associated group---at
its core (B+06b; \citealt{2006MNRAS.369.1375T}).
Major mergers are prone to scramble the memories of their progenitors' prior properties,
but the future assembly of the NGC~1407 cluster appears to involve low mass ratios 
of $\sim$1:5 to 1:10.
Thus it is likely that the NGC~1407 group will remain relatively unscathed at the core of the 
resulting cluster, and we can make a fair comparison of its current dynamical properties
with those of existing cluster cores (such as M87 in Virgo and NGC~1399 in Fornax) 
to see if these systems are part of the same evolutionary sequence.

NGC~1407, like many bright elliptical galaxies, 
is endowed with a swarm of GCs visible as bright sources 
extending far out into its halo, beyond the region of readily observable galaxy light.
GCs are invaluable as tracers of major episodes of star formation in galaxies
\citep{1995AJ....109..960W,2000A&A...354..836L,2006ARA&A..44..193B}.
In NGC~1407,
\citet[hereafter C+07]{2007AJ....134..391C} have used them
to study the galaxy's early metal enrichment history,
and to search for signs of recent gas-rich mergers.
Here our focus is on using GCs as bountiful kinematical tracers, useful for probing
both the dynamical properties of their host galaxy's outer parts,
and the mass distribution of the surrounding group.

The use of GCs as mass tracers is well established:
see \citet{2006ARA&A..44..193B} and \citet{2009gcgg.book..433R} for reviews,
and \S\ref{sec:massprof} of this paper for specific examples.
GCs are so far less widely exploited as {\it orbit tracers}---using
the internal dynamics of a globular cluster system (GCS)
to infer its formation history and its connections
with other galaxy subcomponents.
Studies of GCS rotation have implicitly treated the GCs as proxies
for the halo field stars or for the DM particles (e.g., \citealt{1998AJ....116.2237K}),
but the connections between these different entities have not yet been established.
\citet[hereafter B+05]{2005MNRAS.363.1211B} presented pioneering work in this area,
deriving predictions for GCS, stellar, and DM kinematics from galaxy merger
simulations.  We will make some comparisons of these predictions to the data.

The 10-meter Keck-I telescope with the LRIS multi-slit spectrograph \citep{1995PASP..107..375O}
has produced a legacy of the first large samples of extragalactic GC
kinematics outside of the Local Group 
(\citealt{1997ApJ...486..230C,2000AJ....119..162C,2003ApJ...591..850C}, hereafter C+03).
Now we present the first use of DEIMOS on Keck-II
\citep{2003SPIE.4841.1657F} to study extragalactic GCS
kinematics, representing a significant advance in this field.
Its 80~arcmin$^2$ field-of-view is one of the most expansive available on a large telescope,
and its stability, red sensitivity, and medium resolution make it highly efficient
for acquiring accurate velocities for GCs (which are intrinsically red objects) even
in gray skies, using the Ca~{\small II} absorption line triplet (CaT).
Our results here inaugurate
the SAGES Legacy Unifying Globulars and Galaxies Survey
of the detailed photometric and spectroscopic properties of
$\sim$~25 representative galaxies and their GCSs.

NGC~1407 is a moderately-rotating E0 galaxy, neither boxy nor disky, 
with a ``core-like'' central luminosity profile, a weak central AGN,
and absolute magnitudes of $M_B = -21.0$ and $M_K=-24.8$ (B+06b). 
Its GCS displays a typical color bimodality of ``blue'' and ``red'' GCs
(\citealt{1997AJ....113..895P};
\citealt{2006MNRAS.366.1230F};
\citealt{2006ApJ...636...90H}, hereafter H+06a;
\citealt{Harris09}).
Its overall specific frequency of GCs per unit luminosity was previously
found to be a modest $S_N \sim 4$, 
but a new deep wide-field study has produced a richer estimate of $S_N \sim 6$ 
(L.~Spitler et al., in preparation, hereafter S+09).
NGC~1407 hosts bright, extended X-ray emission that extends as far out as 80~kpc into the
surrounding group (B+06b; Z+07),
although there is nothing about its X-ray properties that marks it as a particularly
unusual galaxy (\citealt{2004MNRAS.350.1511O}, hereafter OP04).
We adopt a galaxy stellar effective radius of \Reff$=57''=$~5.8~kpc (see \S\ref{sec:assumpt}),
with a distance to the galaxy and group of 
20.9~Mpc \citep{2006MNRAS.366.1230F}---discussing
later the effects of the distance uncertainty.

We present the spectroscopic GC observations in \S\ref{sec:obs},
and convert these into kinematical parameters in \S\ref{sec:kin}.
In \S\ref{sec:mass}, we analyze the group mass profile using GC dynamics
and X-ray constraints, and review independent mass constraints more generally.
We place our findings for NGC~1407 into the wider context of galaxy groups
in \S\ref{sec:context}, and \S\ref{sec:conc} summarizes the results.

\section{Observations and Data Reduction}
\label{sec:obs}

Here we present the spectroscopic observations used in this paper,
including new data from Keck/DEIMOS, and published data from Keck/LRIS.
The DEIMOS observations are outlined in \S\ref{sec:deimosobs},
and the data reduction in \S\ref{sec:deimosred}.
The combined DEIMOS+LRIS sample is described in \S\ref{sec:comb}.
We also make use of integrated-light stellar kinematics data for NGC~1407 from 
\citet[hereafter S+08]{2008MNRAS.385..667S}
and photometric data of the galaxy and its GCS from S+08 ({\it HST}/ACS $BI$ imaging)
and from S+09 (Subaru/Suprime-Cam $g'r'i'$ imaging; \citealt{2002PASJ...54..833M}),

\subsection{Keck/DEIMOS observations}
\label{sec:deimosobs}

\begin{figure*}
\centering{
\includegraphics[width=6.0in]{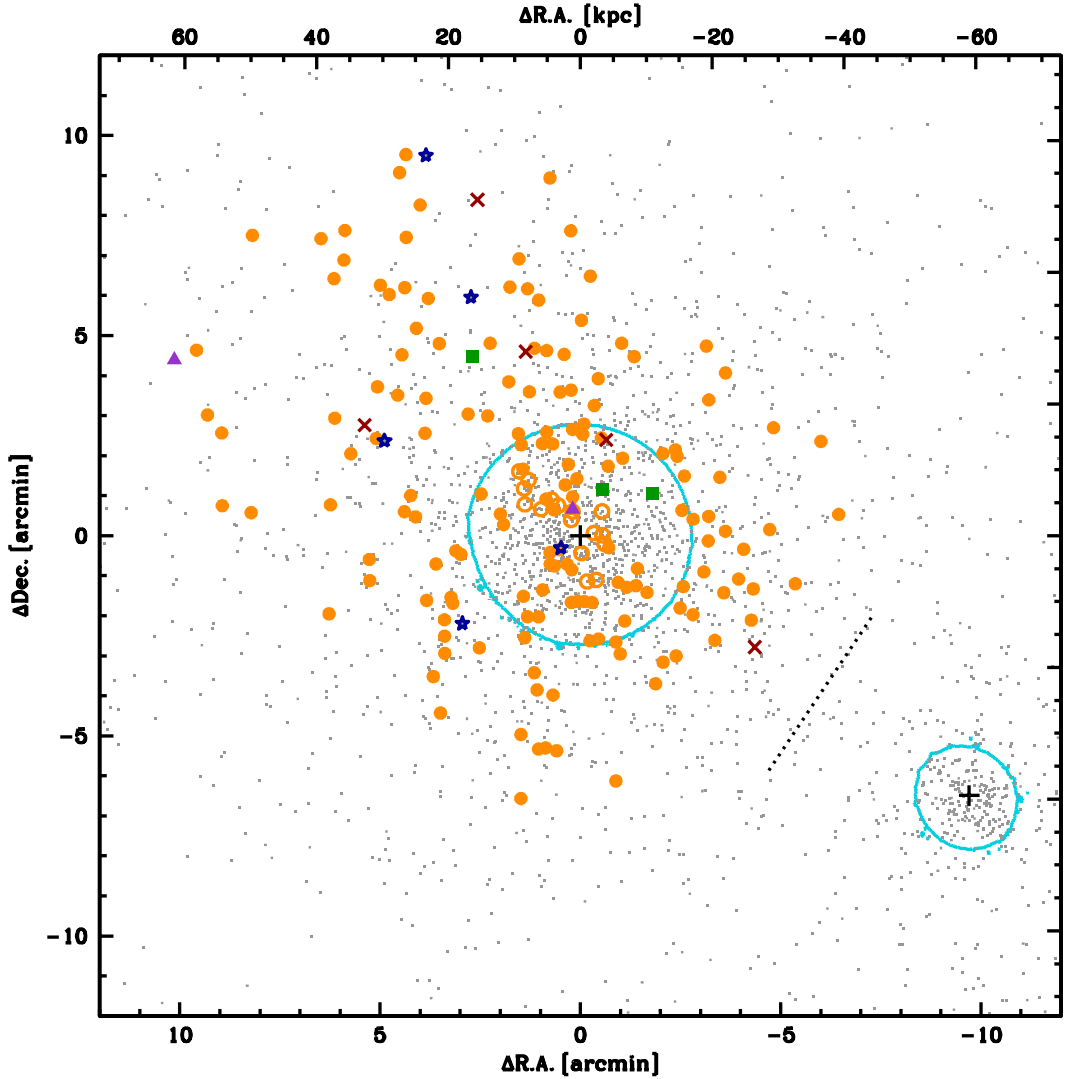}
}
\caption{
Overview of spectroscopic observations around NGC~1407.
Filled symbols show results from DEIMOS, and open symbols from LRIS.
Orange circles show positions of kinematically confirmed GCs,
and green squares show dwarf-globular transition objects.
Purple triangles show borderline velocity results, 
red $\times$ symbols mark background galaxies, and
Galactic stars are shown by blue starlike symbols.
Small gray points show the underlying distribution of GC candidates from Suprime-Cam,
down to the same limiting magnitude as the spectroscopically-confirmed GCs;
the irregular distribution of spectroscopic GCs is caused by mask and slit
placement and geometry, and by poor-quality regions in the imaging.
The central round contour shows the $\mu_I=23.2$ mag~arcsec$^{-2}$ stellar isophote of
NGC~1407, corresponding to $\sim 3$~\Reff{} ($2.85'$).
The lower-right contour shows a similar isophote for NGC~1400.
The dotted line shows the approximate boundary of predominance between the GCSs of
the two galaxies.
\label{fig:spat} 
}
\end{figure*}

The DEIMOS observations were carried out as part of Keck program
2006B\_U024D (PI: J.~P. Brodie).
Candidate GCs were selected from Subaru/Suprime-Cam images of NGC 1407
taken in $g'r'i'$ bands in very good seeing conditions (0.5$''$--0.6$''$).  
The NGC~1407 GC catalog is detailed in S+09 and was
supplemented with the NGC~1407 {\it HST}/ACS GC list of \citet{2006MNRAS.366.1230F}.
An astrometric solution was derived for the Subaru $g'$-band image using
unsaturated USNO-B stars in the field \citep{2003AJ....125..984M}.
Objects detected on the Subaru imaging were tagged as GC candidates if 
they met color limits of $0.4 < g'-i' \leq 1.4$ and 
$0.3 < g'-r' \leq 0.9$, and if the color values were jointly consistent 
with the clear linear sequence of GCs in $g'-i'$ vs. $g'-r'$ color space.

For the multi-slit mask design, GC candidates were ranked according to their 
expected $S/N$.  No cuts were made on GC color beyond the initial selection of
candidates. 
DEIMOS alignment and
guide stars were also taken from the USNO-B catalog, with a preference
for stars in the recommended magnitude range of $15 < R < 20$.  
To ensure a good alignment, at least two stars
were placed at each end of a given DEIMOS mask.
Science target slits were aligned close to the parallactic angle, 
and had lengths of at least 5$''$ each. Three masks
provided coverage of the NGC~1407 GC system out to large radii, and were
positioned to avoid potential contamination from the nearby galaxy
NGC~1400 to the southwest.
The GCS is fairly well sampled out to $\sim$~5$'$,
with an extension of $\sim$~10$'$ to the northeast (see Fig.~\ref{fig:spat}).

Spectroscopic observations were made with Keck-2/DEIMOS on 2006 Nov 19--21,
using the ``standard'' DEIMOS setup with the 1200 l/mm grating, centered on
$\sim 7500$ \AA. This central wavelength allowed coverage of the CaT 
at $\sim$~8500--8700 \AA{},
independent of slit position for GCs at the highest expected line-of-sight velocities in the NGC 1407
GC system ($\sim 2500$ km~s$^{-1}$), while capturing H$\alpha$ for a reasonable
fraction of the GCs observed. For this setup, our slit width of $1''$
gives a FWHM of $\sim 1.5$\AA{} ($\sim 50$ km~s$^{-1}$ at the CaT).
The three masks were observed for 1.5--2 hours each in individual
exposures of 30 min. The seeing on all three nights ranged from
0.55$''$ to 0.8$''$.
The large number of high-quality spectra that will we extract in \S\ref{sec:deimosred}
is testimony that
the $\sim 16'\times 5'$ field of view of DEIMOS, along with its Flexure Compensation
System, provide substantial improvements over Keck-1/LRIS ($\sim8'\times6'$ field,
not all of it usable for full spectral coverage).

\subsection{Keck/DEIMOS data reduction}
\label{sec:deimosred}

\begin{figure}
\includegraphics[width=3.4in]{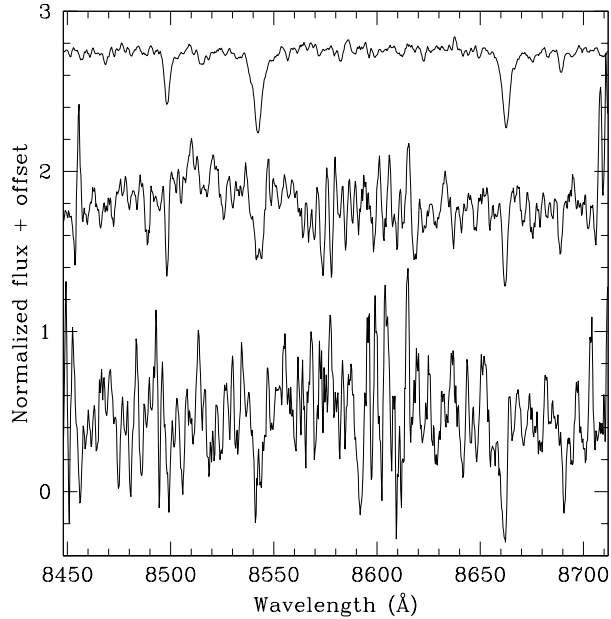}
\caption{
Extracted DEIMOS spectra of NGC~1407 GCs, shifted to zero redshift,
with the CaT region shown,
and with 5-pixel boxcar smoothing applied for visibility.
The top spectrum shows a best-case example (\#136205),
the middle one is typical (\#25581),
and the lower one is a minimally acceptable example (\#192009).
The rest-frame wavelengths of the visible Ca II absorption lines are
8498 \AA{}, 8542 \AA{}, and 8662 \AA{}.
\label{fig:spec}
}
\vspace{0.02cm} % because of running into the text
\end{figure}

The spectra were reduced using the IDL spec2d pipeline written for the
DEEP2 galaxy survey (J.~A. Newman et al., in preparation). 
Images were flatfielded using an internal quartz lamp,
and wavelength calibrated using ArKrNeXe arc lamps. After sky subtraction,
individual one-dimensional spectra were optimally extracted.
Some sample extracted spectra are shown in Fig.~\ref{fig:spec}.

Radial velocities were derived through cross-correlation around the CaT
region with a library of standard stars spanning a wide range of spectral types
(mid F to late K) taken both on this run and during previous DEIMOS runs with the
same setup. 
The derived velocities varied with spectral template choice by only $\sim$1--2~km~s$^{-1}$.
In all cases the best-fit line-of-sight velocity was verified by
visual inspection of both the CaT region and H$\alpha$ (where possible).
Small ($< 5$~km~s$^{-1}$) shifts in the zero-point velocities of the three masks
were corrected using the location of the telluric A band. 

\begin{figure}
\includegraphics[width=3.4in]{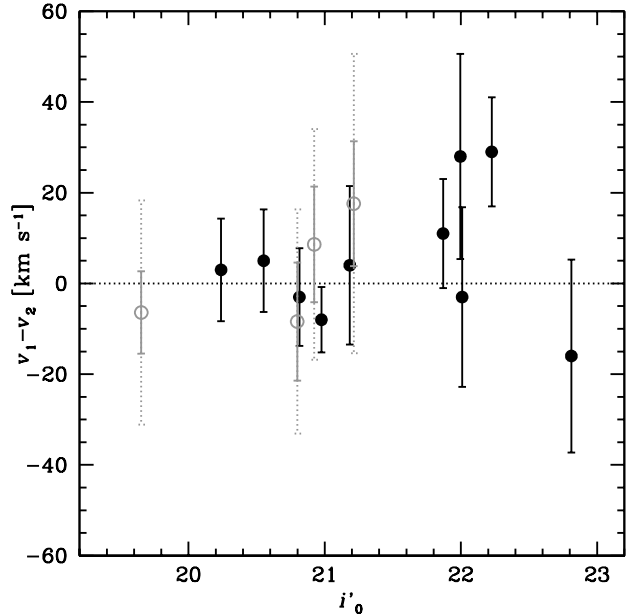}
\caption{
Comparisons of duplicate velocity measurements, as a function
of $i'_0$-band magnitude.
The black filled circles show repeated measurements using DEIMOS.
The gray open circles show duplicate measurements between LRIS and DEIMOS,
where we have corrected for a mean offset of 11~km~s$^{-1}$.
Error bars show the total estimated uncertainties.
In the case of LRIS, the solid bars show statistical uncertainties,
and dotted bars include systematic uncertainties.
\label{fig:comp} 
}
\end{figure}

The velocity uncertainties were estimated from the width of the cross-correlation peak, and
most fall within the range $\sim$~5--15~km~s$^{-1}$, depending on the brightness of the GC. 
To assess the accuracy of these uncertainties, we deliberately included
duplicate observations of GCs on different masks. 
In Fig.~\ref{fig:comp}, we show the results of the 10 repeated velocity measurements
(9 GCs and 1 star).
The velocities are nicely consistent 
within their stated uncertainties, to our faintest magnitude limit of $i'_0 \sim 23$.

Taking duplicates into account, we obtain velocities of 170 objects using DEIMOS,
and list their properties in Table~\ref{tab:data}.
We include two objects serendipitously obtained from mask alignment holes,
and two faint blue objects that are borderline cases in our conservative spectral
selection criteria.
These latter two (\# 131881 and \# 160394) will be tentatively incorporated in our analyses,
while checking that their inclusion does not affect any results.
For a further 40 objects---generally the
faintest in our masks---we could not obtain accurate line-of-sight velocities.
Many of these showed peaks in the cross-correlation consistent with
velocities expected for NGC 1407 GCs, but because of their low $S/N$ no
clear visual confirmation of the CaT could be ascertained. 

\begin{table*}
\caption{Photometry and spectroscopy of globular cluster candidates around NGC~1407.
}\label{tab:data}
\begin{center}
\noindent{\smallskip}\\
\begin{tabular}{l c c c c c c c c c r}
\hline \hline
ID & RA & Dec & $R_{\rm p}$ & PA & $i'_0$ & $(g'-i')_0$ & $(r'-i')_0$ & $v_{\rm p}$ & $\Delta v_{\rm p}$ & Type \\
   & [J2000] & [J2000] & [$''$] & [$^\circ$] & & & &  [km~s$^{-1}$] & [km~s$^{-1}$] \\
\noalign{\smallskip}
\hline
70163 & 03:40:14.257 & -18:41:32.54 & 406.1 & 175.1 & 22.94 & 0.77 & 0.22 & {\it 0.178?} & --- & gal \\ 
71768 & 03:40:18.080 & -18:41:21.86 & 403.9 & 167.3 & 21.93 & 1.08 & 0.36 & 1792 & 9 & GC \\ 
75581 & 03:40:08.063 & -18:40:55.52 & 371.4 & 188.2 & 21.64 & 1.00 & 0.31 & 1750 & 13 & GC \\ 
81722 & 03:40:14.309 & -18:40:10.38 & 324.4 & 173.7 & 22.59 & 0.85 & 0.29 & 1439 & 13 & GC \\ 
82094 & 03:40:16.221 & -18:40:07.78 & 326.0 & 168.9 & 22.38 & 1.24 & 0.45 & 1497 & 13 & GC \\
82266 & 03:40:15.476 & -18:40:06.01 & 322.3 & 170.7 & 22.16 & 0.82 & 0.25 & 1686 & 17 & GC \\
84789 & 03:40:18.073 & -18:39:45.97 & 311.1 & 163.4 & 22.35 & 1.16 & 0.39 & 1788 & 12 & GC \\
88899 & 03:40:26.549 & -18:39:13.77 & 338.5 & 141.8 & 21.31 & 1.07 & 0.36 & 1524 & 7 & GC \\
{\it etc.} \\
\hline
\hline
\end{tabular}
\end{center}
\tablecomments{
The first column gives the Suprime-Cam identification number, followed by
the LRIS identification number from C+07 in parentheses
where appropriate.
Tabulated properties include the projected distance from NGC~1407 ($R_{\rm p}$),
the position angle on the sky (North through East),
the $i'_0$-band magnitude, and two colors---as well as the
heliocentric line-of-sight velocity
$v_{\rm p} \pm \Delta v_{\rm p}$
(if a galaxy, the redshift $z$ is listed instead).
The final column indicates whether the object is classified as
a star, globular cluster (GC), 
intra-group globular cluster (IGC),
dwarf-globular transition object (DGTO), or background galaxy (gal);
objects with ``?'' have uncertain identifications, and objects in (parentheses)
have marginal signal-to-noise.
The photometric values have been corrected for Galactic extinction
\citep{1998ApJ...500..525S}.
The center of the galaxy NGC~1407 is at 
03:40:11.808 -18:34:47.92 (J2000),
and the systemic velocity is 1784~km~s$^{-1}$.
{\it This table is available in its entirety in the electronic
version of the article.}
}
\vspace{0.2cm} % because of running into the text
\end{table*}

Of the good spectra, 8 are clearly low-redshift galaxies (based on emission lines or
extendedness), and another 4
are Galactic stars (see \S\ref{sec:select} for star/GC demarcations).
There are 161 GC candidates confirmed to be kinematically associated with NGC~1407,
and since 7 of the stars and galaxies had non-GC-like colors but
were observed for lack of another target, our effective contamination rate was
a mere 3\%.
This high yield can be attributed to the richness of the NGC~1407 GCS, 
and to the excellent multi-color Suprime-Cam imaging and pre-selection.

One potential worry when observing the CaT is the flurry of bright sky lines 
in the wavelength vicinity.
However, we have checked carefully that either Ca2 or Ca3 is always clear of
a sky line (and usually Ca1 also), except for narrow velocity ranges around
$-150$~km~s$^{-1}$, $+225$~km~s$^{-1}$, and $+1550$~km~s$^{-1}$.
The latter value may be of some concern to this study, 
as we may selectively lose a few faint GCs around this velocity despite the
good sky subtraction of the DEIMOS pipeline.

\subsection{DEIMOS+LRIS combined sample}
\label{sec:comb}

To increase our sample size of GC velocities around NGC~1407,
we include the 20 velocities obtained by C+07 using Keck/LRIS.
These LRIS velocities also allow us to make an external check on our
DEIMOS velocities, using 4 GCs in common: see Fig.~\ref{fig:comp}.
Given the stated LRIS systematic uncertainties, the duplicate measurements are fully consistent, 
although different instruments, telescopes, and reduction procedures were used.
In fact, the systematic uncertainties of $\sim 20$~km~s$^{-1}$ seem to be overly conservative,
and we use only the statistical uncertainties, 
increasing them by 77\% and subtracting 11~km~s$^{-1}$
from the velocities in order to match the DEIMOS data.
The final LRIS uncertainties are typically $\sim 7$~km~s$^{-1}$.
In conclusion, the internal and external checks give us great confidence in the reliability of the 
Keck data.
Velocity accuracies of 5--10~km~s$^{-1}$ are not essential for the current study of NGC~1407, but they
do demonstrate that DEIMOS will be suitable for studying the dynamics of very low-mass galaxies.

After combining duplicate measurements, we arrive at a sample of 
190 GC candidate velocities, including 
172 GCs and 5 other objects associated with NGC~1407
(see \S\ref{sec:select} for selection).
These velocities, and associated source properties, are presented in
Table~\ref{tab:data}.

\section{Kinematics results}
\label{sec:kin}

Having assembled a catalog of GC candidates, we select the probable GCs
and examine their basic photometric and kinematical distributions in
\S\ref{sec:select}.
We study the rotation, velocity dispersion, and kurtosis properties of the GCS
in \S\ref{sec:rot}, \S\ref{sec:disp}, and \S\ref{sec:kurt}, respectively.

\subsection{GC selection and velocity distribution}
\label{sec:select}

\begin{figure}
\includegraphics[width=3.4in]{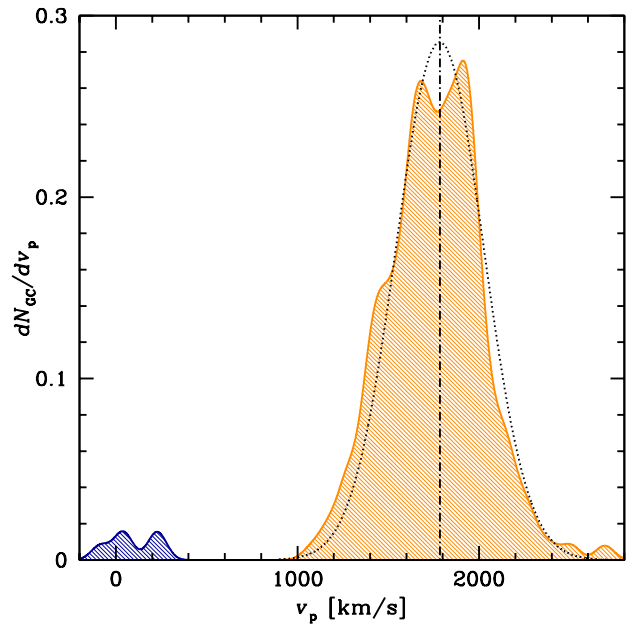}
\caption{
Distribution of GC candidates with velocity, smoothed by the measurement
uncertainties and an additional 50~km~s$^{-1}$ for clarity.
The dot-dashed line marks the systemic velocity of NGC~1407.
The dotted line shows a Gaussian with $\sigma_{\rm p}=$~241~km~s$^{-1}$.
\label{fig:losvd}
}
\end{figure}

\begin{figure*}
\center{
\includegraphics[width=5.0in]{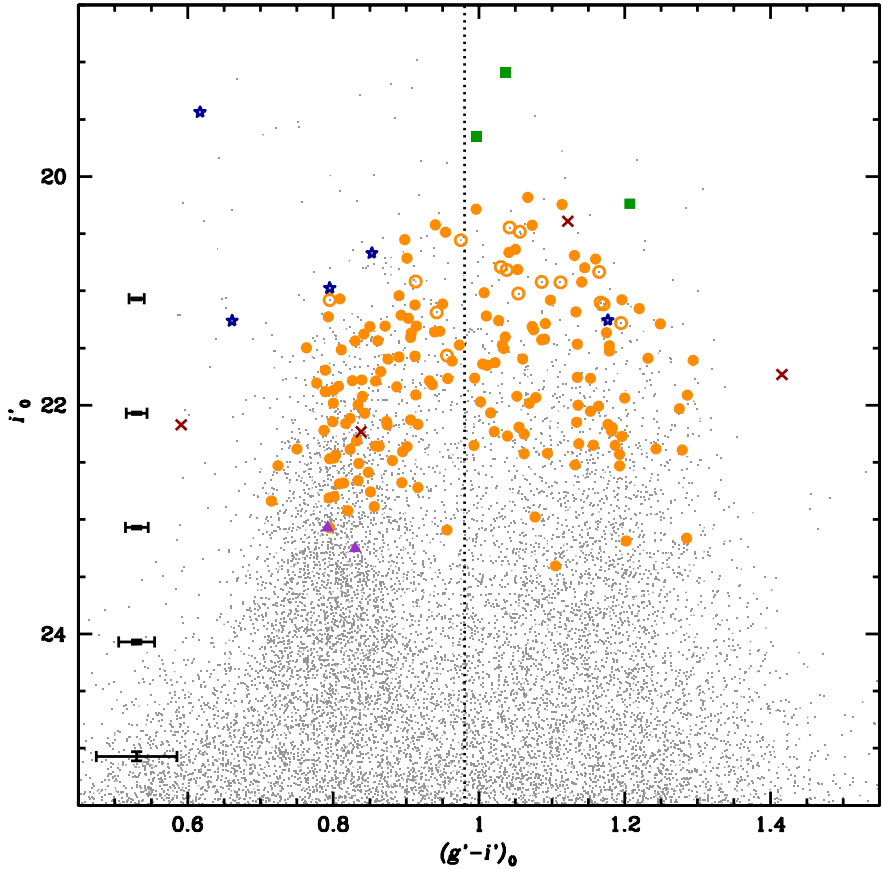}
}
\caption{
Color-magnitude diagram of GC candidates around NGC~1407.
The symbols are as in Fig.~\ref{fig:spat}, 
except now the small gray points show GC candidates only
within a 10$'$ radius of NGC~1407.
The typical photometric uncertainties are shown as error bars on the left.
The dotted line shows the adopted boundary between metal-poor and metal-rich GCs:
$(g'-i')_0=0.98$.
\label{fig:cmd} 
}
\end{figure*}

As all of the GC candidates have colors within the expected range,
we first select the GCs based on velocity.
This selection is trivial, since there is a large gap between the
lowest-redshift obvious GC (1085~km~s$^{-1}$) and the highest-redshift
obvious star (239~km~s$^{-1}$; see Fig.~\ref{fig:losvd}).
We see no objects obviously associated with NGC 1400 
(at a distance of $\sim 12'$ and with a systemic velocity of $\sim$~560~km/s), 
although it is in principle possible for some of the low-velocity NGC~1407 GCs
or the ``stars'' to be extreme-velocity GCs belonging to NGC~1400.
It is more probable that the density of NGC~1400 GCs is simply very low 
in our spectroscopically-surveyed region (Fig.~\ref{fig:spat} shows the
photometrically-estimated cross-over point between the GCS surface densities
of NGC~1407 and NGC~1400),
and further observations will be needed to determine the total radial extent of this galaxy's GCS.

We thus derive an initial sample of 177 GCs, including 2 cases that
are borderline because of their low $S/N$.
The color-magnitude diagram of the identified objects is shown in
Fig.~\ref{fig:cmd}, superimposed on the distribution of all
sources in the region from the Suprime-Cam imaging.  
Our spectroscopy samples the GCs about three magnitudes down the luminosity 
function, with almost equal numbers coming from the blue and the 
red subpopulations.
The dividing color has been set at $(g'-i')_0=0.98$ by inspection of the
color-magnitude diagram.
Assuming that the GCs are uniformly very old and that the color differences reflect
metallicity differences (as C+07 confirmed for a subsample of bright NGC~1407 GCs), 
we will in the rest of this paper refer to the blue and red GCs as ``metal-poor'' and ``metal-rich'',
respectively.
The GC color extrema and red/blue boundary of $(g'-i')_0=(0.70, 0.98, 1.30)$ 
correspond to [Fe/H]$\sim$~($-2.3$,$-1.1$,$+0.3$) for a 12 Gyr age
\citep{2003MNRAS.344.1000B},
although the absolute values of these metallicities should not be considered certain.

An immediately apparent feature of the spectroscopic GC sample is a ``blue tilt'' 
of the metal-poor peak---a systematic trend of color with GC luminosity that
has recently been discovered in many bright galaxies, including NGC~1407 
(e.g., \citealt{2006AJ....132.2333S}; H+06a; \citealt{2006ApJ...653..193M,2008MNRAS.389.1150S,Harris09}).
This is now one of several detections of
such a trend using ground-based photometry
(see also \citealt{2007MNRAS.382.1947F,2008ApJ...681.1233W}), which argues against
claims that the blue tilt is an artifact of {\it HST} photometry
\citep{2008AJ....136.1013K,2009ApJ...693..463W}.

Our data further support a
convergence of the GC color distribution (including the metal-rich GCs)
to a unimodal peak at high luminosities (H+06a).  
The bright end of the GCLF is in many galaxies known to include extended objects 
such as $\omega$~Cen, G1 in M31,
and ultra-compact dwarf galaxies (UCDs)--more generally ``dwarf-globular transition objects''
(DGTOs; e.g.,
\citealt{1999A&AS..138...55H};
\citealt{2000PASA...17..227D,2005ApJ...627..203H,2007MNRAS.378.1036E,2007ApJ...668L..35W}).
As the nomenclature implies, some of these objects may have origins and 
properties that are distinct from ``normal'' GCs; e.g., the DGTOs could be the
remnants of dwarf galaxies that were tidally stripped down to bare nuclei.
It has been suggested that NGC~1407 harbors a sizable population of extended ``GCs'' (H+06a),
and it is important for our current study to ascertain whether
our kinematics sample includes DGTOs, and whether their dynamics 
distinguishes them from the rest of the GCs.

We therefore re-examine the sizes of our candidate GCs in their original discovery images.
The ACS data (near the galaxy's center)
can discern the locus of GC sizes near $\sim$~3--4~pc, and
we do not support the conclusion of H+06a;
their ``extended'' objects seem to be simply examples of bright GCs with normal
sizes that are resolved at a distance of 20--30 Mpc.
We can identify only one spectroscopic object that is a clear outlier 
(\#136205, with \Reff$=$~8~pc), which we designate as a DGTO.
The Suprime-Cam data are much less well suited for measuring sizes without a
very careful PSF-based analysis.  Again, there is one clear outlier (\#160583),
which was used for DEIMOS mask alignment but turned out to have a velocity corresponding 
to the NGC~1407 group.  
It is a known galaxy candidate named APMUKS(BJ) B033807.82-183957.5 \citep{1990MNRAS.243..692M}, and
is well-resolved in the Suprime-Cam images, with \Reff~$\sim$~25~pc.
Its color of $(g'-i')_0=1.04$ and luminosity of $M_{i'_0}=-12.5$ make this object
likely to be a DGTO.

The two clear DGTOs around NGC~1407 are also the brightest two objects in our 
kinematics sample.  They have somewhat red colors, and the second-ranked object
was studied by C+07, who found nothing unusual about its spectroscopic properties
(age $\sim 11$~Gyr, [$Z$/H]$\sim -0.8$, [$\alpha$/Fe]$\sim +0.3$).
Another object, \#134086, corresponds to the ``galaxy''
APMUKS(BJ) B033759.53-184331.2, but is unresolved in the ACS images, 
and we retain it as a GC.  This gives us a first-pass sample of 175 GCs. 

\begin{figure}
\includegraphics[width=3.4in]{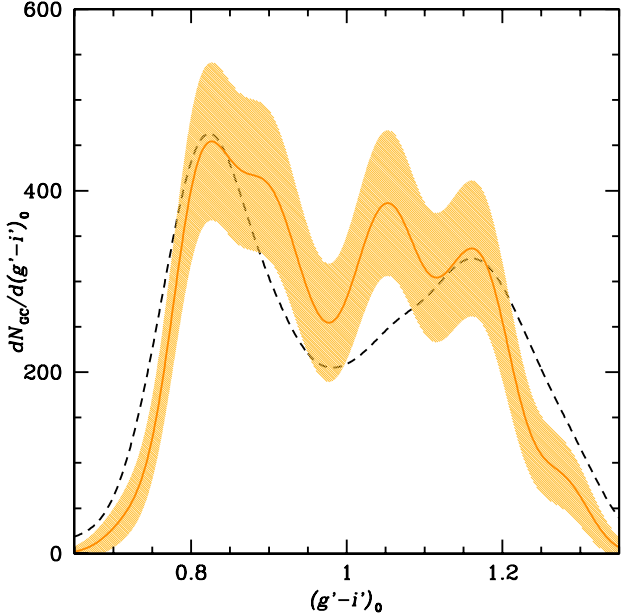}
\caption{
Color distribution of GCs around NGC~1407.
The solid curve shows the 172 spectroscopically-confirmed GCs,
with shaded region indicating the approximate 1~$\sigma$ 
uncertainties from Poisson noise.
The dashed curve shows the underlying Suprime-Cam distribution
of GC candidates with the same range of radius, colors, and magnitude,
and renormalized to match the total number of confirmed GCs.
For clarity, a smoothing kernel has been applied to these curves
with a width of $(g'-i')_0=0.03$, in addition to the measurement errors.
}
\label{fig:colhist} 
\end{figure}

The metal-poor and metal-rich GCs are sampled fairly 
equally at all radii,
which may be marginally inconsistent with the photometric inference that
the metal-poor system is more radially extended;
the inconsistency may be due to some hidden bias in the spectroscopic target selection.
The distribution of confirmed GC colors is compared to the underlying 
distribution in Fig.~\ref{fig:colhist}.
The distributions are very similar, except for an excess of mildly
metal-rich confirmed GCs that appears as a third color peak.  
This is probably an effect of 
the observational bias toward brighter objects, which show a convergence toward
color unimodality (cf. Fig.~\ref{fig:cmd})
and which we will explore further in S+09.
To a good approximation, we can conclude that the spectroscopically
confirmed GCs are representative of the underlying GCS, although in \S\ref{sec:rot} and
\S\ref{sec:disp} we will double-check the kinematics for any magnitude dependence.

The azimuthal distribution of GC measurements is statistically uniform over the
radial range of 106$''$--271$''$.
The mean and median values of the GC velocities in this annulus are
1769 and 1778~km~s$^{-1}$, respectively.
The mean velocity of NGC~1407 from LRIS long-slit stellar kinematics is
1784~km~s$^{-1}$ (after subtracting the LRIS/DEIMOS systematic offset), which we
adopt as the systemic velocity $v_{\rm sys}$.
For comparison, the NED\footnote{http://nedwww.ipac.caltech.edu/ .
The NASA/IPAC Extragalactic Database (NED) is operated by the Jet Propulsion Laboratory, California Institute of Technology, under contract with the National Aeronautics and Space Administration.}
 velocity is 1779 km~s$^{-1}$.

\begin{figure}
\includegraphics[width=3.4in]{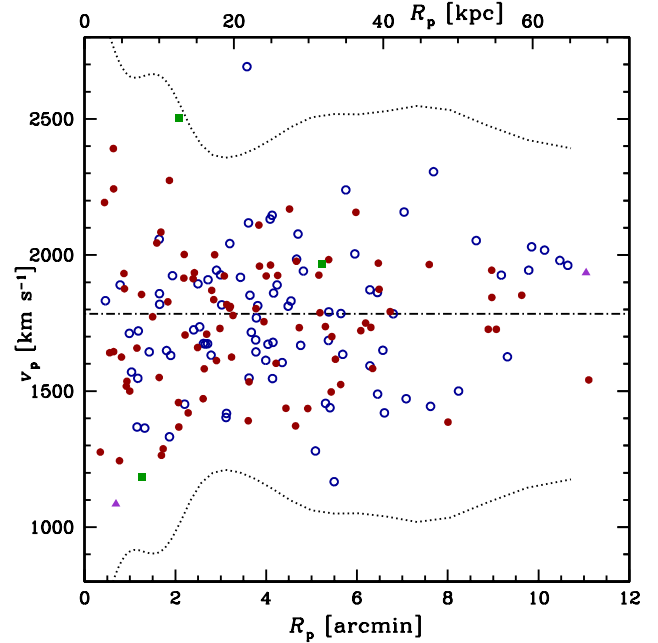}
\caption{
Velocity distribution of GC candidates with radius.
Blue open circles are blue (metal-poor) GCs, and red filled circles
are red (metal-rich) GCs.  The two purple filled triangles are
uncertain (metal-poor) GC detections.
The three DGTOs are shown by green squares.
The systemic velocity is marked by a dot-dashed line, and the
$\pm 3 \sigma$ ``envelope'' 
by dotted lines.
\label{fig:vels} 
}
\end{figure}

The overall distribution of GC candidate velocities was shown in
Fig.~\ref{fig:losvd}.  Given the uncertainties, the main peak of GCs is
consistent with a simple Gaussian profile.
We next inspect the velocity distribution with radius 
(Fig.~\ref{fig:vels}), seeing no obvious asymmetries.

\begin{table*}
\begin{center}
\caption{Kinematical properties of NGC~1407 GCS in various subsamples.}\label{tab:rot}
\noindent{\smallskip}\\
\begin{tabular}{l c c c c c c c c}
\hline
\hline
Subsample & $N_{\rm GC}$ & $v_{\rm rot}$ & $\theta_0$ & Significance & $\sigma_{\rm p}$ & $v_{\rm RMS}$ & $v_{\rm rot}/\sigma_{\rm p}$ & $\kappa_{\rm p}$ \\
& & [km~s$^{-1}$] & [$^\circ$] & & [km~s$^{-1}$] & [km~s$^{-1}]$ \\
\noalign{\smallskip}
\hline
\noalign{\smallskip}
{\it All radii}\\
All, 0.35$'$--11.1$'$ & 172 & $30^{+20}_{-30}$ & $47^{+55}_{-47}$ & 47\% & $240^{+14}_{-12}$ & $241^{+14}_{-12}$ & $0.13^{+0.08}_{-0.13}$ & $-0.32 \pm 0.37$ \\
Blue, 0.35$'$--11.1$'$ & 86 & $41^{+25}_{-41}$ & $129^{+49}_{-58}$ & 44\% & $232^{+20}_{-16}$ & $234^{+21}_{-16}$ & $0.18^{+0.11}_{-0.18}$ & $-0.32 \pm 0.51$ \\
Red, 0.35$'$--11.1$'$ & 86 & $58^{+32}_{-42}$ & $10^{+34}_{-43}$ & 68\% & $243^{+21}_{-16}$ & $247^{+21}_{-17}$ & $0.24^{+0.13}_{-0.18}$ & $-0.28 \pm 0.51 $ \\
\noalign{\smallskip}
\hline
\noalign{\smallskip}
{\it Small radii}\\
All, 0.35$'$--1.73$'$ & 33 & $99^{+62}_{-80}$ & $-178^{+37}_{-48}$ & 57\% & $291^{+44}_{-30}$ & $298^{+45}_{-31}$ & $0.34^{+0.22}_{-0.28}$ & $-0.61 \pm 0.80$ \\
\noalign{\smallskip}
\hline
\noalign{\smallskip}
{\it Azimuthally complete}\\
All, 1.77$'$--4.52$'$ & 75 & $79^{+29}_{-37}$ & $-12^{+24}_{-25}$ & 93\% & $201^{+19}_{-15}$ & $209^{+20}_{-15}$ & $0.35^{+0.15}_{-0.19}$ & $-0.38\pm0.55$ \\
Blue, 1.77$'$--4.52$'$ & 38 & $12^{+33}_{-12}$ & $-115^{+107}_{-98}$ & 3\% & $198^{+27}_{-19}$ & $198^{+28}_{-19}$ & $0.06^{+0.17}_{-0.06}$ & $-0.26 \pm 0.75$ \\
Red, 1.77$'$--4.52$'$ & 37 & $149^{+40}_{-49}$ & $0^{+17}_{-20}$ & 98\% & $190^{+27}_{-19}$ & $220^{+31}_{-22}$ & $0.79^{+0.22}_{-0.28}$ & $-0.41 \pm 0.76$ \\
\noalign{\smallskip}
\hline
\noalign{\smallskip}
{\it Large radii}\\
All, 4.55$'$--11.1$'$ & 64 & $71^{+34}_{-55}$ & $84^{+44}_{-32}$ & 74\% & $237^{+24}_{-19}$ & $242^{+25}_{-19}$ & $0.30^{+0.14}_{-0.23}$ & $-0.36 \pm 0.59$ \\
Blue, 4.55$'$--11.1$'$ & 36 & $112^{+47}_{-112}$ & $102^{+43}_{-33}$ & 78\% & $257^{+36}_{-26}$ & $272^{+38}_{-28}$ & $0.44^{+0.19}_{-0.44}$ & $-0.59\pm0.77$ \\
Red, 4.55$'$--11.1$'$ & 28 & $61^{+40}_{-61}$ & $11^{+46}_{-54}$ & 47\% & $194^{+32}_{-22}$ & $198^{+33}_{-22}$ & $0.32^{+0.21}_{-0.32}$ & $-0.24\pm0.86$ \\
\noalign{\smallskip}
\hline
\noalign{\smallskip}
{\it Outside center}\\
All, 1.77$'$--11.1$'$ & 139 & $52^{+26}_{-30}$ & $32\pm29$ & 83\% & $222^{+15}_{-12}$ & $225^{+15}_{-13}$ & $0.23^{+0.12}_{-0.13}$ & $-0.34\pm0.41$ \\
Blue, 1.77$'$--11.1$'$ & 74 & $54^{+29}_{-54}$ & $122^{+43}_{-45}$ & 62\% & $233^{+22}_{-17}$ & $237^{+22}_{-18}$ & $0.23^{+0.13}_{-0.23}$ & $-0.32\pm0.55$ \\
Red, 1.77$'$--11.1$'$ & 65 & $115^{+32}_{-38}$ & $2^{+17}_{-19}$ & 99\% & $194^{+20}_{-15}$ & $211^{+21}_{-17}$ & $0.59^{+0.17}_{-0.21}$ & $-0.39\pm0.59$ \\
\noalign{\smallskip}
\hline \hline
\end{tabular}
\end{center}
\end{table*}

There is one GC (\#111003) with an ``outlier'' velocity (2692~km~s$^{-1}$), 
well separated from the rest of the
distribution.  Calculating a smoothed profile of velocity dispersion with radius (see \S\ref{sec:disp}),
we define a 3~$\sigma$ envelope outside of which there is unlikely to be any object
(if the distribution is Gaussian).  The rogue GC is far outside of this envelope.
It is a fairly bright, metal-poor object, which is borderline extended in the
Suprime-Cam images, and whose velocity should be reliable.
We conjecture that it belongs to either a loose intra-group population of GCs rather than to the 
GCS immediately surrounding NGC~1407
(cf. the ``vagrant'' GC in the Fornax cluster; \citealt{2008A&A...477L...9S})---or
else to a sparse, dynamically distinct subpopulation of GCs 
that were propelled outward by three-body interactions, or were stripped from
infalling satellites
(e.g., \citealt{2007MNRAS.379.1475S,2009ApJ...692..931L}).
This object is omitted from the following kinematical analyses\footnote{Neither this
object nor any other GC candidate has a velocity that would place its pericenter
near the virial radius as found in \S\ref{sec:mass}, which implies that they are
all associated with the NGC~1407 group.}.

There is a second object (\#135251, 2505~km~s$^{-1}$) 
that lies just below the upper 3~$\sigma$ boundary.
It is one of the brightest and reddest GC candidates around NGC~1407, appearing
almost as an outlier in the color-magnitude diagram (Fig.~\ref{fig:cmd});
from the Suprime-Cam images, it may be unusually extended.
Noticing that one of the previously-identified DGTOs also had an extreme 
(blue-shifted) velocity, we suspect that there is a sizable population of
unrecognized DGTOs concentrated toward the galactic center, with a high velocity dispersion.
Indeed, a disproportionate share of the borderline extended objects seem to be
red objects with high relative velocities, residing at small radii.
Therefore we assign this third object to the DGTO category, and keep in mind that
the GCS kinematics inside $\sim$2$'$--3$'$ ($\sim$~15~kpc)
may still be ``contaminated'' by a subpopulation of DGTOs
(cf. the case of NGC~1399, where a dozen DGTOs have been found within $\sim$~15~kpc;
\citealt{2008MNRAS.389..102T}).

We also note that \#131881 has the lowest velocity in the sample (1085~km~s$^{-1}$), 
but we originally classified it as a borderline spectrum, so we remove it from our sample.
Our final sample of GCs thus has 172 objects. 

\subsection{Rotation}
\label{sec:rot}

As a rough measure of the amount of rotation in the NGC~1407 GCS, we perform
a least-squares fit to the velocities as a function of azimuth,
fitting the expression:
\begin{equation}
v(\theta) = v_{\rm sys} + v_{\rm rot} \sin (\theta_0 - \theta ) ,
\end{equation}
where $v_{\rm rot}$ is the rotation amplitude, and $\theta_0$ is the direction
of the angular momentum vector.
The physical interpretation of this model is discussed in
\citet[hereafter C+01; note the different sign convention]{2001ApJ...559..828C}.
C+01 and some other studies have used 
error-weighted least-squares rotation fitting (equivalent to $\chi^2$ fitting).
However, this approach is not so appropriate for a system with an 
intrinsic dispersion, and it is preferable to use 
straightforward least-squares fitting (equivalent to minimizing the 
rotation-subtracted dispersion).
The differences become important when 
the true underlying rotation is weak ($v_{\rm rot}/\sigma_{\rm p} \lsim 0.4$).
Further refinements are possible (e.g., \citealt{1998AJ....116.2237K,2007AJ....134..494W}),
but for now we will keep things simple using least-squares fitting.
The procedures we describe below for evaluating statistical uncertainties
and significance follow \citet{1998AJ....115.2337S}.

Using the full data set of 172 GCs,
we find $v_{\rm rot} = 30^{+20}_{-30}$~km~s$^{-1}$ and
$\theta_0 = 47^{+55}_{-47}$~degrees, 
where the uncertainties have been estimated through Monte-Carlo simulated recovery of
the best-fit models.
An additional test with random reshuffling of the GC data point position angles
shows a 53\% probability of a rotation amplitude this high happening just by chance,
i.e., there is no significant rotation detected.
The position angle is close to the galaxy's major axis (PA $\simeq$~55--65 degrees; S+08; S+09),
implying that the GCS may rotate {\it around} rather than {\it along} the major axis.

\begin{figure}
\includegraphics[width=3.4in]{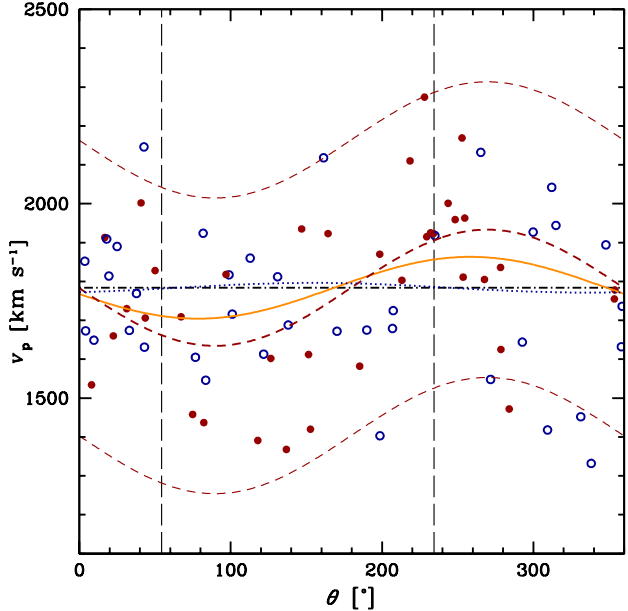}
\caption{
Rotation of NGC~1407 GCS, illustrated by velocities plotted against position angles.
The symbols are as in Fig.~\ref{fig:vels}, with GCs in 
{\it only the azimuthally-complete region} (1.77$'$--4.52$'$) plotted.  
The best-fit sine curves are shown for all GCs ({\it solid}),
metal-poor GCs ({\it dotted}), and metal-rich GCs ({\it dashed}).
For illustration, the $\pm$~2~$\sigma_{\rm p}$ boundaries of the metal-rich GCs
are also shown, corresponding to the dispersion of this subsample.
The horizontal line shows the systemic velocity, and the vertical lines mark
the galaxy photometric major axis
($\sim55^\circ$ at 3$'$; S+09).
\label{fig:rot}
}
\end{figure}

Before drawing any conclusions, we must recognize that the rotation field
can be complicated by a mixture of metal-poor and metal-rich GCs with
very different rotation patterns, by variations with radius, and
by potential DGTO contamination in the innermost regions.
Therefore we fit the rotation for different subsamples of metallicity
and radius, as summarized in Table~\ref{tab:rot}.
The most secure result is for the azimuthally-complete region of
$R_{\rm p}=$~1.77$'$--4.52$'$ (11--27 kpc), where there is highly significant rotation
close to the major axis,
produced by the metal-rich GCs (see Fig.~\ref{fig:rot}).
At larger radii, there is not a clear rotation signature for the metal-rich GCs,
but the metal-poor GCs appear to rotate around an orthogonal direction
intermediate to the major and minor axes.
The combination of the misaligned metal-rich and metal-poor rotation vectors
outside the center
produces a possible overall rotation around the major axis
(with $v_{\rm rot}/\sigma_{\rm p} \sim 0.1$; see 
\S\ref{sec:disp} for the calculation of the velocity dispersion
$\sigma_{\rm p}$).
Looking in more detail at the kinematics, the rotation of both
subpopulations is driven by the region $R_{\rm p} \sim 2'$--$6'$ (13--38~kpc).

In addition, we find evidence for a cold moving group of metal-poor GCs
at $\sim 10'$ ($\sim$~60~kpc) to the Northeast, with
$v_{\rm rot} = 224^{+13}_{-12}$~km~s$^{-1}$ and
$\sigma_{\rm p} = 14^{+10}_{-6}$~km~s$^{-1}$
(this feature is visible in Fig.~\ref{fig:vels}).
These GCs have a wide range of luminosities but a relatively narrow spread in
color [$(g'-i')_0 = 0.89\pm0.06$], suggesting a shared formational history.
One speculative interpretation is that the cold group traces
the recent disruption of a disk galaxy
(cf. \citealt{2008ApJ...684.1062F}).

We have further investigated dependencies of rotation on GC luminosity, and
while it at first glance appears that the brighter GCs rotate more strongly and with
a different axis than the faint GCs, on closer inspection, this may be simply due
to spatial biasing with magnitude of our selected GCs.
The re-introduction of the DGTO velocities (see \S\ref{sec:select})
would change some of the rotation vectors noticeably.
Given the focus of this paper on the mass analysis, we will defer closer examinations
of rotational properties to a future paper with a more complete data set.

We would like to make one final comparison, 
between the rotation of the GCs and the field stars in NGC~1407.  
The general hypothesis is that the metal-poor GCs in galaxies
are closely related to their metal-poor halo stars (which are observationally
inaccessible at the distance of NGC~1407), while the metal-rich GCs are
related to their old metal-rich bulge stars (which correspond to the bulk
of the diffuse starlight in an old elliptical like NGC~1407;
see \citealt{2005AJ....130.2140P,2007MNRAS.382.1947F};
\citealt{2008MNRAS.384.1231B}).
Unfortunately, the existing stellar kinematics data in NGC~1407
extend to only $\sim40''$ along the major axis (S+08), 
which is the region with only minimal numbers of measured GC
velocities, and furthermore with possible DGTO contamination.
Comparing the stellar and GC rotation fields will thus require
further efforts, either by extracting galaxy rotation directly
from the same DEIMOS slits used for the GCs 
(cf. \citealt{2008MNRAS.385...40N,2008MNRAS.385.1709P}; \citealt{Proctor09}),
or by observing planetary nebulae (PNe) as proxies for metal-rich
halo stars \citep{2006IAUS..234..341R}.
We do note that the field stars and metal-rich GCs in the overlap
region were found by \citet{2008MNRAS.385..675S} to have similar ages, metallicities,
and $\alpha$-element enhancements.
If the metal-rich GCs are directly associated with the field stars,
then the increase of $v_{\rm rot}$ from $\sim$~50~km~s$^{-1}$ in the
central parts (S+08) to $\sim$~100~km~s$^{-1}$ in the outer parts may support the
hypothesis that the ``missing'' angular momenta of elliptical galaxies 
(e.g., \citealt{1983IAUS..100..391F}) can be found in their halos 
(e.g., \citealt{1996ApJ...460..101W}; B+05).

\subsection{Velocity dispersion}
\label{sec:disp}

We construct a projected GCS velocity dispersion profile from these data
out to a distance of 60~kpc from the central galaxy.
Technically, we compute the RMS velocity profile, 
$v_{\rm RMS} \sim \sqrt{\Sigma_i (v_i-v_{\rm sys})^2/N}$,
which includes the (small) contribution from rotation, but we will 
refer to it freely as the ``velocity dispersion''.
The true dispersion $\sigma_{\rm p}$ is also calculated by
subtracting the contribution from the rotation (\S\ref{sec:rot}),
and both $v_{\rm RMS}$ and $\sigma_{\rm p}$ are 
shown in Table~\ref{tab:rot}.
We compute the binned dispersions and its uncertainties
using standard formulas \citep{1980A&A....82..322D}, and 
complementarily construct smoothed dispersion radial profiles
using maximum likelihood fitting to a Gaussian 
line-of-sight velocity distribution, weighted at each point by neighboring data
(see \citealt{2006A&A...448..155B}).
We also use this maximum likelihood technique to fit power-law functions to the dispersion
profiles:
\begin{equation}
v^2_{\rm RMS}(R_{\rm p}) = v^2_{\rm RMS,0} \times (R_{\rm p}/R_{\rm 0})^{-\gamma_{\rm p}}.
\label{eqn:PL}
\end{equation}

\begin{figure}
\includegraphics[width=3.4in]{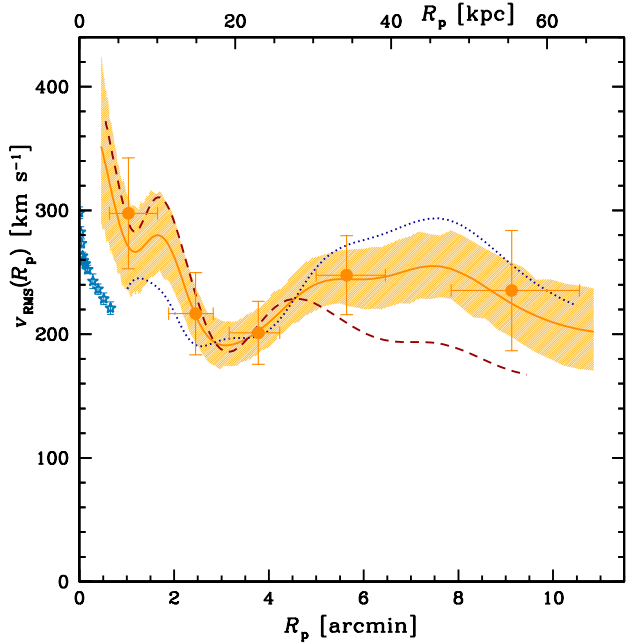}
\caption{
Projected RMS velocity profile of NGC~1407.
The circles with error bars show binned GCS values with no color selection.
The solid curve shows a smoothed GCS profile using a variable-width
Gaussian kernel (see \citealt{2006A&A...448..155B} for details).
The shaded region indicates the 1-$\sigma$ uncertainty boundaries
for the full GCS sample,
as computed by Monte Carlo simulations.
The dotted and dashed lines show the profiles for the metal-poor and
metal-rich GC subsamples, respectively.
The star symbols with error bars show long-slit stellar kinematics data from 
S+08, as updated in \citet{Proctor09}.
\label{fig:disp}
}
\end{figure}

With the full data-set of 172 GCs, there 
is a striking transition in the dispersion profile (Fig.~\ref{fig:disp})
from a rapid decline inside 1.8$'$
(11 kpc---confirming the finding of C+07) to a fairly constant value outside of this radius
($\gamma_{\rm p}=-0.07\pm0.26$).
There is also the hint of another decline outside of 7.5$'$ (50 kpc), but it is not significant.
The constancy of the dispersion profile out to $\sim$~10~\Reff{} is immediately a
strong indication of a massive DM halo around NGC~1407.

To better understand the dispersion profile, we analyze the metal-rich and metal-poor 
GCs separately (Fig.~\ref{fig:disp}).
The dispersions of the metal-rich and metal-poor clusters are similar over
the range of $R_{\rm p} \sim$~3$'$--4.5$'$ ($\sim$~20--30 kpc).
Inside this range, the metal-rich GCs have a significantly higher dispersion
than the metal-poor GCs, and at larger radii, their dispersion is lower.
This difference was already visible by eye in Fig.~\ref{fig:vels},
and translates into very different 
slopes with radius,
with $\gamma_{\rm p} = -0.26\pm0.23$ 
({\it increasing}) for the metal-poor GCs, 
and $\gamma_{\rm p} = +0.54\pm0.19$ 
({\it decreasing}) for the metal-rich GCs.

Such a difference in dispersion slope is {\it qualitatively} expected from the 
differences in spatial density profiles of the two subpopulations (the blue GCS is
more extended than the red GCS), 
but explaining this situation {\it quantitatively} may be a challenge
(investigated in \S\ref{sec:modalt}).
The steepness of the inner slope (driven by the metal-rich GCs) would
have been even steeper if the probable DGTOs had been left in the sample 
(see \S\ref{sec:select}), which we surmise may still include contamination from a population of as yet
unresolved, metal-rich DGTOs on fast orbits at small galactocentric radii
(we will return to this issue in \S\ref{sec:gckin}).
Within the uncertainties of our limited data set, we do not find any trend of
dispersion with GC luminosity.

Since we are unsure about the nature of this steep inner slope,
we will omit the inner dispersion data point from our mass profile analyses (\S\ref{sec:dyn}), 
assuming that it is affected by a distinct subpopulation of objects with very
different orbits than the rest of the GCs.
Note that the dispersion of the metal-rich GCs is 
inconsistent
with the field stars' dispersion in the narrow region of
overlap (Fig.~\ref{fig:disp}).

\subsection{Velocity kurtosis}
\label{sec:kurt}

Other measures of line-of-sight velocity distribution (LOSVD) 
beyond rotation and dispersion can
provide key information about the orbital structure.
In particular, measures of departures of the LOSVD from a Gaussian shape,
such as provided by the fourth-order velocity moment, are important diagnostics
that can be crudely used to distinguish between radial and tangential orbits.
Here we compute a basic projected reduced kurtosis statistic 
that is approximated by:
\begin{equation}
\kappa_{\rm p} = \left[\frac{1}{N} \sum_{i=1}^{i=N} \frac{(v_i - v_{\rm sys})^4}{v_{\rm RMS}^4}\right] - 3 \pm
 \sqrt{\frac{24}{N}} ,
\end{equation}
with a more complicated bias-corrected expression for small-number statistics
given by $G_2$ in \citet{Joanes98}\footnote{In principle, $\kappa_{\rm p}$
can be compromised by velocity measurement uncertainties, 
rotation, and small-number statistics.  However, we have modeled these
effects analytically and with Monte Carlo simulations, and determined that
they are negligible for this data-set, except for the small number statistics which
may bias the results by $\sim -0.1$.}.
For a Gaussian LOSVD, $\kappa_{\rm p}=0$, which describes isotropic orbits in a
logarithmic potential.
The kurtosis is only useful for regions over which $v_{\rm RMS}$ is
approximately constant, so we exclude the inner regions of the GCS 
($R_{\rm p} < 1.73'$) from our analysis.

For the overall GCS outside the central regions, we find $\kappa_{\rm p}=-0.30 \pm 0.41$,
i.e. a suggestion of negative kurtosis.
The ``flat-topped'' behavior of the platykurtic LOSVD can be marginally 
seen by eye in Figs.~\ref{fig:losvd} and \ref{fig:vels}, 
where the velocities do not peak strongly around $v_{\rm sys}$ 
(particularly at large radii).
The LOSVD is still statistically consistent with being Gaussian
(the kurtosis estimator is unfortunately not a very powerful measure of
LOSVD shape), but a strongly-peaked (leptokurtic) distribution {\it does}
appear to be ruled out, with important implications for the orbits as we will
see in the next section.
For the metal-poor GCs, we find $\kappa_{\rm p}=-0.24 \pm 0.55$, and
for the metal-rich GCs, we find $\kappa_{\rm p}=-0.30 \pm 0.59$,
so there is no overall difference detected in the LOSVD shapes between the
two subsystems.

\begin{figure}
\includegraphics[width=3.4in]{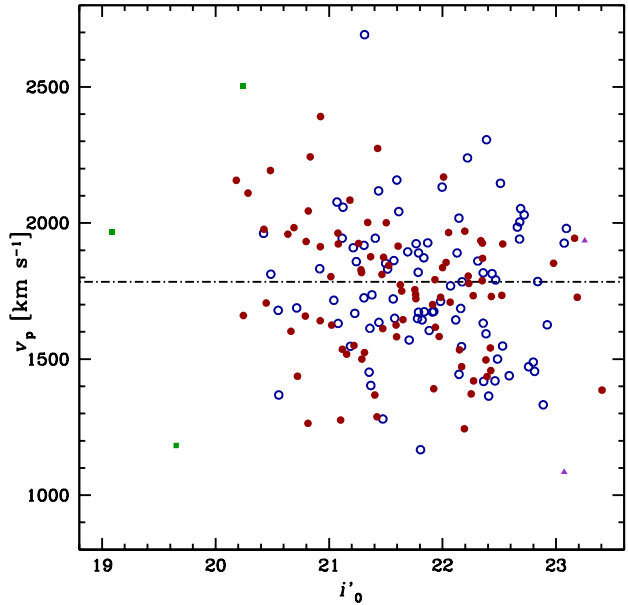}
\caption{
Distribution of GC candidate velocities versus magnitude.
Symbols are as in Fig.~\ref{fig:vels}.
Note the paucity of bright GCs near the systemic velocity.
\label{fig:kurt}
}
\vspace{0.02cm} % because of running into the text
\end{figure}

We would like to further consider the behavior with radius $\kappa_{\rm p}(R_{\rm p})$
but the data are inadequate for detecting significant trends.
We do see signs of the kurtosis becoming stronger with GC luminosity:
see Fig.~\ref{fig:kurt}, where there appears to be a dearth of velocities
near $v_{\rm sys}$ for $i'_0 \lsim 21.3$.
This effect is driven by bright, central objects, and does {\it not}
impact the result of negative kurtosis outside the central regions.

One caveat about our entire kurtosis analysis is that a Gaussian LOSVD was
assumed in \S\ref{sec:select} to remove outliers; 
for example, restoring \#111003 to the data-set
would imply a higher $v_{\rm RMS}$ and a significantly {\it positive} $\kappa_{\rm p}$.
More robust measures of LOSVD shapes and outlier removal will eventually require
detailed self-consistent dynamical modeling.
We will return to the dynamical implications of the kurtosis in \S\ref{sec:modalt},
and to the issue of the luminosity dependence in \S\ref{sec:gckin}.

We also note that there are no faint GCs ($i'_0 > 22.5$)
with $v \simeq 1550$~km~s$^{-1}$,
which may be due to systematic loss near this velocity from sky line collisions
(\S\ref{sec:deimosred}).
But even if this is the case, we would expect to have lost only about one GC
to the effect, which should not compromise our results.
Additionally, a few GC candidates near the extreme mask edges experienced wavelength-dependent
vignetting that was not recognized during the original mask design.
However, we have verified that this issue would not have caused any
velocity bias in the recovered GC spectra.

\section{NGC 1407 mass profile}
\label{sec:mass}

Reaching the main focus of our paper, we consider various aspects of
the mass profile of the NGC~1407 system.
In \S\ref{sec:dyn}, we model the dynamics of the GCS to derive
a set of permitted mass models.
In \S\ref{sec:xray}, we present an X-ray-based mass analysis, and
in \S\ref{sec:indep} we compare independent mass constraints in NGC~1407 and in other galaxies.

\subsection{GCS dynamics}
\label{sec:dyn}

Here we model the dynamics of the GCs around NGC~1407, beginning with
a summary of the modeling assumptions in \S\ref{sec:assumpt},
then results with no DM in \S\ref{sec:noDM}, and a $\Lambda$CDM-motivated
model in \S\ref{sec:fid}.
Alternative models that better reproduce the data are explored in
\S\ref{sec:modalt}.

\subsubsection{Modeling assumptions}\label{sec:assumpt}

Our approach is to construct a sequence of cosmologically-motivated
galaxy+halo mass models,
calculate their projected GCS RMS velocity profiles $v_{\rm RMS}(R_{\rm p})$,
and compare these predictions to the data.
For this initial dynamical model of the GCS of NGC 1407, we make
several simplifying assumptions.
First is that the system is spherically symmetric.
The NGC~1407 outer galaxy isophotes have an ellipticity of $e \sim 0.05$ (S+08),
supporting sphericity as a reasonable approximation---which seems to be the
case in general for massive ellipticals at group and cluster centers, although
flattening may become important at radii of $\gsim$~20~kpc
\citep{1991AJ....101.1561P,1996ApJ...461..146R}.

Secondly we assume a particular form for the GCS velocity dispersion
tensor that is nearly isotropic near the central galaxy,
and becomes increasingly radially-anisotropic at larger 
galactocentric radii (details below).
We also assume that the system is in dynamical equilibrium and that the solutions to the
Jeans equations are physical and stable
(these assumptions are supported by analyses of simulated
DM halos with no recent major merger; 
\citealt{2005MNRAS.364..367D}, hereafter D+05).

Thirdly we represent the cumulative mass profile $M(r)$ by a constant mass-to-light
ratio ($\Upsilon_*$) \citet{1968adga.book.....S}
model for the galaxy's field stars, plus a DM halo characterized by an ``NFW'' profile, 
expressed as a circular velocity that is independent of distance and
bandpass:
\begin{eqnarray}
v^2_{\rm c}(r) &\equiv& \frac{G M(r)}{r} \nonumber \\
&=& \frac{G\Upsilon_* L_*}{r} \times \left\{ 1- \frac{\Gamma\left[(3-p)m,(r/a_S)^{1/m}\right]}{\Gamma\left[(3-p)m\right]}\right\}  \label{eqn:NFW} , \nonumber \\
&&+ v_s^2 \left[ \frac{r_s}{r} \ln(1+r/r_s)-\frac{1}{1+r/r_s} \right] , 
\end{eqnarray}
where $m$ and $a_{\rm s}$ are the S\'ersic index and scale radius,
and $p$ is a function of $m$ (\citealt{2005MNRAS.363..705M}, hereafter M{\L}05),
while $v_s$ and $r_s$ are the characteristic velocity and spatial scales for
the DM halo (\citealt{1996ApJ...462..563N}).
The $v_{\rm c}$ profile of the halo reaches a maximum of $0.465 v_s$
at a radius of $2.16 r_s$.

Our S\'ersic stellar mass model is obtained by fitting a preliminary combined photometric
profile ($I$-band ACS $+$ $i'$-band Suprime-Cam) outside of a radius
of 1$''$, resulting in fit parameters corresponding to
($m=4.32$, \Reff$=57'' = 5.8$~kpc, $M_I=-23.3$).
As is typical for giant ellipticals, these values are fairly uncertain
(total luminosity by at least 20\%, and \Reff{} by at least 30\%)
because of the large fraction of light contained in the faint stellar halo.
One obtains a far superior fit to the galaxy light using a Nuker model \citep{1995AJ....110.2622L}
or a core+S\'ersic model (e.g., \citealt{2004AJ....127.1917T}),
but the former does not converge to a finite luminosity, and the latter is
not simple to incorporate in the modeling equations.
At the radii corresponding to our GC velocity measurements (to $\sim 10'$), the total luminosity
of the S\'ersic model deviates from that of the core+S\'ersic model by up to 15\%, but as we will
see, the total mass in this region is dominated by DM, and in any case, the
luminosity profile is not well constrained outside of $2'$.

The galaxy mass profile is uniquely determined by the stellar mass-to-light ratio
$\Upsilon_*$, which we estimate from
a stellar populations analysis (Z+07) to be $\Upsilon_{*,I}=2.12$~$\Upsilon_{\odot,I}$ 
($\Upsilon_{*,B}=4.45$~$\Upsilon_{\odot,B}$), assuming a Kroupa IMF\footnote{This
value for $\Upsilon_*$ may seem low relative to historical findings in
other bright ellipticals
[e.g., $\Upsilon_{*,I}\sim$~3~$\Upsilon_{\odot,I}$ in \citet{2006MNRAS.366.1126C}, and
$\Upsilon_{*,B}\sim$~6~$\Upsilon_{\odot,B}$ in \citet{2001AJ....121.1936G}],
and for NGC~1407 in particular, the stellar populations models of
\citet[hereafter H+06b]{2006ApJ...646..899H} imply
$\Upsilon_{*,B}=5.7\pm0.5$~$\Upsilon_{\odot,B}$.
However, the Z+07 value is meant to characterize the stars outside of 
0.5~\Reff{} rather than in the oft-studied central regions
($\sim 0.1$~\Reff{} in the case of H+06b);
another study averaging over 1~\Reff{} found $\Upsilon_{*,B}\sim 4$ for NGC~1407
\citep{2009arXiv0901.3781T}.
The full range of possible values Z+07 found for different radii and modeling assumptions
was $\Upsilon_{*,B}=3.8$--7.1~$\Upsilon_{\odot,B}$ for a Kroupa IMF, and
up to 11.2~$\Upsilon_{\odot,B}$ for a Salpeter IMF.}.
The adopted NGC~1407 galaxy mass is thus $M_* = 2.0\times 10^{11} M_{\odot}$;
we will examine the impact of $\Upsilon_*$ on our results in \S\ref{sec:modalt} and \S\ref{sec:exX}.

\begin{figure}
\includegraphics[width=3.4in]{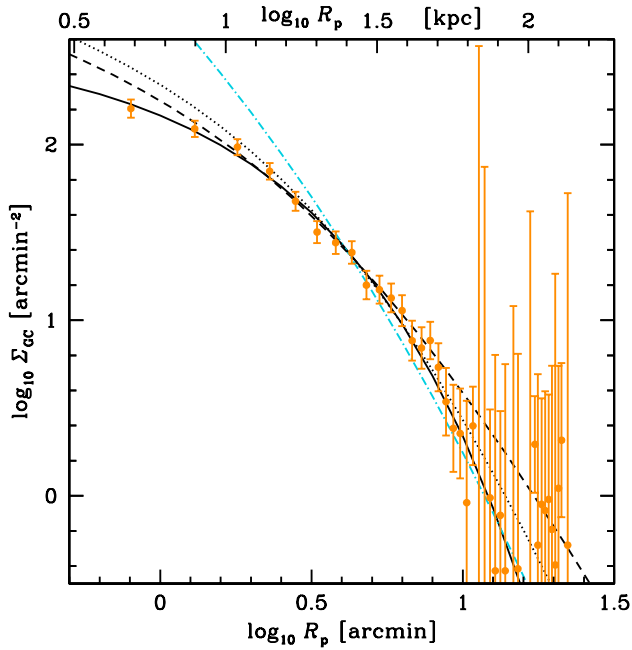}
\caption{
NGC~1407 GCS surface density profile.
Points with error bars show Suprime-Cam data with a background contamination
level subtracted (S+09).
The solid curve is our best-fit S\'ersic model.
For comparison, the NGC~1407 galaxy light model is shown as a blue
dot-dashed line (with arbitrary normalization).
The other curves show profiles corresponding to the $\Lambda$CDM simulations of D+05,
for overdensities of 2.5~$\sigma$ ({\it dashed}) and 4~$\sigma$ ({\it dotted}).
\label{fig:gcs}
}
\end{figure}

The density profile for the GCS is obtained by a S\'ersic model fit 
to preliminary surface density data $\Sigma_{\rm GC}(R_{\rm p})$ from Suprime-Cam, 
over the radial range of 0.8$'$--23$'$ (S+09).
As shown in Fig.~\ref{fig:gcs},
the fit is characterized by the parameters 
($m=1.39\pm0.09$, \Reff$=3.52'\pm0.07'=21$~kpc), demonstrating
that the radial extent of the GCS of NGC~1407 is far greater than that of its field starlight.
Dividing the GCS into subpopulations, the metal-poor component is more extended than
the metal-rich, with $(m \sim 1.3$, \Reff~$\sim 4.3')$
and $(m \sim 1.2$, \Reff~$\sim 2.8')$, respectively.
We have also verified that these density fits are fair representations of
the bright GC subpopulation probed in our kinematics sample $(i'_0 < 23)$.
The uncertainty in the GCS density profile is not explicitly
included in our dynamical modeling, but we will explore the 
effects of reasonable variations in this profile as needed.

The model GCS dispersion profile $v_{\rm RMS}(R_{\rm p})$
is calculated with the spherical Jeans equations
as summarized in M{\L}05\footnote{We use the
\citet{1997A&A...321..111P} function $p(m)$ in the 
GCS density profile since in this case it appears to 
provide a better match than the \citet{1999MNRAS.309..481L} 
function to the S\'ersic surface density profile.}.
In using these equations, we are also assuming that
the net rotation can be simply folded into $v_{\rm RMS}$ along
with the velocity dispersion, since the observed rotation in NGC~1407 is not
dynamically important
($v_{\rm rot}/\sigma_{\rm p} \sim 0.4$: see \S\ref{sec:rot}).
We generally attempt to fit the dispersion
data outside of 1.8$'$ only, since the central values may be
skewed by DGTO contamination (see \S\ref{sec:disp}).
We will return to this assumption in \S\ref{sec:modalt}.

\subsubsection{Constant mass-to-light ratio models}\label{sec:noDM}

\begin{figure}
\includegraphics[width=3.4in]{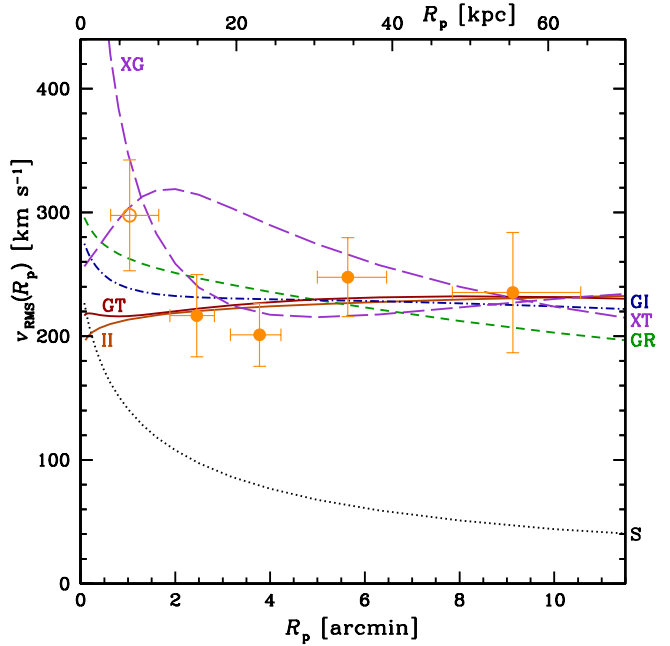}
\caption{
NGC~1407 GCS projected RMS velocity profile.
Points with error bars show the measurements, while curves show model predictions
that are described in more detail in the main text.
The black dotted curve (S) shows a stars-only galaxy (no DM).
The green short-dashed curve (GR) is a model with a $\Lambda$CDM halo and a
theoretically-motivated GC anisotropy profile (Eq.~\ref{eqn:beta}).
The dark blue dot-short-dashed curve (GI) is a similar model with an isotropic GC system.
The orange solid curve (II) is a power-law fit to the data.
The red solid curve (GT) is a $\Lambda$CDM halo with tangential anisotropy.
The purple long-dashed curves come from the X-ray based mass model:
with either $\beta=-0.5$ (XT) or an anisotropy that varies from $\beta=1$ in
the center to $\beta=-\infty$ at large radii (XG; see Fig.~\ref{fig:beta}).
\label{fig:models1} 
}
\vspace{0.2cm} % because of running into the text
\end{figure}

Our first model is a galaxy with no DM halo, or a ``constant-$\Upsilon$'' model;
see Table~\ref{tab:models} for the details of the main models
that we will discuss.
In this case we keep things simple by adopting an isotropic
velocity dispersion tensor: 
$\beta(r)=0$, where $\beta\equiv 1- \sigma_\theta^2/\sigma_r^2$
specifies the bias of specific kinetic energy in the tangential
direction $\sigma^2_\theta$ relative to the radial direction 
$\sigma^2_r$.
This model's predicted $v_{\rm RMS}(R_{\rm p})$ for the GCs 
in the outer parts is much lower than the observed values
[see model S in Fig.~\ref{fig:models1}, with comparison data corresponding
to the $v_{\rm RMS}(R_{\rm p})$ profile from \S\ref{sec:disp}].
In principle, the data could be roughly fitted by setting the
distance to 40~Mpc, the stellar mass-to-light ratio to 
$\Upsilon_{*,I} \sim$~7~$\Upsilon_{\odot,I}$ 
($\Upsilon_{*,B} \sim$~15~$\Upsilon_{\odot,B}$),
and the outer GC orbits to nearly circular ($\beta \sim -10$),
but this solution is highly contrived and implausible.

\begin{table*}
\begin{center}
\caption{Summary of noteworthy mass models for the NGC~1407 group.}\label{tab:models}
\noindent{\smallskip}\\
\begin{tabular}{l c c c c c c c c r r c }
\hline
Label & $\Upsilon_*$ & $\beta(r)$ & $v_s$ & $\rho_s$ & $r_s$ & $r_{\rm vir}$ & $c_{\rm vir}$ & $M_{\rm vir}$ & $\tilde{\chi}^2_{\rm GC}$ & $\tilde{\chi}^2_{\rm X-ray}$ & Model description \\
& [$\Upsilon_{\odot,B}$] & & $\left[\frac{\rm km}{\rm s}\right]$ & $\left[\frac{10^6 M_\odot}{{\rm kpc}^3}\right]$ & [kpc] & [Mpc] & & [$10^{13} M_{\odot}$] & ($N_{\rm dof}$) & ($N_{\rm dof}$) \\
\hline
\noalign{\smallskip}
S & 4.45 & 0 & --- & --- & --- & 0.18 & --- & 0.03 & 43.5 (4) & 46.0 (7) & Constant-$\Upsilon$ \\
\noalign{\smallskip}
GR & 4.45 & Eq.~\ref{eqn:beta} & $1600^{+600}_{-400}$ & $0.83^{+1.23}_{-0.50}$ & $230^{+290}_{-110}$ & $1.35^{+0.65}_{-0.37}$ & $5.8^{+2.7}_{-2.0}$ & $14.2^{+31.9}_{-8.7}$ & 2.6 (2) & 10.9 (7) & Radial $\Lambda$CDM \\ 
\noalign{\smallskip}
GI & 4.45 & 0 & $1300^{+600}_{-200}$ & $1.2^{+1.7}_{-0.8}$ & $170^{+200}_{-80}$ & $1.13^{+0.52}_{-0.28}$ & $6.7^{+3.1}_{-2.3}$ & $8.4^{+17.7}_{-5.0}$ & 1.1 (2) & 11.1 (7) & Isotropic $\Lambda$CDM \\ 
\noalign{\smallskip}
II & --- & 0  & --- & --- & --- & --- & --- & --- & 0.8 (2) & 81.7 (7) & Isotropic GC inversion \\
\noalign{\smallskip}
GT & 4.45 & $-0.5$ & $1200^{+400}_{-200}$ & $1.5^{+2.1}_{-1.0}$ & $140^{+160}_{-70}$ & $1.00^{+0.45}_{-0.24}$ & $7.3^{+3.4}_{-2.5}$ & $5.7^{+12.1}_{-3.2}$ & 0.8 (2) & 11.6 (7) & Tangential $\Lambda$CDM \\ 
\noalign{\smallskip}
XR & 4.45 & Eq.~\ref{eqn:beta} & $1300\pm100$ & $11^{+4}_{-3}$ & $52^{+13}_{-9}$ & $0.88^{+0.09}_{-0.06}$ & $17\pm2$ & $3.9^{+1.3}_{-0.9}$ & 8.4 (4) & 0.8 (5) & X-ray based NFW  \\
\noalign{\smallskip}
XT & --- & $-0.5$ & --- & --- & --- & $0.76^{+0.44}_{-0.04}$ & --- & $2.6^{+7.5}_{-0.5}$ & 5.6 (4) & 0.0 (2) & Tangential X-ray based \\
\noalign{\smallskip}
XG & --- & Fig.~\ref{fig:beta} & --- & --- & --- & $0.76^{+0.44}_{-0.04}$& --- & $2.6^{+7.5}_{-0.5}$ & 0.7 (4) & 0.0 (2) & GC/X-ray consonance \\ 
\noalign{\smallskip}
C & 4.45 & $-0.5$ & $1300\pm200$ & $2.4^{+1.8}_{-0.9}$ & $120\pm40$ & $1.07^{+0.18}_{-0.17}$ & $9.1^{+2.3}_{-1.7}$ & $7.1^{+4.1}_{-2.8}$ & 4.0 (2) & 7.2 (5) & Consensus  $\Lambda$CDM \\ 
\noalign{\smallskip}
GG & 4.45 & $-0.5$ & $1400^{+200}_{-100}$ & $0.71^{+0.60}_{-0.28}$ & $230^{+100}_{-70}$ & $1.26\pm0.21$ & $5.4^{+1.6}_{-1.1}$ & $11.5^{+6.7}_{-4.7}$ & 0.7 (2) & 14.0 (7) & GCs + galaxies\\ 
\noalign{\smallskip}
\hline
\hline
\end{tabular}
\end{center}
\tablecomments{
The characteristic NFW halo density is
$\rho_s \equiv v_s^2/(4\pi G r_s^2)$.
The reduced $\chi^2$ statistic, 
$\tilde{\chi}^2 \equiv \chi^2/N_{\rm dof}$,
is provided along with the number of degrees of freedom $N_{\rm dof}$.
}
\vspace{0.2cm} % because of running into the text
\end{table*}

The discrepancy between the velocities measured and predicted for a simple stellar galaxy
is strong evidence for either non-classical gravity behavior 
(e.g., \citealt{2006MNRAS.367..527B,2008MNRAS.387.1470A}),
or for an unseen mass component---or for both.  
It is beyond the scope of this paper to treat alternative gravities,
so we will simply presume to be witnessing the effects of a massive DM halo.
Note that the missing mass is not attributable to the hot gas in the system,
whose mass of $\sim 10^{10} M_{\odot}$ (Z+07) is less important than 
even the stellar component.

\subsubsection{Fiducial $\Lambda$CDM models with anisotropy}
\label{sec:fid}

We next examine a fiducial set of models including DM halos.
These models are parameterized by the virial mass $M_{\rm vir}$
and the halo concentration $c_{\rm vir}\equiv r_{\rm vir}/r_s$, which are correlated
according to the predictions of $N$-body cosmological simulations by the relation:
\begin{equation}
c_{\rm vir} = 17.88 \times (M_{\rm vir}/10^{11} M_{\odot})^{-0.125} ,
\label{eqn:cM}
\end{equation} 
with a scatter in $c_{\rm vir}$ (at fixed $M_{\rm vir}$) of 0.14 dex
(though the scatter may be smaller for high-mass, relaxed halos;
\citealt{2001MNRAS.321..559B}; \citealt{2002ApJ...568...52W};
see \S3.2 in \citealt{2005MNRAS.357..691N}). 
Here we have adopted a virial overdensity of $\Delta_{\rm vir}=101$,
a power-spectrum normalization of $\sigma_8=0.9$, and
a Hubble constant of $H_0=$~70 km~s$^{-1}$~Mpc$^{-1}$.
The relation (\ref{eqn:cM}) was derived from \citet{2001MNRAS.321..559B}
for galaxy-mass halos, but turns out to also be valid for higher masses,
differing by at most 5\% 
from the updated relation of \citet{2007MNRAS.381.1450N} over the range 
$M_{\rm vir}=(2$--$70)\times10^{13} M_{\Sun}$.
Lower values for $\sigma_8$ would imply systematically lower concentrations,
but the effect is probably smaller than the intrinsic scatter of the concentrations
\citep{2008MNRAS.390L..64D}.

For simplicity, we do not include any back-reaction of the baryons on the DM profile (such as
adiabatic contraction: e.g.,
\citealt{1986ApJ...301...27B}; \citealt{2004ApJ...616...16G}), 
which would be only a small correction at the outer DM-dominated radii that we are probing (see also H+06b).
We also do not include the most recent refinements to the NFW profile
used in equation~\ref{eqn:NFW}
(e.g., \citealt{2004MNRAS.349.1039N,2008MNRAS.387..536G}).

The anisotropy profile is motivated by numerous theoretical
studies that generically predict that DM halos, and the
baryonic tracers within them, are radially anisotropic
in their outer regions
(e.g., \citealt{1982MNRAS.201..939V,1996MNRAS.281..716C,1999MNRAS.309..610D,2000ApJ...539..561C,2001ApJ...557..533F,2003ApJ...593..760V,2004MNRAS.351..237R}; 
\citealt{2005MNRAS.361L...1W,2005Natur.437..707D}; M{\L}05;
D+05; \citealt{2006MNRAS.365..747A,2006NewA...11..333H,2006MNRAS.369..958S,2006MNRAS.370..681N,2007MNRAS.376...39O,2007MNRAS.376.1261M}; \citealt{2008ApJ...689..919P}).
D+05 in particular explored the
dependence of the anisotropy of tracers on their concentrations within
their host halos, and we adopt their equation (8) to represent
the anisotropy profile of the NGC~1407 GCS\footnote{The D+05 anisotropy profile
has the rotational component subtracted, unlike our modeling convention,
but this difference impacts their predictions at the level of $\Delta \beta \sim -0.02$ at most.}:
\begin{equation}
\beta(r)=\beta_0 \frac{r}{r+r_0} .
\label{eqn:beta}
\end{equation}

\begin{figure}
\includegraphics[width=3.4in]{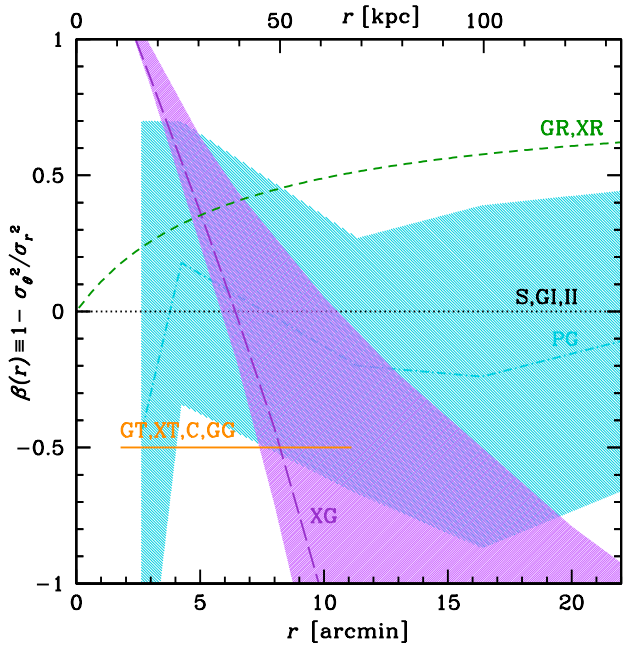}
\caption{
Velocity dispersion anisotropy profiles for GCs in NGC~1407. 
The black horizontal dotted line shows isotropy.
The green short-dashed line shows a theoretically-motivated profile
(Eq.~\ref{eqn:beta}).
The orange horizontal solid line shows the dynamics-based estimate of the GC anisotropy
(\S\ref{sec:modalt}).
The purple long-dashed line with surrounded shaded region
shows the profile necessary for agreement with the X-ray analysis
(\S\ref{sec:mult1}).
The blue dot-dashed line with shaded region shows the cosmological simulations of
metal-poor GCs from \citet[see \S\ref{sec:gckin}]{2008ApJ...689..919P}, 
scaled to the virial radius of NGC~1407.
\label{fig:beta} 
}
\end{figure}

To assign values to the asymptotic anisotropy $\beta_0$ and the break
radius $r_0$, we first compare our GCS surface density profile to the
parameterized projected profiles of the D+05 simulations corresponding to DM
subsets characterized by an overdensity of $n$ times the RMS background
fluctuation (see their equation 1).
We ignore the central cusp in these density profiles since D+05 did
not report the natural variation around a typical log slope of $-1.2$,
and since the central GCS number density may be altered considerably
by destruction processes.
Focusing on the outer density profiles, we find that $n \sim 4$ is needed
to reproduce the steep decline seen in NGC~1407
(see Fig.~\ref{fig:gcs}).
The data include both metal-rich and metal-poor GCs, so this fit
is not appropriate for determining GC formation redshift as proposed by D+05
and \citet{2006MNRAS.368..563M}, but rather provides a heuristic reasonable assumption
for the anisotropy profile.
The log density slope at $r_0$ of the $n=4$ simulation model is $-3.3$,
and finding the location in our GCS density model with the same slope,
we arrive at $r_0=6.33'=$~38.5~kpc.
As specified by D+05 equation 8, $n=4$ corresponds to $\beta_0=0.8$;
Fig.~\ref{fig:beta} shows the resulting anisotropy profile.
The values of $\beta_0$ and $r_0$ are not very sensitive to our choice for $n$.

Now to find the best-fit fiducial model, we combine a $\chi^2$
fit of the model to the $v_{\rm RMS}(R_{\rm p})$ data
(excluding the innermost point which we regard as
questionable---see \S\ref{sec:select} and \S\ref{sec:disp})
with a $\chi^2$ fit to Eq.~\ref{eqn:cM} including its scatter.
This fitting method is equivalent to a Bayesian analysis where
the model priors are given by Eq.~\ref{eqn:cM}.
We find a virial mass 
$M_{\rm vir} \simeq 1.4\times 10^{14} M_{\odot}$, with a
corresponding virial radius of $r_{\rm vir} = $1.0--2.0 Mpc---which
is somewhat larger than the optical size of the group of 
0.4--0.6 Mpc (B+06b).
As seen in Fig.~\ref{fig:models1} (with best fit reported in Table~\ref{tab:models}
as model GR), this ``standard'' $\Lambda$CDM model
is roughly consistent with the GCS data
(including the central $v_{\rm RMS}$ point)
but it does not match particularly well the
constant $v_{\rm RMS}$ profile outside of the center
($-0.3 \lsim \gamma_{\rm p} \lsim +0.2$).  The model
predicts a fairly steep decline of $v_{\rm RMS}(R_{\rm p})$
(slope $\gamma_{\rm p} \sim +0.3$), 
despite a rising $v_{\rm c}(r)$.
This behavior is due partially to the strong radial anisotropy,
and partially to the very low S\'ersic index of the GCS density profile, 
whose flat core and sharp outer edge mimic the effects of radial anisotropy.

\subsubsection{Model alternatives}\label{sec:modalt}

\begin{figure*}
\includegraphics[width=3.4in]{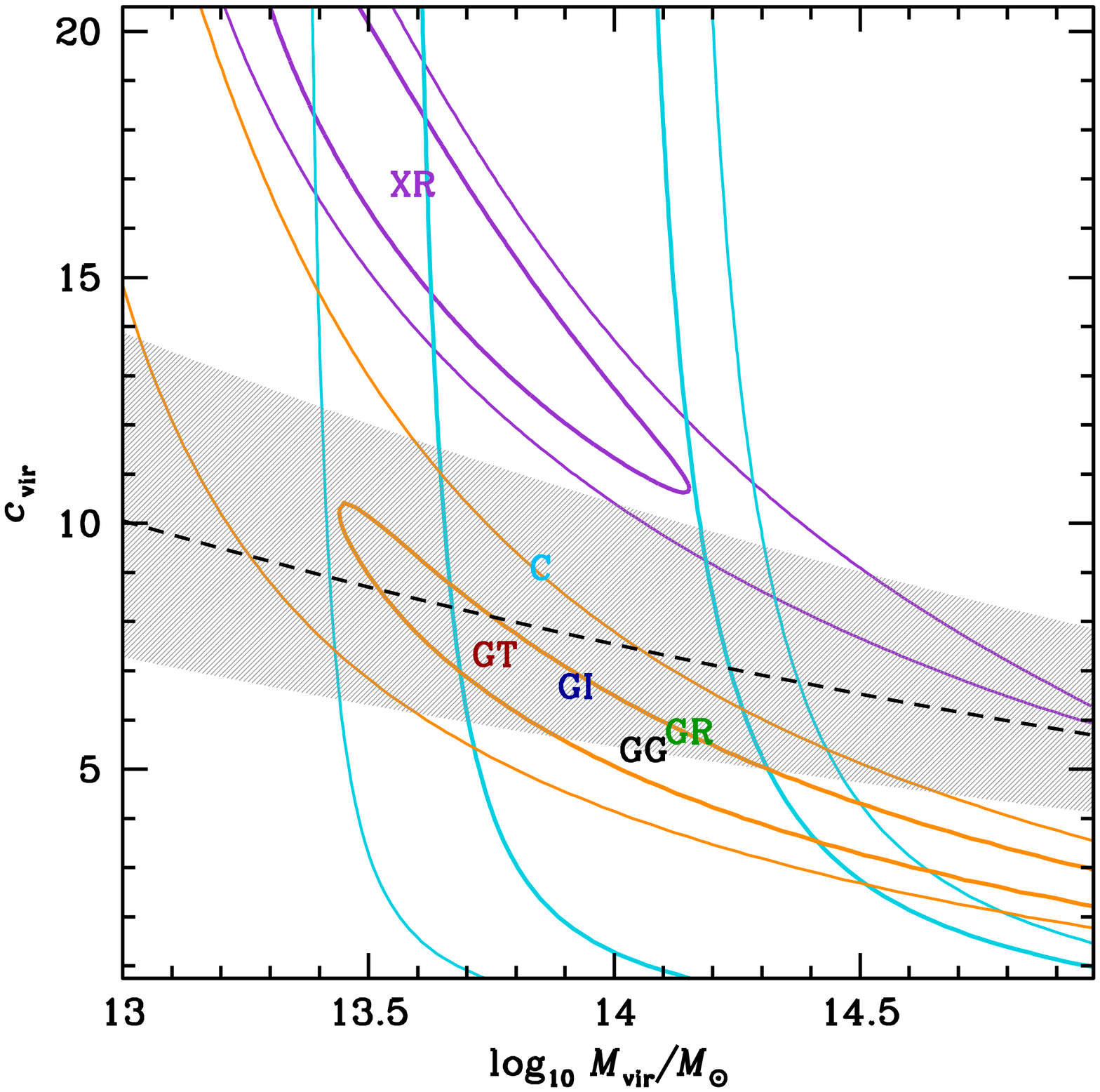}
\includegraphics[width=3.4in]{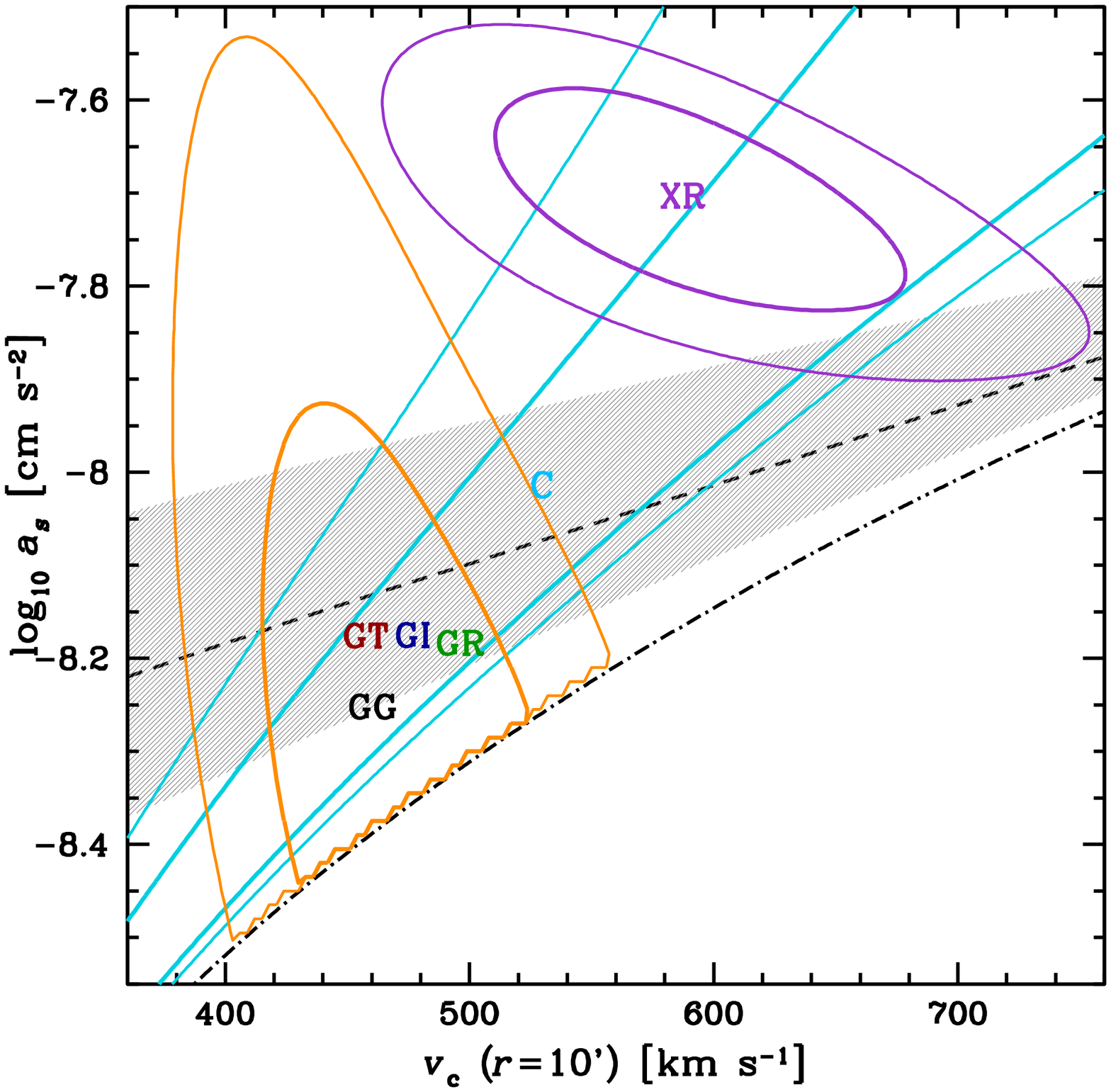}
\caption{
Mass profile solutions for NGC~1407.
Thick and thin solid contours show the 1~$\sigma$ and 2~$\sigma$
confidence levels, respectively (as found by $\Delta \chi^2$ tests);
none of the contoured solutions includes a concentration prior.
The orange contours 
toward the bottom of the panels
correspond to the tangentially anisotropic GCS solutions,
the purple contours 
toward the top
to the X-ray solutions (model XR), and
the blue near-vertical contours to the group galaxies solutions.
The mean relation and scatter from cosmological simulations are shown
by the dashed curve and shaded region (see text).
Additional letters label various best-fit solutions as described in the text
and in Table~\ref{tab:models}.
The left panel shows the virial mass and concentration, and
the right panel shows the circular velocity inside the GC-constrained region of $10'$ (60~kpc)
as well as the characteristic halo acceleration.
The dot-dashed line in the right panel shows the physically-meaningful limit of
the NFW model (i.e. $c_{\rm vir}=0$).
The plots demonstrate that the GCS dynamics are consistent with a ``normal'' 
high-mass $\Lambda$CDM halo, 
while the X-ray based mass model is mildly inconsistent with $\Lambda$CDM
and strongly inconsistent with the GC results.
\label{fig:cM} 
}
\vspace{0.2cm} % because of running into the text
\end{figure*}

Suspecting that at least one of the modeling assumptions is awry,
we next try excluding any {\it a priori} constraint on the concentration.
Without this constraint, there is a severe degeneracy between the virial
quantities $M_{\rm vir}$ and $c_{\rm vir}$, since even with an NFW profile
assumed, we cannot make reliable extrapolations from data extending to only
$\sim$~5\% of $r_{\rm vir}$.
As seen in the left panel of Fig.~\ref{fig:cM},
the DM halo could have a high mass and a low concentration, 
or alternatively a low mass and a high concentration.
The right panel illustrates some physical quantities more directly constrained
by the data: the circular velocity inside $10'$, and the characteristic acceleration of
the halo $a_s \equiv 0.193 v_{\rm s}^2/r_s$ 
(which corresponds to a constraint on the projected mass density $\rho_s r_s$,
or equivalently to the gravitational lensing convergence $\kappa$);
these quantities are more tightly constrained than other halo parameterizations
(e.g., \citealt{2002ApJ...572...34A}).
The GC data naturally prefer very low-concentration, high-mass halos
in order to try to reproduce the flat $v_{\rm RMS}(R_{\rm p})$
profile despite the low S\'ersic index and the presence of radial anisotropy.
The best-fit solution is implausible
($M_{\rm vir} \sim 10^{17} M_{\odot}$, $c_{\rm vir} \sim 10^{-2}$),
and the fit is still not very good
($\chi^2_{\rm GC}=3.0$ for 2~d.o.f.).

Another possibility is that our adopted $\beta(r)$ profile is too extreme, 
and so we return to the standard $c_{\rm vir}$-$M_{\rm vir}$ correlation,
with a simple $\beta=0$ isotropic assumption. 
There is little theoretical justification for isotropy on scales of $\sim$~50 kpc,
but there are empirical indications of isotropic GC systems in other massive
group-central ellipticals 
(e.g., \citealt{2001ApJ...553..722R}, hereafter RK01; C+01; C+03; 
\citealt{2008A&A...488..873S}; \citealt{Johnson09}, hereafter J+09;
see summary in \citealt[hereafter H+08]{2008ApJ...674..869H}).
With isotropy, we find a better fit to the data, with naturally a lower $M_{\rm vir}$ preferred since
mass is no longer ``hidden'' by radial anisotropy
(see model GI in Table~\ref{tab:models} and in Figs.~\ref{fig:models1} and \ref{fig:cM}).
The resulting virial radius of 0.8--1.7~Mpc is closer to
the optical group radius.

\begin{figure}
\includegraphics[width=3.4in]{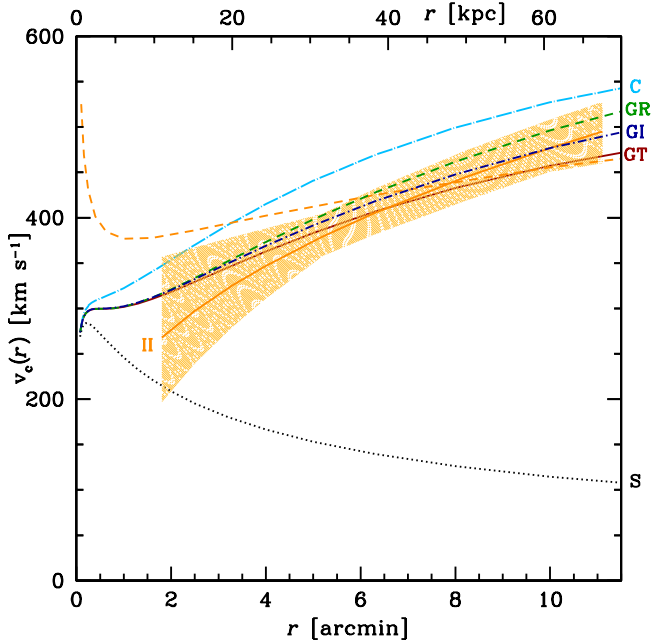}
\caption{
Mass modeling results for NGC~1407 from GCS dynamics, 
expressed as circular velocity profiles.
Individual models are labeled by letters and are described in the text and in
Table~\ref{tab:models}.
The orange shaded region around model II (direct mass inversion) reflects the 
$68\%$ statistical uncertainties on the dispersion profile.
The orange dashed line shows model II where the dispersion data at all
radii have been included.
\label{fig:models2} 
}
\vspace{0.02cm} % because of running into the text
\end{figure}

Given that the isotropic model provides an improved fit, we use this simplifying
fact as a starting point to derive an estimated mass profile directly from the data,
using no theoretical priors such as Eq.~\ref{eqn:NFW}.
We wish to deproject the $v_{\rm RMS}(R_{\rm p})$ profile and use it
in the isotropic Jeans equation to find the $v_{\rm c}(r)$ profile.
We start with the standard Abel deprojection equation for velocity dispersion
(e.g., \citealt{1987gady.book.....B}, eq.~4-58b) and differentiate it, 
integrating by parts and using Leibniz's Rule to find:
\begin{eqnarray}
v_{\rm c}^2(r) & = & -\frac{r}{\nu}\frac{d(\nu\sigma_r^2)}{dr}  \nonumber \\
& = & \frac{r^2}{\pi\nu} \int^\infty_r \frac{dR_{\rm p}}{\sqrt{R_{\rm p}^2-r^2}} \frac{d}{dR_{\rm p}} \left[ \frac{1}{R_{\rm p}} \frac{d(\Sigma v^2_{\rm RMS})}{dR_{\rm p}}\right] ,
\end{eqnarray}
where $\Sigma(R_{\rm p})$ and $\nu(r)$ are the projected and deprojected
density profiles of the GC system, as described in \S~\ref{sec:assumpt}.
Now parameterizing the observed $v_{\rm RMS}(R_{\rm p})$ by Eq.~\ref{eqn:PL} 
with ($v_{\rm RMS,0}=225$~km~s$^{-1}$, $R_{\rm 0}=4.24'$, 
$\gamma_{\rm p}=-0.07$; see Fig.~\ref{fig:models1}), we find a steeply rising mass profile,
shown as model II in Fig.~\ref{fig:models2}.
This model is similar to but steeper than the isotropic fiducial 
$\Lambda$CDM-based model, which does not reproduce the observed
increasing behavior of $v_{\rm RMS}(R_{\rm p})$.
Also shown in Fig.~\ref{fig:models2} is the mass profile if the central dispersion
points were not excluded
($v_{\rm RMS,0}=234$~km~s$^{-1}$, $R_{\rm 0}=3.78'$, $\gamma_{\rm p}=0.24$):
the central mass would be significantly higher, with
a density cusp inside $\sim$~5~kpc.

Given that the fiducial anisotropy profile does not seem to be correct, and
isotropy is an arbitrary assumption, we next investigate what the data themselves
can tell us about the anisotropy.
This transpires in two ways: 
by considering the metal-poor and metal-rich GC populations
as independent dynamical subsystems that must yield consistent results for the mass profile;
and by making use of the observed kurtosis to directly constrain the anisotropy.

We begin by considering the system to be scale free, i.e., $\nu(r)$
and $\sigma^2_r(r)$ vary with radius as simple power-laws with exponents
$-\alpha$ and $-\gamma$ respectively, while $\beta$ is constant with radius.
Then we can connect the circular velocity and projected dispersion by
a constant relation:
\begin{equation}
v_{\rm c}(r=R_{\rm p}) = k \sigma_{\rm p}(R_{\rm p}) ,
\label{eq:keqn}
\end{equation}
where $k$ is a complicated function of $\alpha$, $\beta$, and $\gamma$
\citep{2005Natur.437..707D}.
We select a fixed intermediate radius $r=R_{\rm p}=4.24'$,
where $\alpha_{\rm MP} \simeq 2.7$, $\alpha_{\rm MR} \simeq 3.4$,
$\gamma_{\rm MP} \simeq -0.4$, and $\gamma_{\rm MR} \simeq +0.4$.
Now using either the metal-poor or metal-rich GCs for the right-hand side of
equation~\ref{eq:keqn} must yield the same answer for the left-hand side,
so we search for combinations of $(\beta_{\rm MP},\beta_{\rm MR})$
that satisfy the equation 
$k_{\rm MP}/k_{\rm MR} = \sigma_{\rm p,MR}/\sigma_{\rm p,MP}$.
We find that a fully isotropic solution ($\beta_{\rm MP}=\beta_{\rm MR}=0$)
is not preferred, because the steeper density {\it and} dispersion profiles
of the metal-rich GCs should depress $\sigma_{\rm p,MR}$ relative to
$\sigma_{\rm p,MP}$ even more than is observed.
In fact, strong tangential anisotropy ($\beta \lsim -1$) is implied for both
the metal-poor and metal-rich GCs.
The degeneracies and uncertainties among the parameters do not allow us to
determine more than this (if we assume $\beta_{\rm MP}=\beta_{\rm MR}$
for simplicity, then these formally take the value of $-3.6$).

We now turn to constraints from the kurtosis, which as higher-order LOSVD moments
can provide direct information about the orbital types.  It is beyond the scope of this 
paper to explore this theme in detail, but we can take advantage of one simplification
discussed in \citet[eqn B10]{2009MNRAS.393..329N}.
If $\gamma=0$, $\beta$ is a constant, 
and a simplified distribution function is adopted,
then $\beta$ can be directly estimated
by deprojecting $\kappa_{\rm p}$ and $\Sigma_{\rm p}$---with no dynamical
modeling necessary.
We therefore take this approach for the overall GCS of NGC~1407, since
$\gamma_{\rm p} = -0.07 \pm 0.26$.
With $\kappa_{\rm p}=-0.30\pm0.41$ (\S\ref{sec:kurt}),
we find that $\beta = -0.5^{+0.6}_{-1.3}$,
where the uncertainty is based only on the observed uncertainty in $\kappa_{\rm p}$,
and does
not include the uncertainties in $\Sigma_{\rm p}$ nor in the deviations of the galaxy from
the simplified ($\gamma,\beta)$ model.
Thus we find an independent indication that the GC system is tangentially anisotropic overall.

Given the two separate indicators of tangential anisotropy,
we now fit a $\Lambda$CDM model to the GC data with $\beta=-0.5$ assumed.
The best-fit model (``GT'') provides a slightly better fit than the isotropic model
(see Fig.~\ref{fig:models1} and Table~\ref{tab:models}), and
we therefore have three different lines of evidence for tangential anisotropy---none
of which is statistically significant on its own, but in combination are suggestive
of a reliable result.
This solution implies a slightly lower virial mass of $\sim 6\times 10^{13} M_\odot$
(Figs.~\ref{fig:cM} and \ref{fig:models2}) than the isotropic model.

We can next take this best-guess model
as a starting point to estimate the anisotropy profiles
of the metal-poor and metal-rich subsystems.
Using the method of \citet[Appendix]{1983ApJ...266...58T}, we take the model $v_{\rm c}(r)$
and the observed $\sigma_{\rm p}(R_{\rm p})$ (parameterized in our case by a power-law fit),
then invert them to an intrinsic $\sigma_r(r)$, and finally solve the Jeans equation to
find $\beta(r)$
(see also \citealt{1997MNRAS.289..685H}).

\begin{figure}
\includegraphics[width=3.4in]{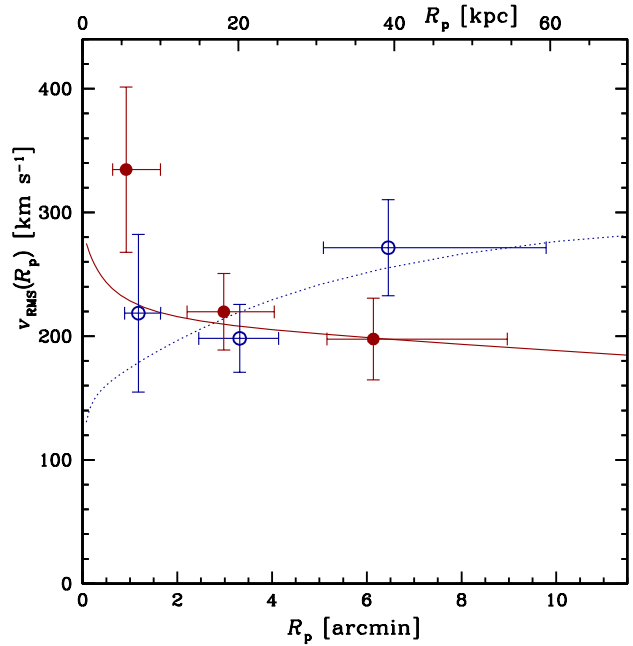}
\caption{
Projected RMS velocities of NGC~1407 GCs, where curves show model predictions
and points with error bars show the data.
The blue dotted curve and open points show the metal-poor GCs, while the red
solid curve and filled points show the metal-rich GCs.
The model is the best-fit overall GC solution (``GT''), with $\beta_{\rm MP}=-4$
and $\beta_{\rm MR}=0$.
\label{fig:redblue}
}
\end{figure}

This inverse approach can be sensitive to assumptions about $v_c$ and $\sigma_{\rm p}$
outside the regions constrained by the data, so we use it as a starting guess for 
simplified $\beta(r)$ functions, re-solving the Jeans equations and comparing the
model $\sigma_{\rm p}$ to the data.
We adopt the simplest assumption of constant anisotropy, and
find that $\beta_{\rm MP} \sim -4$ (tangentially-biased orbits)
provides a good description for the metal-poor GCs (see Fig.~\ref{fig:redblue}).
The metal-rich GCs on the other hand appear to be roughly isotropic
($\beta_{\rm MR} \sim 0$) outside $\sim 2'$.
The high metal-rich velocity dispersion
observed at smaller radii suggests the orbits become very radial in this region---although
even $\beta_{\rm MR}=+1$ is not quite enough to reproduce the data,
and to compensate, very tangential orbits would be needed in a transition region at $\sim 1.5'$.
These peculiarities could be caused by an admixture of DGTOs concentrated in the central
regions as discussed in \S\ref{sec:disp}, and will require more detailed modeling in the future.
Note that changing our assumed distance and $\Upsilon_*$ to higher values
would have little impact on the inferred values for $\beta$, $M_{\rm vir}$,
and $c_{\rm vir}$.

In summary, we can conclude that the GC kinematical information is compatible
with a standard $\Lambda$CDM halo with size and mass parameters similar to a very high-mass galaxy group.
The GCs appear to reside on somewhat tangential orbits overall, which may be
decomposed into very-tangential orbits for the metal-poor subpopulation,
and isotropic orbits for the metal-rich GCs.
The anisotropy results are not yet robust, and 
in \S\ref{sec:indep} we will explore additional constraints
that may clarify the mass profile of the NGC~1407 system.

For now,  assuming this anisotropy result is correct,
we report a range 
for the mass inside the GCS-constrained region of $10'$ (60 kpc)
to be (2.7--3.2)$\times 10^{12} M_{\odot}$, implying a mass-to-light ratio relative
to the galaxy NGC~1407 of $\Upsilon_I=$~(29--34)~$\Upsilon_{\odot,I}$, or
$\Upsilon_B=$~(61--71)~$\Upsilon_{\odot,B}$.
The systematic uncertainty from plausible ranges on the anisotropy is ~$\sim$~20\%;
the uncertainties from the GCS density profile and the distance are at the level of
$\sim$~10\% and $\sim$~20\%, respectively.
The GCs with the lowest binding energies have apocenters of $\lsim 15' \sim 90$~kpc,
which does not necessarily mean that the GCS ends beyond this radius
(cf. Fig.~\ref{fig:gcs}), but rather is
consistent with our model of the GCs on near-circular orbits.

\subsection{X-ray analysis}
\label{sec:xray}

We next consider a fully independent constraint on the mass profile
of the NGC~1407 group, making use of X-ray emission from hot gas 
trapped in its potential well.
As a first approximation to estimating the group mass in this way,
we take the {\it ROSAT}-derived X-ray temperature of 1.02 keV (OP04),
and use the mass-temperature relation
of \citet{2006ApJ...640..691V} to estimate 
$M_{500}=3.3\times10^{13} M_{\odot}$
(i.e. for an overdensity of $\Delta=500$).
An alternative relation from \citet{2006MNRAS.372.1496S} 
that  may be more valid in the group regime
yields $M_{500}=1.7\times10^{13} M_\odot$.
With reasonable NFW extrapolations to the virial radius,
we thus infer $M_{\rm vir} \sim$~(2--6)~$\times10^{13} M_\odot$.

\begin{figure*}
\includegraphics[width=3.4in]{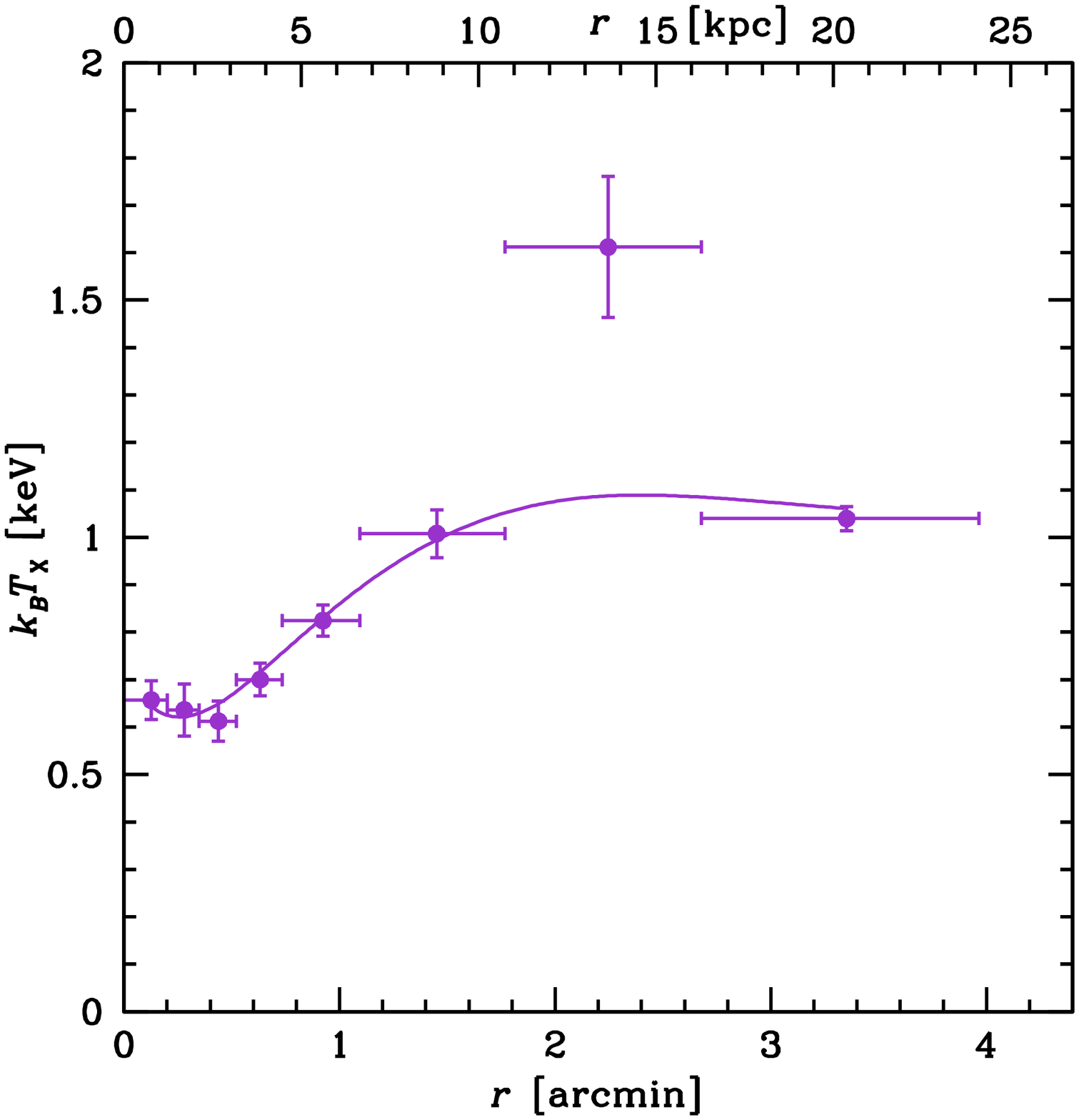}
\includegraphics[width=3.4in]{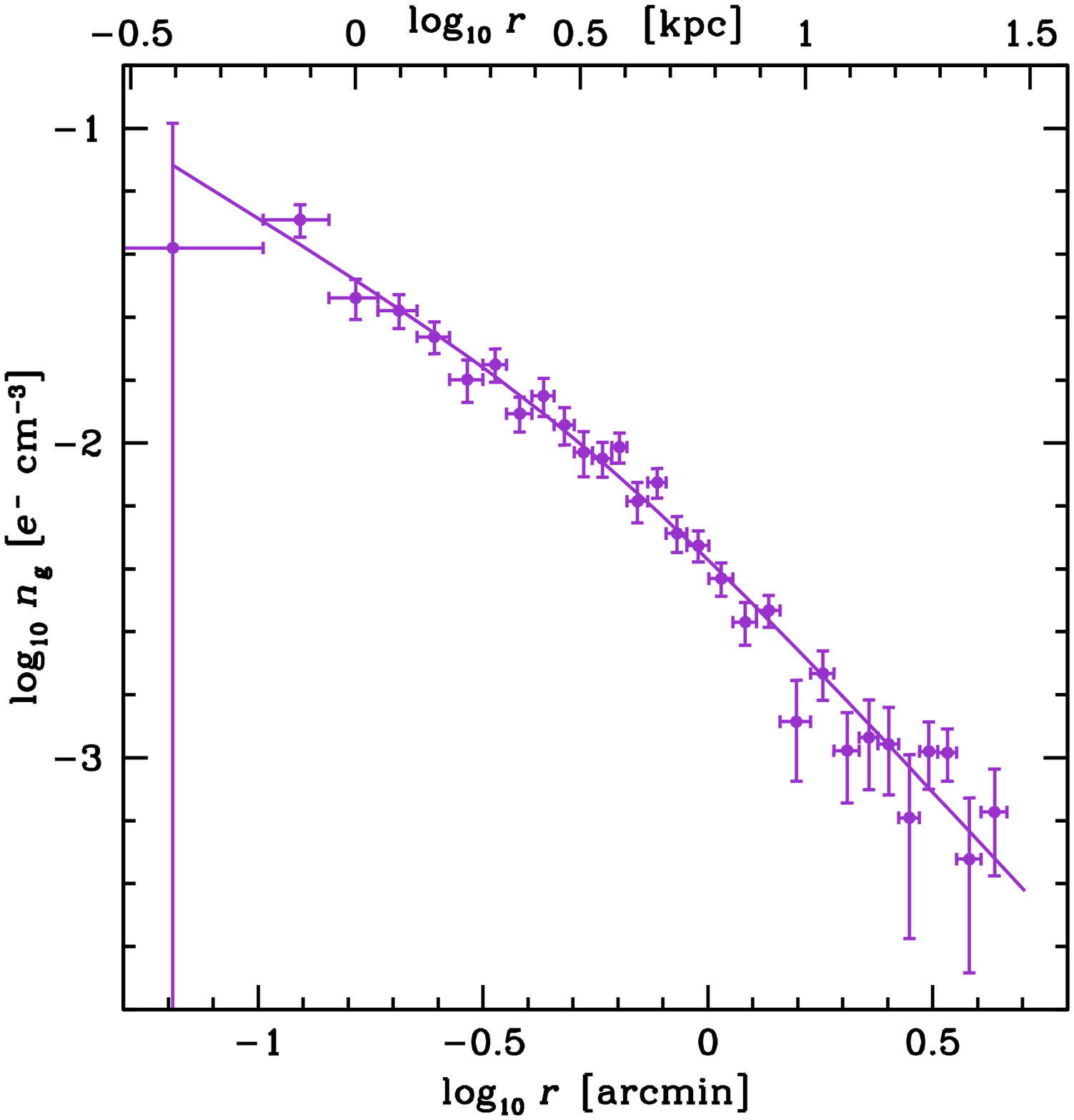}
\caption{
Deprojected properties of the hot gas in NGC~1407, from {\it Chandra}
X-ray analysis.
The points with error bars show the measurements in bins,
where the horizontal bars show the bin boundaries in radius, and
the vertical bars show the 68\% uncertainties 
recovered from 100 Monte Carlo realizations of the procedure.
{\it Left:} Temperature profile, computed in coarse bins, with
a curve showing the smoothed interpolation to the binned profile.
Although the fitting was done in log-log space, the plot is linear
for more direct comparison with the GC dispersion profile
(e.g., Fig.~\ref{fig:models1}).
{\it Right:} Gas density profile, computed in fine bins,
with a curve showing the cusped beta model.
\label{fig:tx}
}
\vspace{0.2cm} % because of running into the text
\end{figure*}

A firmer result on the mass profile requires a careful analysis 
based on high-resolution X-ray data.
NGC 1407 was observed by \Chandra\ \citep{2002PASP..114....1W} on 2000 August 16 for 
$\sim$49~ksec (Obsid 791) using the ACIS instrument with the S3 chip at the focus. 
A new level 2 events file was produced from the level 1 events file using version 
3.4 of the \Chandra\ Interactive Analysis of 
Observations\footnote{http://cxc.harvard.edu/ciao/} (\ciao), with \caldb\ version 3.3.0. 
Bad pixels, and events with \asca\ grades of 1, 5, and 7, were removed from the analysis,
and the appropriate gain map and time-dependent gain correction were applied. 
We eliminated flares by extracting lightcurves from CCDs 5 and 7, and one from the 
front-illuminated CCDs (2,3,6 and 8), and filtered these lightcurves using the Markevitch 
script lc\_clean. 
The net remaining exposure time was 30.4~ksec.
Background subtraction was performed using the appropriate blank-sky 
data set\footnote{http://cxc.harvard.edu/contrib/maxim/bg/}, which was rescaled using 
the ratio of count rates at particle-dominated energies. We used the \ciao\ tool 
wavdetect to identify point sources in the data, and spectra and response files were 
extracted from the cleaned data using the \ciao\ analysis threads.

The X-ray data analysis 
techniques are described in detail in J+09;
a summary of 
the essential features is given here. Our aim is to produce a high-resolution mass profile, 
which is achieved through a two-stage analysis, by firstly analyzing temperature on a 
coarse grid, and then computing gas density on a fine grid. We extract a series of spectra 
from coarsely spaced concentric annuli, whose spacings are determined by a requirement 
of 1500 net counts in each annulus. We find that this criterion is sufficient to robustly 
constrain the temperature and to perform a successful deprojection of the spectra. We fit
absorbed \apec\ models to the spectra from each coarsely spaced annulus using \xspec\ version 
11.3.2t, ignoring energies below 0.5~keV and above 7.0~keV. 
We allow for the presence of unresolved point sources in the emission by 
including a power-law component of fixed index = 1.56
\citep{2003ApJ...587..356I}
following a de Vaucouleurs $R^{1/4}$ density law with \Reff$=1.17'$ (Z+07).
We use the \textsc{projct} model in \xspec\ under the assumption of
spherical symmetry to deproject the spectra, 
fitting a single global abundance value across all radial bins for the
sake of deprojection stability.
This yields three-dimensional model parameters 
at a series of characteristic coarse radii. We fit smoothing spline models to the profiles 
using the \textsc{smooth.spline} algorithm from the \textsc{r project} statistical 
package\footnote{http://www.R-project.org/ \citep{R2006}}.
This allows the interpolation of the fitted parameters at a series of characteristic fine 
radii, described below.

It is worth noting that the temperature profile of NGC 1407 shows typical
cool-core behavior, and we would expect an associated abundance gradient
\citep{2007MNRAS.380.1554R}. 
In our current modeling, we fit the abundance at a constant level,
which affects the APEC
model normalization and thus the gas density, owing to the
dominance of line emission at low temperatures. 
The primary effect of allowing for a gradient might be to decrease the inferred mass 
at the level of $\sim 20\%$ (J+09).

The resulting temperature profile is shown in Fig.~\ref{fig:tx} (left), where the vertical
error bars are recovered from 100 Monte Carlo realizations of the
procedure, and are used to weight the smoothing spline fit 
(by the inverse square of the temperature uncertainty).
The most notable feature is a sharp peak of $\sim$~1.6~keV in the binned profile
between 2$'$ and 3$'$. Although this feature appears significant, it may be an
artifact of some instability in the deprojection procedure; comparable
analyses by H+06b and by Z+07 found a peak of only $\sim$~1.3~keV.
The large error bar means
that this point is down-weighted in the fit, but in any case, our Monte
Carlo simulations do take large $T_{\rm X}$ excursions into account in the reported
mass profile uncertainties.

We next extract a series of spectra from finely spaced concentric 
annuli, and again fit absorbed \apec\ models in each annulus. We use
the \textsc{projct} model in \xspec\ to yield the deprojected model 
parameters. However, in this stage we only fit for the \apec\ model 
normalization, which directly yields the gas density. The temperature, abundance and Galactic 
column in each shell are fixed at the interpolated value from the smoothing spline fits in the 
coarse stage. This process allows us to measure the gas density on a finer radial scale. 

To determine the mass profile of NGC 1407, we use the smoothing
spline function fitted to the coarse 3D temperature profile, and fit a
cusped ``beta model'' (e.g., H+06b) to the fine 3D gas density
profile ($r_{\rm core}$ = $33.0^{\prime\prime}$, $\beta_{\rm X}$ = 0.5; 
see Fig.~\ref{fig:tx}, right).
This allows the 
determination of the circular velocity profile under the assumption of hydrostatic equilibrium,
using the equation: 
\begin{equation}
v_{\rm c}^2(r) = - \frac{{k_{\rm B}}T_{\rm X}(r)}{{\mu}m_{p}}\left(\frac{d~\ln~n_{\rm g}}{d~\ln~r}+\frac{d~\ln~T_{\rm X}}{d~\ln~r}\right),
\label{eqn:Mr}
\end{equation}
where $T_{\rm X}(r)$ and $n_{\rm g}(r)$ 
are the gas temperature and density profiles
\citep{1980ApJ...241..552F}.

\begin{figure}
\includegraphics[width=3.4in]{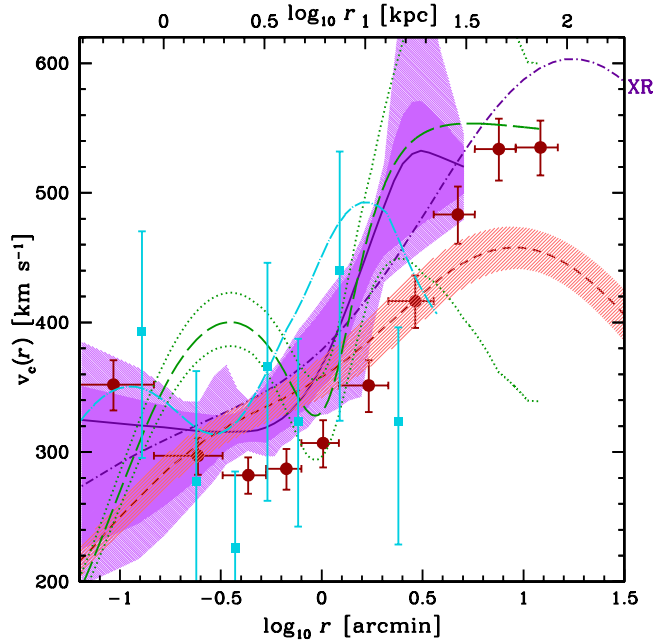}
\caption{
X-ray-based mass modeling results for NGC~1407.
Our new {\it Chandra}-based result is shown as a {\it purple solid curve} with shaded regions
showing the 1~$\sigma$ and 2~$\sigma$ uncertainty regions.
The best-fit stars+NFW solution (model XR) is shown as a {\it purple dot-short-dashed curve}.
The {\it Chandra}-based results of \citet{2006ApJ...636..698F} are shown, including
their double-beta model fit ({\it blue dot-long-dashed curve}) and their deprojection analysis
({\it blue boxes with error bars}).
The {\it Chandra}-based results of H+06b include their $\Lambda$CDM-based
model fit without adiabatic contraction ({\it red short dashed curve with shaded uncertainty region}) 
and their deprojection analysis ({\it red circles with error bars}).
The {\it Chandra}+{\it ROSAT} analysis of Z+07 is shown by the
{\it green long-dashed curve}, with {\it green dotted curves} showing the 90\% uncertainties.
\label{fig:xray}
}
\vspace{0.3cm} % because of running into the text
\end{figure}

We use a Monte Carlo procedure to produce 100 realizations of the mass 
profile of NGC 1407, which are each analyzed as described above. The span of results from these 
random realizations is used to define the confidence regions shown in Fig.~\ref{fig:xray},
where we now adopt a logarithmic radius scale in order to see the models more clearly.
We find the $v_{\rm c}(r)$ profile to be consistent with constancy inside 4~kpc,
and outside of this radius, it rapidly rises until 20~kpc---increasing
the enclosed mass by a factor of 10--15 despite a radius increase of
only a factor of 5---whereafter it may level off again.
It is the rapid increase of $T_{\rm X}$ with radius (Fig.~\ref{fig:tx})
that produces the $v_{\rm c}$ increase; reinstating the highest $T_{\rm X}$
point in the modeling would produce an even steeper (and possibly unphysical)
$v_{\rm c}$ profile.
We will further discuss the features of the mass profile in the next section.

\subsection{Comparing independent results}
\label{sec:indep}

Having derived mass profiles from both GC and X-ray data, we now explore
the systematic uncertainties by making independent modeling comparisons.
In \S\ref{sec:exX} we compare our own X-ray result for NGC~1407 to literature 
X-ray studies of this system,
and in \S\ref{sec:mult1} we consider three different mass probes in NGC~1407.
In \S\ref{sec:mult2} we survey similar multiple mass probes in other systems.

\subsubsection{Multiple X-ray studies in NGC~1407}
\label{sec:exX}

Several independent analyses of the same {\it Chandra} data for NGC~1407
have been published, whose results we also plot in Fig.~\ref{fig:xray} for reference.
\citet{2006ApJ...636..698F} performed both a non-parametric deprojection analysis, and
a parameterized ``double-beta model'' fit, to produce their mass profiles.
H+06b also made deprojected models, as well as forward fitting of
$\Lambda$CDM-based galaxy+halo models;
their results extend to larger radii than ours by analyzing all of
the available ACIS chips, while we conservatively used the back-illuminated 
S3 chip only.
Z+07 again used the {\it Chandra} data, combining them also with
{\it ROSAT} data at larger radii, using both projected and deprojected 
double-beta models.

As seen in Fig.~\ref{fig:xray}, these various analyses yield qualitatively
similar results for the mass profile, finding a rapid rise in $v_{\rm c}(r)$
followed by a probable leveling-off.
But in quantitative detail, there are differences in the results that are
larger than the claimed uncertainties.
The Z+07 results agree well with ours outside 1$'$, but inside that radius,
have an extra ``hump'' in $v_{\rm c}$ that is not seen in any other analysis.
The hump is produced by their double-beta model for the gas density, 
a parameterization for which we see no motivation from our higher-resolution
density profile.
The H+06b $v_{\rm c}(r)$ deprojected result has a similar shape and amplitude to ours, 
but is oddly offset by a factor of $\sim$~2 in radius---perhaps owing to
the different abundance treatments.
We will examine these differences in more detail in another paper, and here 
note that there can be substantial systematic effects in deriving mass profiles from
X-ray data, depending on the parameterizations and assumptions in the modeling.

Since the rapid $v_{\rm c}(r)$ rise between 5 and 20~kpc 
seems to be a robust conclusion from the X-ray models, 
we next consider the implications of this feature.
Similar ``kinks'' have been found in the X-ray-based mass profiles of many other
galaxy groups and clusters (see Z+07 for a summary), 
suggesting a transition between a DM halo belonging to the central galaxy, 
and a larger-scale halo pertaining to the entire group or cluster.
Such a galaxy-group halo transition would pose a problem to CDM theory, which predicts that
halos should relax into smooth mass profiles similar to the NFW profile; 
any kink would be unstable and transient.
Indeed, H+06b did not find a good fit for their $\Lambda$CDM-based models, 
which as seen in Fig.~\ref{fig:xray} is probably due to an inability of
such models to reproduce the kink.
The model solutions that come closest are forced to higher DM halo concentrations than 
expected in $\Lambda$CDM (H+06b Fig.~4).

We next try to fit our own galaxy+halo model sequence, using our own X-ray data analysis.
In contrast to the forward-modeling performed by H+06b, we use a less direct method of
fitting the model $v_{\rm c}(r)$ to our deprojected X-ray profile.
The fit is performed at 31 points in radius, but since there were effectively
7 independent data points in the original coarse-binned data, we renormalize
the $\chi^2$ statistic by a factor of $7/31$.
Our best-fit solution (model XR; see Table~\ref{tab:models}) 
is shown in Fig.~\ref{fig:xray}. 
Again, the NFW-based model has difficulty reproducing the kink, and requires a DM
halo that is more concentrated than is typical for $\Lambda$CDM 
(see Fig.~\ref{fig:cM}).

To see if we can find a better solution,
we re-examine our modeling assumptions.
The distance to the galaxy, and $\Upsilon_*$, are both
key parameters that could plausibly have higher values (see \S\ref{sec:assumpt}).
However, increasing either of them causes the X-ray fit to
deteriorate rapidly (driven by the constraints in the central
regions).  {\it Decreasing} either of these parameters 
(e.g., to $\Upsilon_{*,I}=$~1.52~$\Upsilon_{\odot,I}$) yields better fits,
but also pushes the halo concentration to higher values.

As a single case, the high concentration of NGC~1407 is not a problem,
but we note that most of the X-ray bright galaxies and groups examined
by H+06b and by \citet{2007ApJ...669..158G} 
were also found to have high concentrations, 
and unconventionally low values for $\Upsilon_*$
(see also the high concentrations in 
\citealt{2007MNRAS.379..209S,2008MNRAS.390L..64D,2008ApJ...685L...9B,2008MNRAS.390.1647B}).
It has been shown that improper modeling of the stellar mass in galaxies
can strongly skew X-ray based inferences about their DM halos
\citep{2005MNRAS.362...95M}.
However, in the cases of NGC~1407 as well as the other galaxies 
studied by H+06b (see also H+08), it
seems that the X-ray data simply do not allow for plausible $\Upsilon_*$ values.
H+06b found best-fits for NGC~1407 corresponding to 
$\Upsilon_{*,B}\sim$~2--3~$\Upsilon_{\odot,B}$,
as compared to their stellar population estimate of
$\sim$~5--6~$\Upsilon_{\odot,B}$ and ours of
$\Upsilon_{*,B}\sim$~4--7~$\Upsilon_{\odot,B}$
(see \S\ref{sec:assumpt};
using our shorter adopted distance would lessen this discrepancy).

\begin{figure}
\includegraphics[width=3.4in]{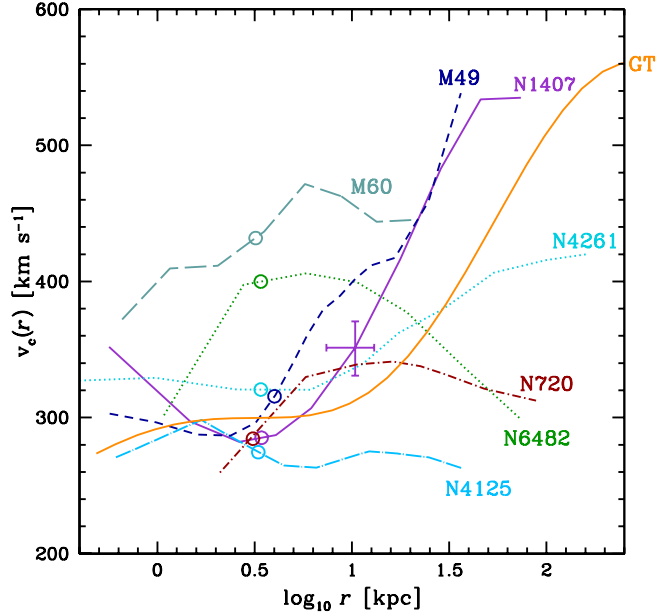}
\caption{
Circular velocity profiles of seven X-ray bright galaxies from H+06b.
Also shown is our best-fit GC-based solution for NGC~1407 ({\it solid orange curve}; model GT).
The {\it open circles} mark 1~\Reff{} (in the $K_s$-band) for each galaxy.
{\it Error bars} show the typical radial binning and
mass uncertainty at one measurement point.
\label{fig:kink} 
}
\end{figure}

We illustrate the situation further in Fig.~\ref{fig:kink} using the deprojected X-ray
results of H+06b for their entire sample (revising the distance to 21~Mpc for NGC~1407).
Presented as circular velocity profiles, several trends are immediately apparent.
Three galaxies (NGC~720, NGC~4125, and NGC~6482) have flat or declining profiles
consistent with galaxy-scale halos (modulo some issues with the innermost data points).
The other four galaxies have increasing profiles suggesting group-scale halos
(cf. also \citealt{2009arXiv0903.2540N}).
However, these increases generally set in much too quickly (by $\sim$~1~\Reff{}),
as can be seen by comparison with the $\Lambda$CDM-based model for NGC~1407.
In order to reproduce these kinks,
a combination of low $\Upsilon_*$ and high $c_{\rm vir}$ is necessary
(though still not sufficient).

The combination of awkward mass profile kinks, and
non-standard stellar and DM mass parameters, leads us to suspect a systematic
problem with some X-ray mass inferences.
The kinks found by H+06b 
seem to coincide with strong increases in $T_{\rm X}$, which are examples
of the common ``cool core'' phenomenon found in galaxy groups and clusters
(e.g., \citealt{2009ApJ...693.1142S}).
Independently of the wider implications of the cool cores, we wish to know
whether there is a problem with applying equation~\ref{eqn:Mr} in these regions.
In the case of NGC~1407, the onset of the kink
coincides not only with a rapid $T_{\rm X}$ increase, but also
with an onset of increasing gas isophote ellipticity 
(from $e=0.0$ at 0.6$'$ to $e=0.3$ at 1.7$'$) and slight asymmetry.
It thus seems plausible that the gas in the kink region is not in hydrostatic
equilibrium, or that there is additional non-thermal pressure support 
inside the kink radius (which might produce mass underestimates in these regions).
Testing these possibilities with independent mass tracers is exactly one of the motivations
of the present GCS kinematics study, as we will see in the following sections.

\subsubsection{Multiple mass probes in NGC~1407}
\label{sec:mult1}

\begin{figure}
\includegraphics[width=3.4in]{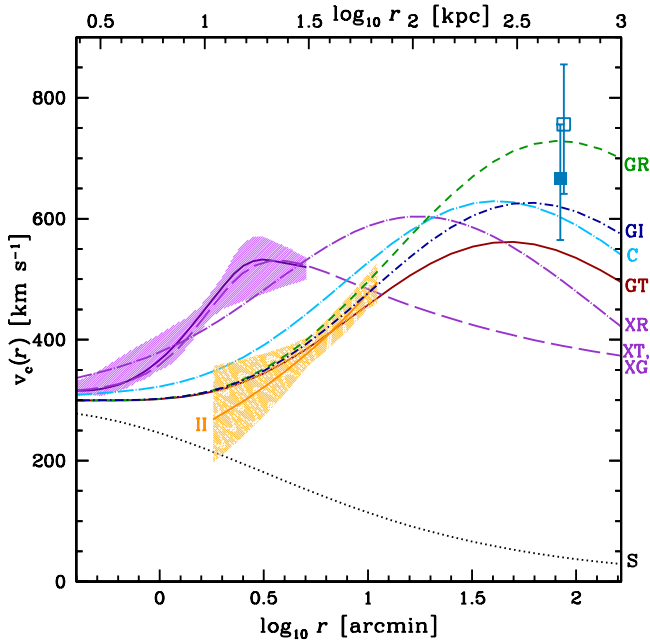}
\caption{
Circular velocity profile of NGC~1407, plotted vs. log radius.
Individual models are labeled by letters and are described in the text and in
Table~\ref{tab:models}.
Further included is our X-ray result (\S\ref{sec:xray}).
The boxes with error bars show estimates from group galaxy dynamics (B+06b), using
the friends-of-friends sample ({\it filled box}) or the entire sample of galaxies in
the projected region ({\it open box}).
\label{fig:models3} 
}
\end{figure}

We next compare results from the available independent probes on the mass profile of the NGC~1407 group.
These constraints include the GC and X-ray analyses that we presented in \S\ref{sec:dyn}
and \S\ref{sec:xray}, as well as the dynamics of galaxies in the group (B+06b), and
are summarized in Figs.~\ref{fig:cM} and \ref{fig:models3}.

B+06b used a friends-of-friends algorithm to identify 19 galaxies belonging to the group,
and a virial estimator to convert the kinematics into a mass.
Their galaxy sample notably excluded NGC~1400, the bright early-type apparently in close 
three-dimensional proximity
to NGC~1407, but with a high peculiar velocity.
B+06b made an alternative mass estimate including
NGC~1400 and another 5 galaxies in the region of the group---yielding both a higher
group mass and a higher group luminosity; B+06a provided yet another updated
mass that is intermediate to the previous two.
The results for both the low- and high-mass group samples are shown
in Fig.~\ref{fig:models3}, implying a mass at 0.5~Mpc of (4--10)$\times10^{13} M_{\odot}$,
or $\Upsilon_B=$~600--1200~$\Upsilon_{\odot, B}$
(the corresponding group velocity dispersion is $\sigma_{\rm p}=$~320--450~km~s$^{-1}$).
\citet{2006MNRAS.369.1375T} identified 10 new dwarf galaxies belonging to the
group, and estimated a virial mass of $6.1\times10^{13} M_{\odot}$.
We plan to revisit this calculation in the future with the addition of even
more group dwarfs to the catalog.
Using the B+06b low-mass results for now,
the permitted regions of $M_{\rm vir}-c_{\rm vir}$ parameter space are shown
in Fig.~\ref{fig:cM}.

The orbit anisotropy of the group galaxies is an additional source of 
systematic uncertainty as with the GC dynamics;  
the virial mass estimator used by B+06b implicitly assumes isotropy.
Somewhat conflicting conclusions have been made about the accuracy of
isotropic mass estimators in galaxy groups and clusters in general
\citep{2006MNRAS.369..958S,2006A&A...456...23B,2006MNRAS.370..427B,2007MNRAS.375..313F,2008ApJ...676..218H}.
In the case of the NGC~1407 group, B+06b found a near-zero skewness for the
galaxy velocities as well as a crossing time of only 0.02~$H_0^{-1}$, supporting
the assumption of the group as virialized, and
some indication of negative kurtosis, suggesting that isotropic
estimators might overestimate the mass.
In any case, the systematic mass uncertainty is probably no larger than the
$\pm40\%$ statistical uncertainty.

Now comparing the results from the three mass probes,
we see that they each independently imply a virial mass for the NGC~1407 group
within the range $M_{\rm vir} \sim$~(4--10)$\times 10^{13} M_{\odot}$
(Fig.~\ref{fig:cM}).
This is an encouraging broad consistency, but we are further interested in
the accuracy of the spatially-resolved mass profiles.
The group galaxies and GC dynamics results are consistent
(assuming fiducial DM models and any of the GC anisotropy profiles 
discussed in \S\ref{sec:fid} and \S\ref{sec:modalt}),
providing no evidence so far of a
major problem with either approach.
But snags appear when considering the X-ray mass profile:
while the virial mass from the group galaxies and the X-rays is compatible
(Fig.~\ref{fig:cM}), the mass profile at smaller radii conflicts both
with the theoretical expectations for a $\Lambda$CDM halo (see also \S\ref{sec:exX})
and with the GC-derived profile.
The best-fitting CDM halo parameters from the GC and X-ray data are inconsistent
at the 2~$\sigma$ level. This discrepancy also exists
{\it independently of the mass model assumed}, as seen in the overlap region
($\sim 2.5'$--5$'$) of Fig.~\ref{fig:models3}, where the unparameterized
mass profile inferred from the GCs is $\sim$~70\% lower than the X-ray result.
This difference is approximately halved when considering instead the deprojected
X-ray profile of H+06b, but the problem persists.

Our GC results are not yet robust, with the orbital anisotropy as the
key remaining source of potential error.
To see if this systematic uncertainty could account for the GCS/X-ray discrepancy,
we construct a toy model where the X-ray-based mass profile is assumed to be 
correct, and investigate the GCS dynamics in the corresponding potential.
To represent the X-ray result, we construct a $v_{\rm c}(r)$ profile
with an arbitrary functional form:
\begin{equation}
\label{eqn:vc}
v_{\rm c}(r)=v_0+\frac{v_1 r^\delta}{1+(r_{\rm c}/r)^\zeta} .
\end{equation}
Model XT/XG has parameters
($v_0=$~312~km~s$^{-1}$, $v_1=$~404~km~s$^{-1}$, $r_{\rm c}=1.94'$, $\zeta=3.1$, $\delta=-0.37$)
and is shown in Fig.~\ref{fig:models3};
the range of plausible values for the asymptotic profile exponent $\delta$ is
between $-0.5$ and $+0.08$.

Assuming $\beta=-0.5$ for the GCS as derived in \S\ref{sec:modalt},
we carry out a Jeans analysis in this mass potential, 
and illustrate the resulting prediction for $\sigma_{\rm p}(R_{\rm p})$
in Fig.~\ref{fig:models1} (model XT).
This model is a poor fit to the data, mainly because
the relatively high X-ray-derived $v_{\rm c}$ at
$\sim$~3$'$ overpredicts $\sigma_{\rm p}$ for the GCs around this intermediate radius.
Setting aside our estimate for the anisotropy, we next investigate what profile $\beta(r)$
would be required to reproduce the GC observations, given the X-ray mass model
(cf. \citealt{2003ApJ...599..992M}).
Some qualitative experimentation reveals that
in order to suppress $\sigma_{\rm p}$ at intermediate radii while 
keeping $\sigma_{\rm p}$ high at larger radii, 
the GC orbits must be very radial in the center, and then rapidly become
tangential at large radii.
More quantitatively, we use the Tonry inversion method (see \S\ref{sec:modalt})
to derive $\beta(r)$ for the GCS, given the observed $\sigma_{\rm p}(R_{\rm p})$
(outside the central regions) and the assumed $v_{\rm c}(r)$.

The result of this mass-anisotropy inversion is shown in Fig.~\ref{fig:beta}:
the GC orbits are near-isotropic in a narrow range around $r\sim 8'$, becoming
very tangential at larger radii, and very radial at smaller radii
(with an unphysical $\beta > 1$ for $r \lsim 3'$).
Interestingly, while the dispersion data inside $1.75'$ were not used in this analysis,
the model does fit them {\it a posteriori} 
(see model XG in Fig.~\ref{fig:models1}, which uses a parameterized approximation to the
numerically inverted anisotropy profile).
This agreement lends credence to the existence of some objects 
(possibly DGTOs; \S\ref{sec:disp}) on very radial orbits near the galaxy center, 
but it is not clear that this scenario would solve the GC/X-ray inconsistency problem
(q.v. {\it dashed curve} in Fig.~\ref{fig:models2}).

These anisotropy conclusions are not substantially affected by 
the extrapolation of the X-ray mass profile (parameter $\delta$
in Eq.~\ref{eqn:vc}), nor by the uncertainties in the GC density
distribution $\nu(r)$;
they are also completely independent of the distance assumed.
Adopting the alternative X-ray based mass profile from H+06b (which notably 
accounts for a metallicity gradient) does not substantially change the
$\beta(r)$ result.
Reconciling the GC and X-ray data therefore
seems to require an {\it ad hoc} GC anisotropy profile with extreme
radial variations, conflicting with the anisotropy inferred from the
GC data themselves.
As we will discuss in \S\ref{sec:gckin}, this profile is also highly implausible
on theoretical grounds.

We therefore have apparently found a fundamental incompatibility between the GC and X-ray results.
However, it is premature at this juncture to pinpoint the culprit, since
there are puzzling quirks in both analyses (the X-ray kink and the central GC dispersion peak).
More exhaustive study of this system will be needed in the future, 
including stronger empirical constraints on the GC anisotropy (see \S\ref{sec:mult2}), 
and a dynamical analysis of the field star kinematics---which 
should provide a crucial cross-check on the central X-ray mass profile
(cf. the study of M60 by H+08, where the mass inside 1~\Reff{} derived from
stellar dynamics may be significantly higher than the X-ray-based mass).

For now, we
wish to estimate the best mass model that jointly fits all of the constraints,
surmising in the spirit of the central-limit theorem that the various systematic
problems will roughly cancel out.
The inconsistencies between the different constraints do suggest that we
not na\"ively use the usual method of simply adding up all of their $\chi^2$ values,
but rather attempt to account for the systematic uncertainties through a weighting scheme.
In particular, 
we do not wish the tighter constraints from the X-ray analysis to dominate
the final solution.

Combining weighted data sets is a non-trivial exercise
(see e.g., \citealt{1996astro.ph..4126P}; \citealt{2002MNRAS.335..377H}),
and for now we adopt the following {\it ad hoc} approach.
We renormalize the $\chi^2$ corresponding to the X-ray and GC constraints
by their best-fit values when each of them is fitted independently, including
the $\Lambda$CDM concentration prior.  
The new $\chi^2$ functions, as well
as the unmodified group-galaxies $\chi^2$
(corresponding to the total mass inside 500~kpc; see Fig.~\ref{fig:models3}),
are summed along with the concentration prior to produce a new total $\chi^2$.

The resulting ``consensus'' solution is a normal $\Lambda$CDM halo
with $M_{\rm vir} \sim 7\times10^{13} M_{\odot}$ (see model C in Table~\ref{tab:models}
and Figs.~\ref{fig:cM} and \ref{fig:models3}).
With this solution, DM becomes dominant over the baryonic mass
outside of $r \sim 1.3'$ ($\sim 1.4$~\Reff~$\sim $ 8~kpc).
The benchmark mass-to-light ratio gradient parameter introduced by \citet{2005MNRAS.357..691N}
is $\dML \sim 2.0$ for NGC~1407, which is by far the highest value reported for any
early-type galaxy, demonstrating the extreme dominance of DM in this system.
Note that the DM scale radius $r_s$ is much larger than the
extent of the GC kinematics data (120 kpc vs 60 kpc), so it
is really the group galaxy kinematics that determines $M_{\rm vir}$,
with $c_{\rm vir}$ then pinned down by the GC and X-ray
constraints\footnote{The crossing time for the outer GCs is only $\sim$~0.2 Gyr, 
and for the group galaxies is $\sim$~0.3 Gyr, so non-equilibrium effects
are unlikely to be the explanation for the high mass inferred.}.
In the near future, it should be possible to push the GC kinematics measurements
out to larger radii, and gain a stronger constraint on $M_{\rm vir}$ without
reference to the group galaxies.

One alternative estimate for the mass profile would be derived from the GCs and
group galaxies only, disregarding the X-ray data and the $\Lambda$CDM concentration prior.
The resulting model GG has a mass of $M_{\rm vir} \sim 1.1\times10^{14} M_{\odot}$
and a slightly low concentration (Fig.~\ref{fig:cM}).

\subsubsection{Multiple mass probes in other systems}
\label{sec:mult2}

More detailed analyses of the current and future data, including direct constraints on
the GCS anisotropy, should provide a more accurate picture of the NGC~1407 mass profile.
But given the apparent conflicts in this case between independent mass constraints, we pause
to review similar comparisons that are available in other early-type galaxies.
Such an approach has become commonplace in the arena of galaxy clusters,
where gravitational lensing, X-rays, and galaxy dynamics may all be used as cross-checks
and to provide stronger constraints when used jointly
(e.g.,
\citealt{1998ApJ...505...74G}; 
\citealt{2004ApJ...604...88S}; 
\citealt{2006MNRAS.370..427B}; 
\citealt{2008MNRAS.384.1567M}). 
But comparable studies in galaxies are in their infancy, as the data are difficult to acquire.

A handful of galaxy halos have been studied using both X-rays and kinematics;
these are mostly cases of central group and cluster galaxies with GC data,
but sometimes also with PN kinematics
(e.g., \citealt{1997ApJ...486..230C,2000A&AS..144...53K}; RK01; C+01; 
\citealt{2002A&A...383..791N}; C+03;
\citealt{2004AJ....127.2094R};
\citealt{2006ApJ...636..698F,2006MNRAS.373..157B,2006A&A...459..391S,2006SerAJ.173...35S,2006ApJ...643..210W};
\citealt{2008A&A...478L..23R}, hereafter R+08;
\citealt{2008AJ....135.2350C}; J+09).
Within the uncertainties, most of these X-ray/kinematics inter-comparisons are 
broadly consistent:
the discrepancy we find in NGC~1407 is one of the largest seen\footnote{\citet{2003ApJ...599..992M}
found a discrepancy between X-ray and dynamics-based mass profiles in the
{\it central} regions of M49.}.
However, the kinematical tracers in these studies have generally been assumed to
be fairly isotropic---which may indeed turn out to be typical
for the GCs in group/cluster central galaxies, but is not yet certain.

Given the freedom to adjust orbital anisotropies, one could make almost any two
data sets agree (e.g., \S\ref{sec:mult1}), 
so to conclusively compare mass results, one must derive independent
constraints on anisotropy.
This is possible by measuring higher-order shapes of the 
LOSVDs, which can be used to constrain the types of orbits
(c.f. our crude kurtosis analysis in \S\ref{sec:modalt}).
Such mass/anisotropy degeneracy breaking techniques are well-established for 
integrated-light stellar kinematics, but can also be applied to discrete velocities 
(e.g., RK01; \citealt{2003Sci...301.1696R,2006ApJ...643..210W,2008MNRAS.385.1729D,2008ApJ...682..841C}).
Although it is commonly supposed that
a few hundred velocities are not sufficient to break this degeneracy,
in fact additional constraints may be brought to bear on the problem.
One is to make use of stellar kinematics including their LOSVDs to pin down the
mass profile out to $\sim$~1--2~\Reff{} (cf. RK01).
Another is to assume that the
underlying spatial profile of the tracer population is fairly well known, e.g.,
the total GC surface density profile represents the same population as the kinematically-measured GCs
(cf. \citealt{1993ApJ...409...75M,1993AJ....106.2229M}).

\begin{figure}
\includegraphics[width=3.4in]{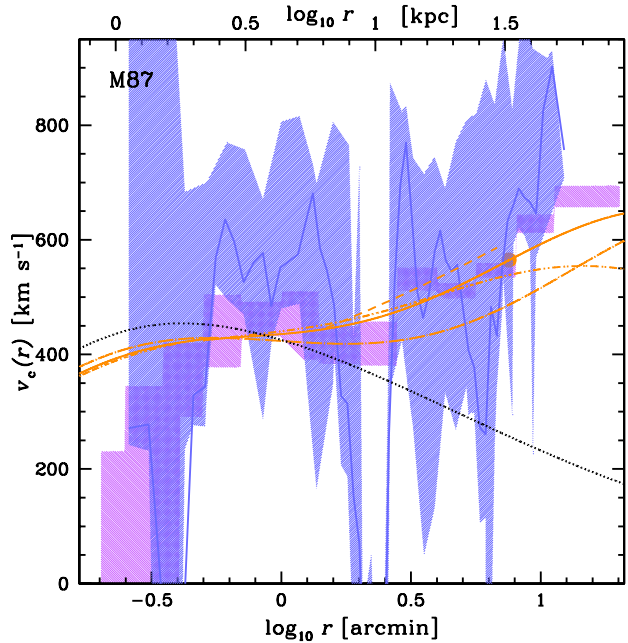}
\caption{
Mass modeling results for M87.
The orange solid curve shows the best fitted $\Lambda$CDM-based orbit model to the stellar
and GCS kinematics; the dot-dashed curves show other acceptable solutions (RK01).
The filled circle shows the outer extent in radius of the kinematics data fitted.
The orange dashed curve shows the result of anisotropic distribution-function based
modeling of the GCS kinematics by \citet{2006ApJ...643..210W}, with the extent of the curve
showing the radial range of the data fitted.
The purple shaded region shows the {\it XMM-Newton}-based mass model of \citet{2002A&A...386...77M}.
The blue jagged solid line with shaded region shows the {\it Chandra}-based model of C+08 (mean molecular weight $\mu=0.61$ assumed),
with $4\times$~radial binning applied to smooth out the highest-frequency variations;
the uncertainty region incorporates the differences from single- and multi-temperature modeling.
The black dotted curve shows a model with no DM.
The figure thus shows that the detailed optical kinematics constraints are 
broadly consistent with the X-ray mass results, but without artificial smoothing,
the X-ray data imply an unphysical mass profile.
\label{fig:m87}
}
\end{figure}

So far, the only galaxy to have its DM content studied using both high-quality X-ray and
optical data (including LOSVD constraints) is M87.
RK01 used nonparametric ``orbit modeling'' of the dynamics of the GCs and field stars in this galaxy, 
deriving a mass profile that may be compared with the subsequent {\it XMM-Newton}
result of \citet{2002A&A...386...77M}.
As seen in Fig.~\ref{fig:m87}, the optical and X-ray results in the region of overlap
are nicely consistent, except at small radii ($<$2~kpc), where the typical
problem of an X-ray core mass underestimate is apparent
(e.g., \citealt{1997ApJ...486L..83B}; H+08).

\citet[hereafter C+08]{2008MNRAS.388.1062C} have recently used new {\it Chandra} data to examine the 
mass profile of M87 in 
more detail, and again found that the optical and X-ray results on the mass profile are 
broadly consistent, but that the X-ray profile contains rapid radial variations
that are not seen (by construction) in the smooth optical models.
These ``wiggles'' appear at a nuisance level when examining the gravitational 
potential, but when differentiated to derive a profile of $v_{\rm c}(r)$ or $M(r)$, 
the wiggles are amplified catastrophically, 
even implying at one location ($\sim$~10 kpc) a {\it negative} enclosed mass
(see Fig.~\ref{fig:m87} and \citealt{2004ApJ...609..638G}).
The wiggles were not seen
in the earlier {\it XMM-Newton} study because of its
use of smooth analytical models to describe the gas density and temperature profiles.
This standard practice avoids amplifying noise in low-$S/N$ data, but in the
case of M87, it appears to have suppressed real features in the data
(see also J+09).

C+08 showed that the apparent M87 mass wiggles could be readily 
explained as a consequence of shock waves propagating through the gaseous interstellar 
medium.  They claimed that one can ignore these localized effects and still safely consider
global changes in the mass potential, taking also into account additional
(apparently minor) corrections to the X-ray based mass such as non-thermal pressure 
support (see also 
\citealt{2003ApJ...592...62P};
\citealt{2004MNRAS.350..609C,2006MNRAS.370.1797P,2006MNRAS.369.2013R,2007ApJ...655...98N,2007ApJ...668..150D,2007MNRAS.380.1409O}).
However, in practice, it is difficult to know {\it a priori} how to smooth out
the X-ray wiggles, and thus the conclusions derived about DM halo properties in detail
can be strongly affected.  NGC~1407 is a case in point,
as its X-ray kink feature could be attributed to a shock, but because we
do not have more extended X-ray data to provide the asymptotic mass
profile, the feature forces the $\Lambda$CDM models toward high concentrations
and low values for $\Upsilon_*$.
However, there is less {\it a priori} reason to suspect AGN-generated shocks in NGC~1407,
as its X-ray image shows no major disturbances, and 
its radio activity is $\sim$~1000 times less than that of M87
\citep{1998AJ....115.1693C}.

M87 has clearly disturbed X-ray isophotes, while there may also exist a class of
fairly undisturbed systems where the X-ray results will be very reliable.
\citet{2008ApJ...680..897D} have introduced a parameter $\eta$ to describe asymmetries
in gaseous X-ray halos.
The correlations of $\eta$ with central radio and X-ray luminosity,
and with outer temperature gradient, suggest that both the ambient medium and the
central AGN cause disturbances in the X-ray gas---even in cases such as N1407 where
the AGN activity appears fairly weak.

Values of $\eta$ were provided for most of the galaxies with X-ray/kinematics mass
comparisons (see references at the beginning of \S\ref{sec:mult2}).
The galaxies are M60, M49, NGC~1399, NGC~4636, and NGC~1407, with 
$\eta = 0.05, 0.08, \leq0.09, 0.09$, and 0.17, respectively.
NGC~1407 has a relatively high value of $\eta$ (despite having less obvious
disturbances than e.g. NGC~4636) and manifests a mass discrepancy,
but there is otherwise no obvious correlation between $\eta$ and the reliability
of the X-ray results. So far, none of the other galaxies clearly shows a large
mass discrepancy, but all of them 
have at least a hint of
a non-equilibrium kink in the mass profile---even in the case of M60, which 
has one of the smoothest gas halos in the \citet{2008ApJ...680..897D} sample.

\begin{table*}
\begin{center}
\caption{Kinematical properties of nearby group/cluster-central galaxies.}\label{tab:gal}
\noindent{\smallskip}\\
\begin{tabular}{l c c c c c c c c c c c}
\hline
NGC & Type & Env. & $N_{\rm GC}$ & $D$ & $M_B$ & $\sigma_0$ & $v_{\rm rot}/\sigma_{\rm p}$ & $v_{\rm rot}/\sigma_{\rm p}$ & $v_{\rm RMS}$ & $v_{\rm RMS}$ &  Ref. \\
& & & & [Mpc] & & [km~s$^{-1}$] & (MPGCs) & (MRGCs) & (MPGCs) & (MRGCs) \\
\hline
\noalign{\smallskip}
1399 & E1 & CCG & 228 & 18.5 & $-21.1$ & $344\pm6$ & $0.19^{+0.11}_{-0.19}$ & $0.11^{+0.10}_{-0.11}$ & $311^{+24}_{-20}$ & $258^{+22}_{-17}$ & D+04, R+04 \\
1407 & E0 & GCG & 172 & 20.9 & $-21.0$ & $273\pm6$ & $0.18^{+0.11}_{-0.18}$ & $0.24^{+0.13}_{-0.18}$ & $234^{+21}_{-16}$ & $247^{+21}_{-17}$ & this paper \\
3379 (M105) & E1 & GCG & 38 & 9.8 & $-19.9$ & $207\pm2$ & --- & --- & --- & --- & B+06c \\
4472 (M49) & E2 & GCG & 127 & 15.1 & $-21.7$ & $291\pm8$ & $0.30^{+0.14}_{-0.22}$ & $0.26^{+0.16}_{-0.26}$ & $373^{+38}_{-29}$ & $215^{+28}_{-23}$ & C+03 \\
4486 (M87) & E3 & CCG & 144 & 14.9 & $-21.5$ & $332\pm5$ & $0.33^{+0.13}_{-0.20}$ & $0.66^{+0.19}_{-0.23}$ & $414^{+39}_{-32}$ & $383^{+47}_{-35}$ & H+01 \\
4636 & E2 & GCG & 88 & 13.6 & $-19.8$ & $203\pm3$ & $0.17^{+0.15}_{-0.17}$ & $0.57^{+0.36}_{-0.57}$ & $209^{+25}_{-20}$ & $180^{+26}_{-18}$ & S+06 \\
4649 (M60) & E2 & GCG & 59 & 15.6 & $-21.3$ & $336\pm4$ & $0.48^{+0.19}_{-0.28}$ & $0.53^{+0.26}_{-0.53}$ & $215^{+31}_{-23}$ & $178^{+45}_{-28}$ & L+08 \\
5128 & E2/S0 & GCG & 170 & 3.9 & $-20.7$ & $120\pm7$ & $0.54^{+0.24}_{-0.27}$ & $0.40^{+0.16}_{-0.20}$ & $75\pm11$ & $159^{+16}_{-13}$ & W+07 \\
\noalign{\smallskip}
\hline
\hline
\end{tabular}
\end{center}
\tablecomments{
The third column denotes cluster-central galaxy or group-central galaxy.
The fourth column is the number of GCs with measured velocities
(after discarding outliers).
The seventh column shows the galaxy central velocity dispersion from 
HyperCat\footnote{http://leda.univ-lyon1.fr// \citep{2003A&A...412...45P}.}.
The eighth through eleventh columns show parameters for the
metal-poor and metal-rich GCs (MPGCs, MRGCs),
as recalculated for this paper.
The data references are
H+01: \citet{2001ApJ...559..812H};
C+03: \citet{2003ApJ...591..850C};
D+04: \citet{2004AJ....127.2114D};
R+04: \citet{2004AJ....127.2094R};
B+06c: \citet{2006A&A...448..155B};
S+06: \citet{2006A&A...459..391S};
W+07: \citet{2007AJ....134..494W};
L+08: \citet{2008ApJ...674..857L}.
}
\end{table*}

On a related note, 4 of the 7 galaxies studied in X-rays by H+06b
have deprojected mass profiles that are inconsistent with the forward-fitted smooth
NFW-based models (see \S\ref{sec:exX}).  
These are again cases where the X-ray approach may be problematic,
but the $\eta$ values for them and the other 4 galaxies are not clearly different.
C+08 did note that $\eta$ may not be a reliable diagnostic of 
hydrostatic equilibrium, and we speculate that a metric based on {\it radial} 
(rather than azimuthal) fluctuations in the X-ray photometry and/or temperature 
may be more useful for this purpose.

The generic reliability of X-ray mass constraints could be further tested by
applying robust dynamical models (as in the case of M87) to the stellar and GC data in
NGC~1407 and in the other galaxies mentioned 
(while also extending the models to consider non-spherical effects).  
These are demanding exercises that we defer to future papers.
One might consider the X-ray analyses more suspect owing to the highly uncertain
physics involved, but even in the relatively well-posed case of gravitational dynamics,
there are questions about the reliability of the modeling results
(e.g., \citealt{2005Natur.437..707D}).
The latter issue can be addressed by cross-checking independent kinematical
constraints, e.g., from PNe, and from the metal-poor and metal-rich GC subsystems
separately.  We hope in the future to pursue these avenues in NGC~1407 as well
as in other galaxies; studies with GCs and PNe will also be useful for testing
the presumed association between metal-rich GCs and field stars.

\section{The group context}
\label{sec:context}

Having determined some properties of the mass distribution and GC kinematics
in the NGC~1407 group, we now attempt to place this system
in the context of what is known and expected for galaxy groups and clusters.
Are the high $\Upsilon$ value and rising $v_{\rm c}(r)$ profile really unusual 
for groups of comparable luminosity?
We may alternatively consider the system as a dim cluster, since its mass approaches
the usual (arbitrary) group/cluster boundary of $10^{14} M_{\odot}$:
is it rare for clusters to have been so inefficient in forming stars?
How do the GCS kinematics compare to those of other systems?

To address these questions, we compile all of the published data 
available for GCS kinematics in group- and cluster-central galaxies
(generically abbreviated as GCGs)
with reasonably large and radially-extended GC velocity catalogs
(see Table~\ref{tab:gal}).
For the sake of consistency,
we re-derive the kinematical properties of these data sets using the same techniques
that we have applied to NGC~1407.
The redshifts of all the galaxies besides NGC~1407 are low enough to cause
potential contamination from Galactic stars,
which can easily distort the kinematical inferences by their accidental inclusion, 
or by over-compensating and excluding some low-velocity GCs.
To be conservative, we have
rejected all objects with velocities below
$v_{\rm sys}$ of the host galaxy\footnote{A similar exercise for NGC~1407 yields
rotation and kurtosis parameters that are formally inconsistent with the full data-set
results, which is a warning that the uncertainties in the kinematical parameters
are probably underestimated in general.}.

We discuss the GCS kinematics and dynamics in \S\ref{sec:gckin}, the mass profiles in
\S\ref{sec:massprof}, the baryon content in \S\ref{sec:bary},
and some implications in \S\ref{sec:impl}.

\subsection{GCS kinematics and dynamics}\label{sec:gckin}

The kinematics of any galaxy should provide clues about its evolutionary history.
However, GCGs 
have fairly similar optical 
properties to their more isolated elliptical galaxy counterparts 
(e.g., \citealt{2007MNRAS.379..867V}; \citealt{2008MNRAS.389.1637B,2008MNRAS.391.1009L}),
which might reflect an early galaxy collapse history decoupled from the
more recently assembled surrounding group.
To probe the assemblies of both the GCGs and their groups,
the large radial range of the GCs could be of service.
The properties we will consider include the rotation (\S\ref{sec:gcrot}),
velocity dispersion (\S\ref{sec:gcdisp}), and orbits (\S\ref{sec:gcorb}) of GCs.

\subsubsection{Rotation}\label{sec:gcrot}

\begin{figure*}
\includegraphics[width=3.4in]{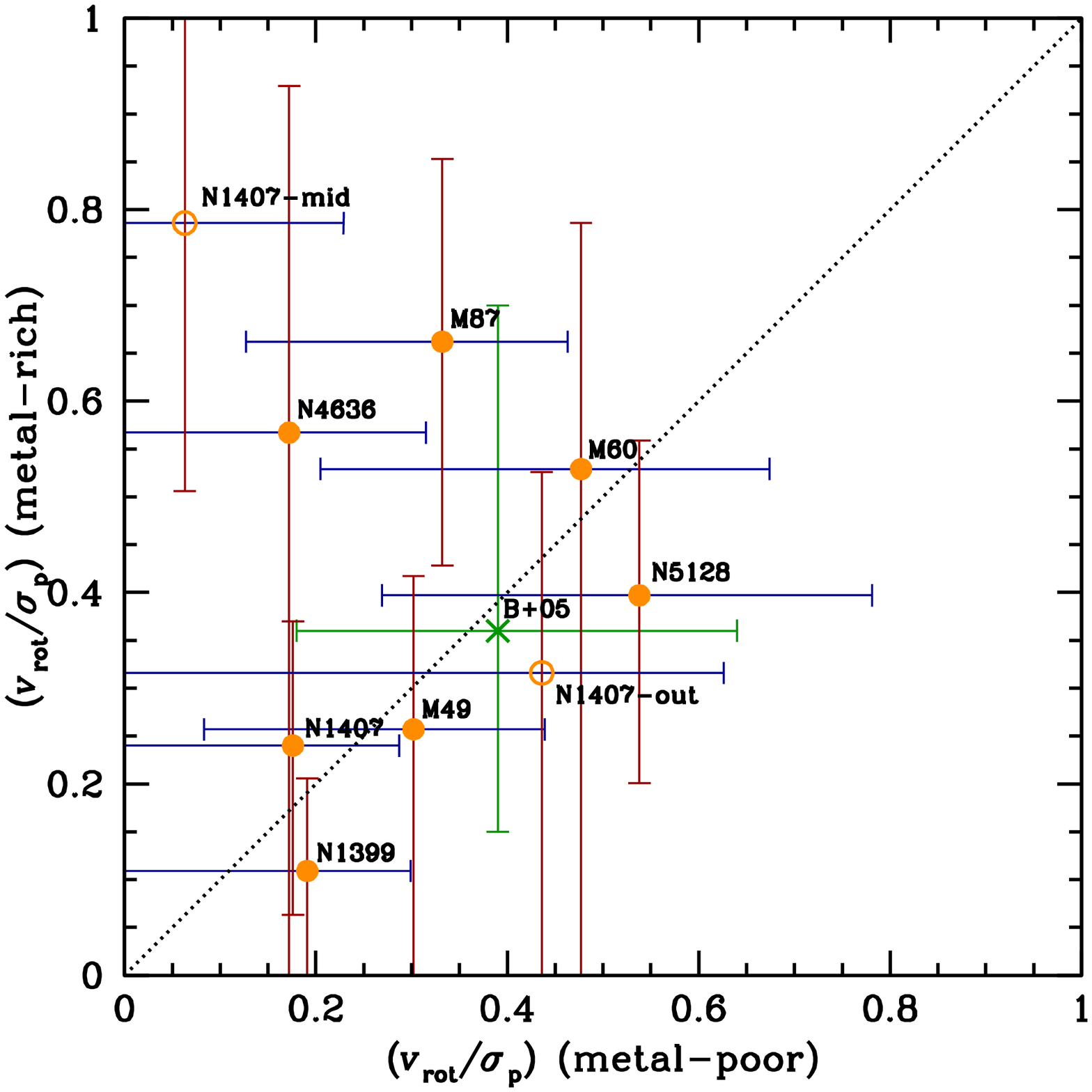}
\includegraphics[width=3.4in]{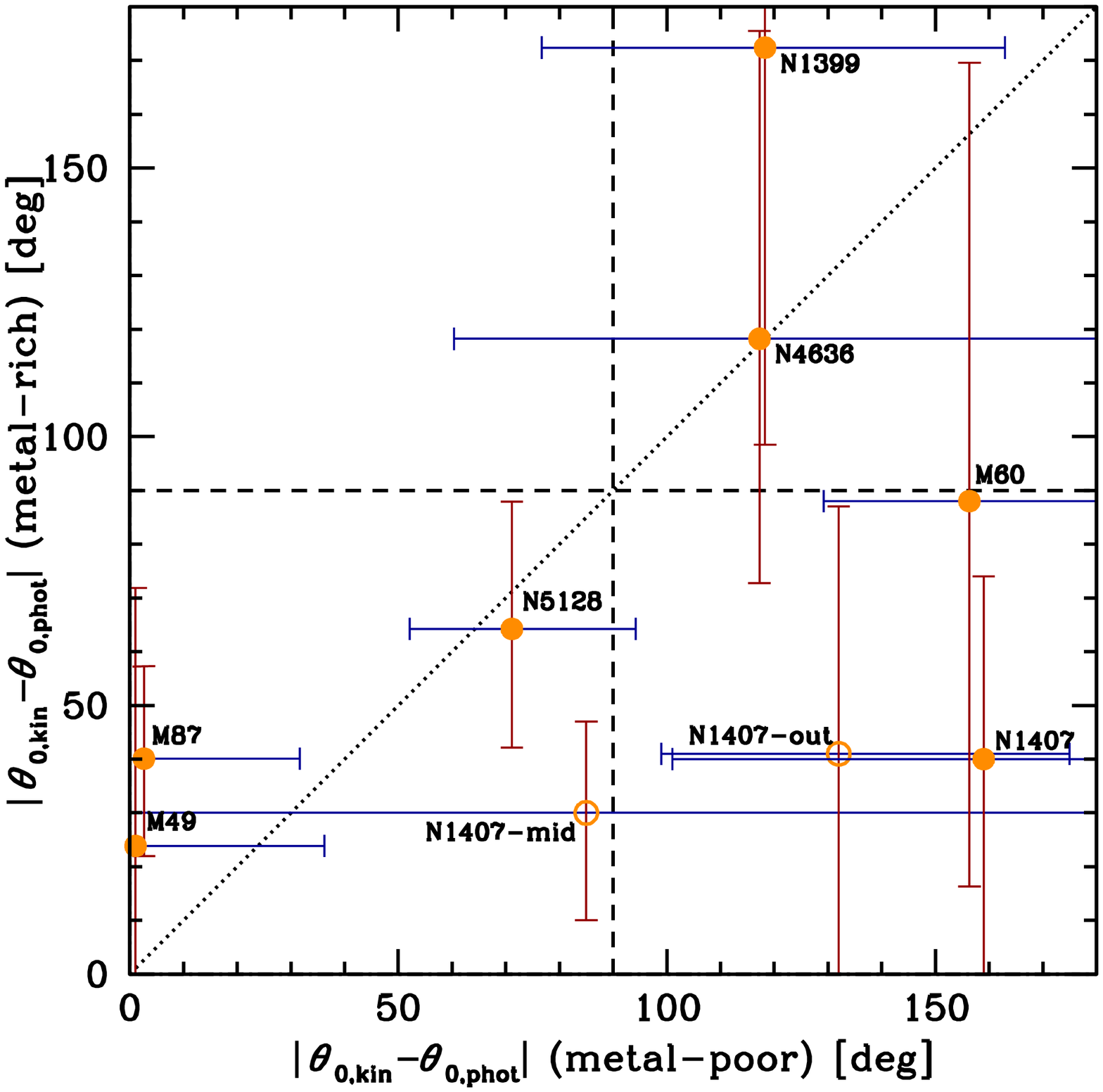}
\caption{GCS rotation
parameters for group/cluster-central galaxies, 
with the metal-poor and metal-rich subsystems compared.  
In the case of NGC~1407, separate points are plotted for the overall
GCS, for intermediate radii (1.77$'$--4.52$'$), and for outer radii
(4.55$'$--11.1$'$).
The dotted diagonal lines mark equal values for the subsystems.
{\it Left:} Rotation amplitude relative to velocity dispersion.
The $\times$ marks the results of the merger simulations of B+05.
{\it Right:} Kinematic misalignment.
The angles $\theta_{0,{\rm kin}}$ and $\theta_{0,{\rm phot}}$ correspond to 
the angular momentum vector and to the photometric minor axis, respectively.
Classic rotation around the minor axis is at $0^\circ$,
rotation {\it along} it at $90^\circ$ (dashed lines), and
counter-rotation at $180^\circ$.
\label{fig:vsig}
}
\end{figure*}

We begin by examining GCS rotation in the galaxy sample, considering
the metal-poor and the metal-rich rotational dominance
parameter ($v_{\rm rot}/\sigma_{\rm p}$) and the kinematic misalignment angle
($\theta_{0,\rm kin}-\theta_{0,\rm phot}$, where the $\pm180^\circ$ degeneracy
in $\theta_{0,\rm phot}$ is broken by reference to the direction of stellar rotation
from the literature).
These parameters are in general very dependent on the sampling of the GC velocity data
by color, magnitude, and position---as illustrated in 
Table~\ref{tab:rot} and Fig.~\ref{fig:vsig}
by the results for different radial bins of NGC~1407 GCs.
However, momentarily throwing caution to the wind, we plot 
all of the available data from Table~\ref{tab:gal}
(except for NGC~3379 which does not have enough GC velocities for this purpose).

These systems are generally consistent with an overall
mild rotation of $v_{\rm rot}/\sigma_{\rm p} \sim$~0.2--0.6 seen equally in
both the metal-poor and metal-rich GCs
(see also discussions in
\citealt{2006AJ....131..814B}; \citealt{2007AJ....134..494W}; and H+08).
The kinematic position angles 
of the metal-poor and metal-rich GCs are in most cases consistent with each other, 
but with no preferred alignment relative to the galaxy isophotes
(i.e. both major- and minor-axis rotation are found, as well as orientations in-between).
Such kinematic misalignments are unsurprising, since the photometric axis of a
triaxial galaxy can be readily twisted in projection
(e.g., \citealt{1991ApJ...383..112F,1998ApJ...493..641R}).

The GCs also appear in some cases to be misaligned relative to the (central)
stellar kinematics, even to the extent of counter-rotation, which would imply
an intrinsic kinematic misalignment.
Furthermore, there are apparent kinematical twists with radius (see \S\ref{sec:rot})---as
in the case of M87, where such a twist may coincide with the onset of the cD envelope and
reflect a transition to the group environment (see C+01).
Note however that a stellar/GC misalignment, or a kinematical twist within a
population, does not necessarily imply radically decoupled sub-components, since subtle
shifts in the orbit populations with radius can dramatically alter the net angular momentum
(e.g., \citealt{2007MNRAS.376..997J,2008MNRAS.385..647V}).

Deciphering the implications for the intrinsic structure and kinematics of these galaxies and 
their GC systems will require improved data and in-depth modeling.
But our current overall impression is that the metal-poor and metal-rich GC subsystems have
similar rotational properties, which might be distinct from the rotation of the field stars.

Considering next the connections to galaxy formation theory, we
plot in Fig.~\ref{fig:vsig} ({\it left}) a theoretical prediction 
for the rotation amplitude from B+05, as
the median value and spread inside 6~\Reff{}
of simulations of dissipationless disk galaxy pair major mergers
(their multiple merger results appear to be fairly similar).
Broadly speaking, their predictions 
($v_{\rm rot}/\sigma_{\rm p} \sim$~0.4 for both metallicity subsystems)
agree with the data.
We will leave it as a future exercise to check their prediction that
the rotation increases with radius.

B+05 also predicted that the GCs would have kinematic misalignments relative
to the galaxy isophotes, with similar rotation axes for the metal-rich and 
metal-poor subsystems---again in broad agreement with the observations.
It is not clear whether kinematic twists with radius,
or kinematic misalignments between the stars and GCs, are 
expected from these merger models.
Intriguingly, B+05 also found that the rotation of a GC system aligns with the rotation
of its host DM halo,
potentially providing an indirect probe of an otherwise inaccessible property.

Detailed kinematics predictions do not exist for other GCS formation scenarios,
such as accretion, quasi-monolithic multi-mode galaxy collapse,
and metal-rich GC formation by major mergers---but with all of these scenarios,
one would probably
expect a difference between the rotation of the metal-poor
and metal-rich GCs that is not seen in the data (see discussion in H+08).
This observation suggests that the GCs were in general formed {\it prior} to the
assembly of their host GCG galaxies by mergers---which affected the
metal-rich and metal-poor GCs equally, 
replacing any pre-existing orderly rotation with merger-induced hot rotation.
A related implication is that the last major merger episodes in these galaxies 
were ``dry'' or ``damp'' (relatively gas-poor; \citealt{2007ApJ...659..188F}),
as also suggested by other studies
(\citealt{2007MNRAS.375....2D,2008MNRAS.385...23L,2008MNRAS.388.1537M};
\citealt{2008ApJ...683L..17T}).

Such a scenario of early GC formation
was considered as part of a recent foray into the modeling of 
GCGs in a cosmological context \citep{2008MNRAS.387.1131B}.
Their prediction of relatively low GCS rotation overall
($v_{\rm rot}/\sigma_{\rm p} \lsim 0.3$) appears to conflict with the observations, but
the simulations did not directly include non-gravitational processes, which could
alter the rotation predictions.
Also, their result reflects {\it all} of the GCs within the virial radius, while
the observations sample the GCs within only $\sim$~0.1~$r_{\rm vir}$.

It will be interesting to see if a different rotation pattern emerges from observations
of non-central ellipticals (e.g., B. Kumar et al., in prep.), and of the general class of
fainter, disky, rotation-dominated ellipticals and lenticulars 
(e.g., \citealt{2007MNRAS.379..401E}).
Such systems may have undergone more sporadic merger histories involving different mass 
ratios and more recent GC formation,
and may have experienced more outer halo tidal stripping than GCGs have.
Measurements of PN kinematics in all of these systems would also allow testing of the B+05
prediction that the GCs and the field stellar halos should
have kinematical similarities
(broadly confirmed in the cases of NGC~5128 and M31; 
\citealt{2004ApJ...602..705P};
\citealt{2007AJ....134..494W}; \citealt{2008ApJ...674..886L}). 

To be robust, the theoretical predictions must also be improved to include
baryonic effects in the mergers, and a full cosmological context.

\subsubsection{Velocity dispersions}\label{sec:gcdisp}

\begin{figure}
\includegraphics[width=3.4in]{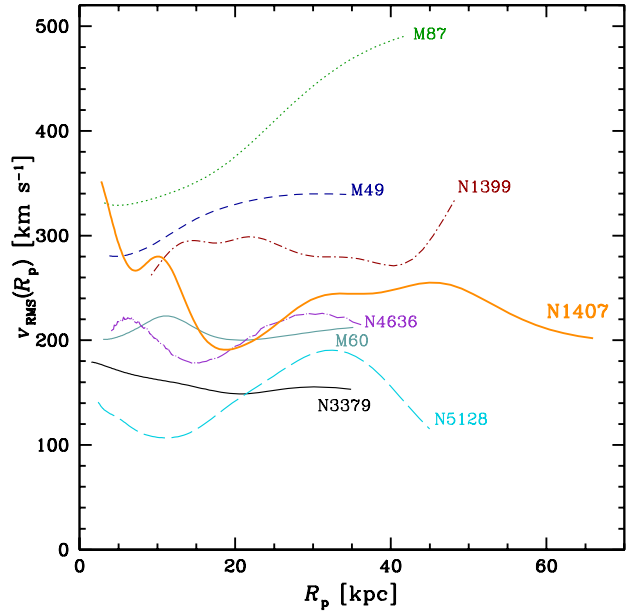}
\caption{
Projected, smoothed GCS RMS velocity profiles of group- and cluster-central
early-type galaxies.
For clarity, average values are shown but no boundaries of uncertainty;
typically the outermost 10--20\% of the radial range shown is based on
few data points and the results there are not robust.
\label{fig:dispcomp}
}
\end{figure}

We next compare the velocity dispersions in the GCG sample.
In Fig.~\ref{fig:dispcomp}, we have applied our smoothing
technique to the data sets listed in Table~\ref{tab:gal} to produce GCS dispersion 
profiles.
We note first of all that our NGC~1407 data out to 60~kpc comprise the most
radially extended set of GC velocities so far published
for an early-type galaxy (a forthcoming data set for NGC~1399 will reach
80~kpc, as previewed in R+08).
Next we see that in GCGs, a fairly constant 
profile is so far a generic property (see also the discussion in H+08).  
Some departures from constancy are also evident, with hints in several
galaxies of a central dispersion drop to $\sim$15--20~kpc, driven by
the metal-rich GCs (with NGC~1407 the most extreme case).
M87 also stands out 
with its rapidly ascending dispersion.
Although there is some indication in the NGC~1407 data of the onset of a large
halo as in M87, the mass scales involved are considerably different
(we will examine their mass profiles more directly in \S\ref{sec:massprof}).

In the context of the B+05 simulations,
the fairly constant GC dispersion profiles in all of these systems
would support their formation in
multiple mergers, since pair mergers were found to have systematically more
steeply declining dispersion profiles.  This result in the simulations is
presumably caused not by mass profile differences but
by some combination of variations in the GCS spatial densities
and orbital anisotropies.
However, none of the B+05 merger remnants shows a dispersion profile quite as
flat as seen in the data---a discrepancy that could very well be due to 
more massive, extended DM halos around these GCGs.
Interpreting the patterns seen in Fig.~\ref{fig:dispcomp} requires dynamical
models to include the effects of mass, anisotropy, and spatial density
(e.g., \S\ref{sec:massprof}).

\begin{figure}
\includegraphics[width=3.4in]{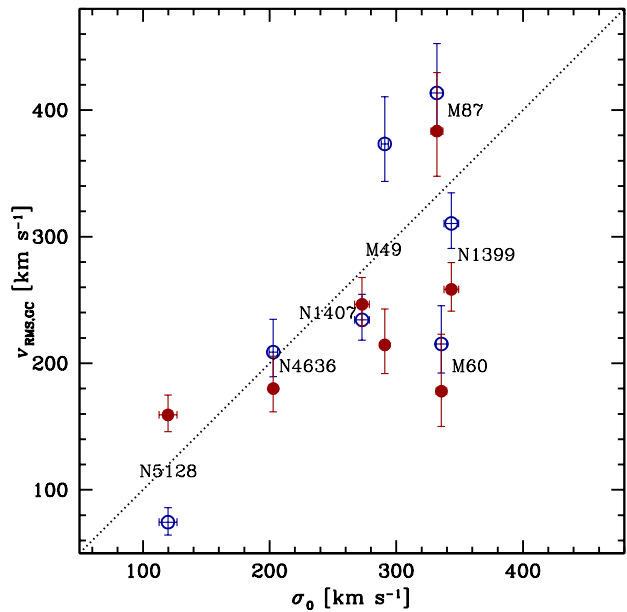}
\caption{
Overall GCS velocity dispersions vs central galaxy velocity dispersions.
Open blue circles show metal-poor subcompoments, and filled red circles show metal-rich.
The dotted line shows a one-to-one relation.
\label{fig:sigs2}
}
\end{figure}

We next make a basic comparison between the global velocity
dispersion amplitudes of the metal-poor and
metal-rich GCs in these systems, and notice some interesting trends (Fig.~\ref{fig:sigs2}).
The $v_{\rm RMS}$ of the metal-poor GCs
seems to roughly track the central velocity dispersion of the host galaxy ($\sigma_0$),
while metal-rich GCs may have a weaker dependence on $\sigma_0$.
One would na\"ively expect the opposite, since the metal-poor GCs extend farther out into
the group halo and should be less coupled to the central galaxy's properties.
Clearer conclusions will require a more in-depth, homogeneous analysis of these and additional
systems.

\subsubsection{Orbital properties}\label{sec:gcorb}

\begin{figure}
\includegraphics[width=3.4in]{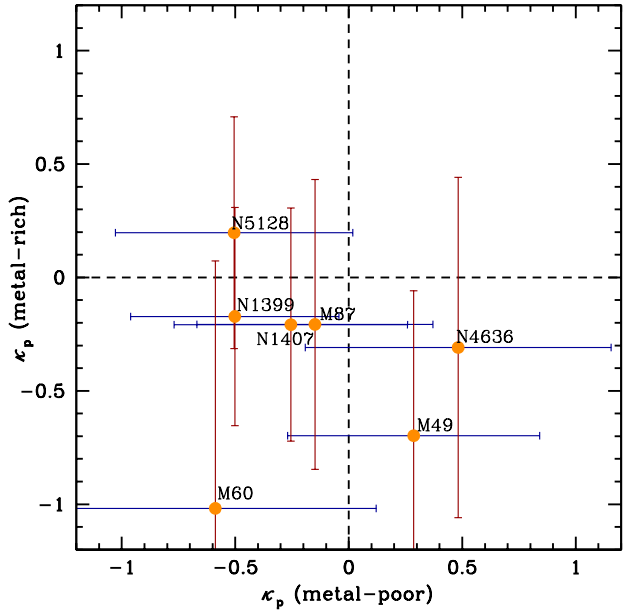}
\caption{
Kurtosis of the line-of-sight velocity distribution, for the metal-poor and metal-rich
GC subpopulations of GCGs.
Dashed lines mark Gaussian distributions ($\kappa_{\rm p}=0$).
Negative and positive kurtosis generally imply tangential and radial
anisotropy, respectively.
\label{fig:kurtsumm}
}
\end{figure}

Lastly, we consider the dynamics of the GCSs, in the context of their
orbital anisotropies.  We have made a tentative determination for NGC~1407
that the overall GCS has somewhat tangential anisotropy, which is strongest
for the metal-poor GCs, with near-isotropy for the metal-rich GCs (\S\ref{sec:modalt}).
Such studies are in their infancy, but there is an emerging consensus
that the overall GCS anisotropies of group-central ellipticals 
are isotropic or modestly tangential
(see overview in H+08).
To summarize the situation, we plot the kurtoses of the metal-poor and metal-rich
systems in Fig.~\ref{fig:kurtsumm}.  In all cases, both subpopulations are consistent
with zero or negative kurtosis ($\kappa_{\rm p} = -0.2 \pm 0.2$ on average), 
suggesting isotropic or somewhat tangentially anisotropic orbits.

This observational result conflicts with the general expectation
from theory that stars and GCs in galaxy halos should be in somewhat radial
orbits owing to infall and merger processes (see references in \S\ref{sec:fid}),
a prediction that has been observationally confirmed for PNe
\citep{2008MNRAS.385.1729D,2008arXiv0804.3350D,2009MNRAS.393..329N}.
In particular, the results for the metal-poor GCs stand in stark contrast to
the proposal of D+05 and \citet{2006MNRAS.368..563M}
that these GCs are associated with DM subsets on radial orbits
(see Fig.~\ref{fig:beta}).

One caveat is that the theoretical predictions have generally considered
individual galaxies rather than GCGs, which could in principle have
distinctly different halo and GCS properties---e.g., perhaps the transition
radius $r_0$ in Eq.~\ref{eqn:beta} would be much larger in such 
systems\footnote{The simulations of GCG halo stars by \citet{2006MNRAS.369..958S}
predict anisotropies that are very similar to the D+05 model in Fig.~\ref{fig:beta} (GR,XR)
in the case of fossil groups, and slightly less radial in the case of non-fossils.}.
There are so far no comparable data on more isolated galaxies, except for
the Milky Way, where the metal-poor GC orbits are only mildly radial on average
\citep{1999ApJ...515...50V}.

The other key consideration is that the present-day population of GCs is
probably a small surviving remnant of a primordial population that was subject
to various destructive effects such as evaporation, dynamical friction, and tidal shocking
(e.g., \citealt{1997ApJ...474..223G,2001ApJ...561..751F}).
In principle, these effects could be correlated with orbit type, so that GCs on
radial orbits were preferentially destroyed, leaving a more isotropic or tangential
distribution for present-day observers (e.g., \citealt{1988ApJ...335..720A,1997MNRAS.291..717M,1998A&A...330..480B}).
However, a simulation of GCs {\it in situ} around an M87 analog indicated that orbit
anisotropies are decreased only within a few kpc of the galaxy center
\citep{2003ApJ...593..760V}---while the GCS anisotropies are {\it observed} to be low
on scales of tens of kpc.

Still, there is much work remaining to be done for the predictions of initial GC orbits,
and the connections of orbits with destruction.
More realistic treatments of shocks and triaxial orbits may imply that GC disruption
is more effective at larger radii
(e.g., \citealt{2004AJ....127.2753D,2005MNRAS.356..899C}).
Most importantly, the evolution of GCs should be followed in the context of
a full cosmological picture of galaxy assembly.
The only work along these lines so far is by \citet{2008ApJ...689..919P}, who
found that tidal shocking is not a dominant effect in metal-poor GC evolution, and that
the GC orbits are set by the orbits of their progenitor satellite galaxies (whether
extant or disrupted),
resulting in near-isotropy out to radii of $\sim$~0.15~$r_{\rm vir}$,
which is not too different from our observational constraints
(Fig.~\ref{fig:beta}).
However, the GCSs in these simulations are much more spatially extended than real systems,
calling into question their mechanism for depositing the GCs and thus their
anisotropy predictions (cf. \citealt{2006MNRAS.365..747A}).
Furthermore, the simulations describe an isolated Milky Way counterpart rather than a GCG,
where the violent merging process should erase whatever
initial orbits the GCs had anyway, leaving again the radial orbit puzzle discussed above.

Keeping in mind that the disruption rates of GCs may depend on their masses and
concentrations, one wonders whether there is any signature dependence
of GC kinematics with their luminosities or sizes.
In fact, there are only a few galaxies where this issue has been examined:
NGC~1399, NGC~4374, and NGC~4636 (R+04; \citealt{2006A&A...459..391S}; 
B. Kumar et al., in prep.),
and now NGC~1407 (see \S\ref{sec:kurt}).
In each case, a plot of GC velocities with magnitude reveals
that the very brightest objects have noticeably
platykurtotic distributions and correspondingly higher dispersions than the
rest of the GCs.  These features are not statistically significant in all
cases (including NGC~1407), but the overall pattern is clear.
One explanation could be that the population of bright GCs on radial orbits was
depleted by disruption effects, leaving them with only near-circular orbits.
This does not necessarily mean that the brightest GCs were most sensitive to
these effects, since a general reduction in mass for all GCs 
on radial orbits would shift them toward lower luminosities,
leaving the bright end of the distribution as the only subpopulation observably depleted
of radial orbits.

In NGC~1407, we also see peculiar behavior at small radii,
where the metal-rich GCs have a sharply increasing velocity dispersion that could
imply a radial orbit distribution
(see \S\ref{sec:disp} and \S\ref{sec:mult1}).
This conclusion seems to contradict the generic theoretical expectation that
central GCs will tend to more {\it tangential} orbits, as
high-velocity radially-plunging objects would be destroyed near their pericenters.
The presence of the DGTO candidates (\S\ref{sec:select}) suggests another possibility:  
that some of the ``normal'' central GCs are really DGTOs that were formed from
the nuclei of tidally-stripped galaxies 
(e.g., \citealt{2001ApJ...552L.105B,2006MNRAS.367.1577F,2007MNRAS.380.1177B,2008MNRAS.385.2136G}).
These DGTOs would now have radially-biased orbits, while a population of
tangentially-biased progenitor galaxies should linger at larger radii, for the
same reasons discussed above for normal GC disruption.
Whether the detailed anisotropy profiles from this theoretical scenario
and the NGC~1407 observations are consistent,
and whether the progenitor population can be found, remains to be seen.
Other open questions are why there would be no signature of a residual
central DGTO population in the GC kinematics of other GCGs,
and why the DGTO population around NGC~1399 is observed to have a {\it decreasing}
velocity dispersion toward the center
\citep{2008MNRAS.389..102T}.

\subsection{Mass profiles}
\label{sec:massprof}

We next examine the mass profiles of the systems from Table~\ref{tab:gal}
(we have not re-derived these profiles as we did for the kinematics properties).
For the sake of homogeneity, these include only results where the halo
masses are constrained by the GCs rather than by any independent constraints
such as PNe and X-rays.
X-ray studies in particular could be selection-biased
toward systems with unusually massive, concentrated DM halos
(e.g., \citealt{2006ApJ...640..691V,2007ApJ...664..123B,2007A&A...473..715F}),
since the resulting high X-ray luminosities of such systems are tempting
observational targets.
The GC studies so far may also have a similar, less direct bias, since
the galaxies hosting large, readily-studied GC populations are also likely 
to have more massive DM halos \citep{2008MNRAS.385..361S}.
Such a bias suits our current analysis of GCGs,
but more general conclusions about halo masses in elliptical galaxies will require
a careful galaxy sample selection independent of the GC content
(e.g., B. Kumar et al., in prep.).

All of these GC-based mass results involved near-isotropic GC systems;
we use model GT for NGC~1407 (see \S\ref{sec:modalt}) rather than the consensus model C.
In the case of M87, the RK01 model has been used rather than the mass model of C+01---which
attempts to fit the overall Virgo cluster dynamics but does
not match well the steep profile increase in the region constrained by the X-rays and GCs.
We do not include NGC~3379 and M60 since clear results are not yet available based
on the GCs alone\footnote{For
NGC~3379, the GCs are compatible with the PN-based result that 
the $v_{\rm c}(r)$ profile declines rapidly with radius
\citep{2006A&A...448..155B,2007ApJ...664..257D}, while for M60, 
the X-ray results imply a roughly constant $v_{\rm c}(r)$ profile to 
$\sim$~30 kpc (see H+08).}.

\begin{figure}
\includegraphics[width=3.4in]{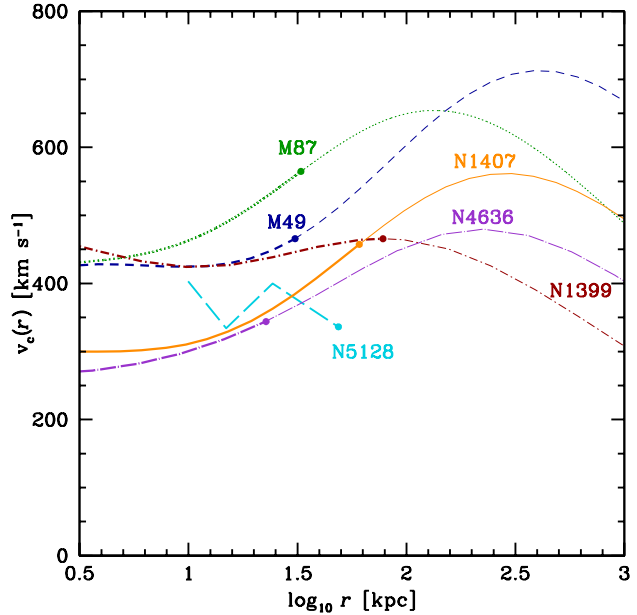}
\caption{
Mass modeling results for group/cluster-central galaxies, based on GC dynamics.
For clarity, no uncertainties are shown.
The linestyles correspond to galaxies as in Fig.~\ref{fig:dispcomp}.
The thicker segments of the curves with terminating points show the approximate radial range
 constrained by the data.
\label{fig:vc}
}
\end{figure}

The summary of these results is shown in Fig.~\ref{fig:vc}.
In most of these systems, an upturn in $v_{\rm c}(r)$ is visible:
this is an intrinsic feature of $\Lambda$CDM halos that is masked in the
central parts by the mass contribution of the galaxy, where it conspires to
produce a roughly constant $v_{\rm c}(r)$ profile
(q.v. \citealt{2001AJ....121.1936G,2003ApJ...595...29R,2007ApJ...667..176G}).
In most cases, the upturn is mild at the radii probed by the GC data,
but in the cases of NGC~1407 and M87, a strong upturn is seen.
Note that the $v_{\rm RMS}(R_{\rm p})$ profiles (cf. Fig.~\ref{fig:dispcomp})
are not straightforward indicators of the $v_{\rm c}(r)$ profiles, 
which can be inferred
only with a dynamical modeling process that includes the slopes of the tracer density
profiles and (ideally) the tracer orbit anisotropies.

Before comparing NGC~1407 and M87 further, we note that their best-fit
GC-based mass profiles are not unique, and in fact the two DM halo profiles
are nearly consistent within the uncertainties.
However, the independent constraints from X-rays (for M87) and group galaxies
(for NGC~1407) do support these best-fit models.
The difference in $v_{\rm c}$ amplitudes between the two systems
implies a factor of 2 difference in DM content within $\sim$~30 kpc
(consistent with the observation that the X-ray temperatures differ by a factor of 2).
The M87 sub-cluster
appears to have a much higher halo concentration than the NGC~1407 group, which
suggests that the NGC~1407 group collapsed at a much lower
redshift than than core of the Virgo cluster did.

To first order, a CDM halo's initial collapse redshift $z_{\rm c}$ is related to its
characteristic density $\rho_s$ (tabulated for NGC~1407 in Table~\ref{tab:models}) 
by $\rho_s \propto (1+z_{\rm c})^3$ \citep{1997ApJ...490..493N,2001MNRAS.321..559B}. 
In more detail, the $\Lambda$CDM toy model of \citet{2001MNRAS.321..559B} implies 
$z_{\rm c} \sim 1.7$ for NGC~1407 (age~$\sim$~10~Gyr) and $z_{\rm c} \sim 4$ for M87,
assuming no alteration of the central DM density through baryonic
effects\footnote{\citet{2006MNRAS.369.1375T} 
used different criteria to suggest $z_{\rm c} \sim 0.8$ for the NGC~1407 group,
while the simulations of \citet{2008MNRAS.386.2345V} predict $z_{\rm c} \sim -0.1$ for
NGC~1407, and $z \sim 0.05$ for M87 and NGC~1399.}.
The halo mass profile of NGC~1399 is still fairly uncertain (see R+08),
but the best-guess central mass profile implies $z_{\rm c} \sim 4$.
These numbers also reflect our result of a ``normal'' DM halo concentration for NGC~1407,
while the other systems have fairly high concentrations, since
at fixed mass, higher-concentration halos will collapse first.
(Note that if the X-ray result for NGC~1407 is correct, then it has $z_{\rm c} \sim 5$.)
Optically, the Fornax core seems somewhat more relaxed than Virgo, with a higher
galaxy density (also somewhat higher than that of the NGC~1407 group; B+06b; 
\citealt{2007ApJS..169..213J}), suggesting an earlier collapse---but
the DM density should provide the more accurate picture.
Thus the NGC~1407 group appears to be at an earlier evolutionary stage than both
Virgo and Fornax (i.e., starting its collapse later).

\subsection{Baryon content}\label{sec:bary}

\begin{figure}
\includegraphics[width=3.4in]{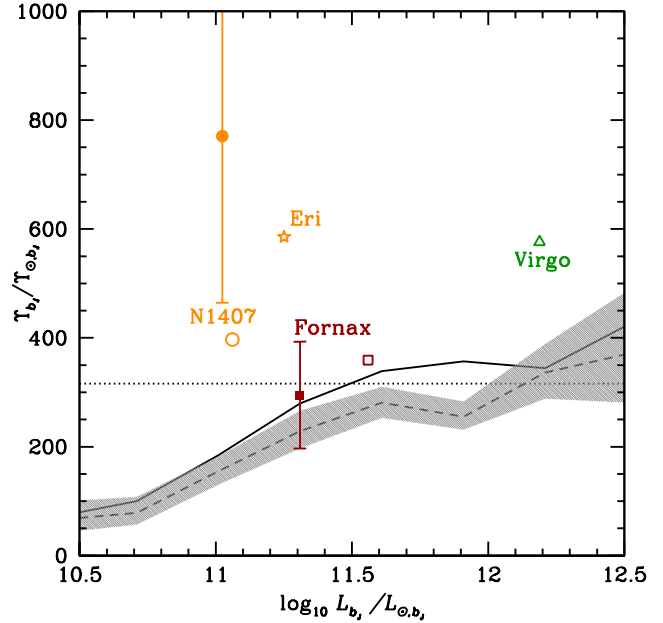}
\caption{
Virial mass-to-light ratio, as a function of group luminosity.  
The dashed line with shaded region shows the empirical results (with uncertainties)
from \citet{2006MNRAS.370.1147E} based on group dynamics.
The solid line shows the semi-empirical result from the same authors, based on 
luminosity-mass function matching.
The horizontal dotted line shows the mean value for the universe, assuming
a matter density of $\Omega_{\rm M}=0.3$;
a Hubble constant of $H_0$=70~km~s$^{-1}$~Mpc$^{-1}$ is adopted.
The solid points with error bars show results for the NGC~1407 group (model C
from this paper) and for the Fornax cluster (\citealt{2001ApJ...548L.139D}; R+08),
based on models of their internal dynamics.
The open points show more uniform results for the same systems and for the Virgo cluster,
i.e., using the same survey data and mass analyses
(from the projected mass-estimator; \citealt{2005ApJ...618..214T}).
The star symbol shows a rough estimate for the eventual Eridanus supergroup/cluster,
extrapolated from B+06b.
\label{fig:ML}
}
\vspace{0.01cm} % 
\end{figure}

We next consider some global baryonic properties of the NGC~1407 group,
comparing them to empirical and theoretical values for large samples of groups 
in the universe.
We start with the average overall mass-to-light ratio $\Upsilon$ for galaxy groups.
There are many possible sources for these values, most of them with
similar findings, and we select the results of \citet{2006MNRAS.370.1147E},
who used two different techniques to derive group and cluster $\Upsilon$ values
based on $\sim$~100,000 2dFGRS galaxies at $z \sim 0.1$
(their halo mass definitions are within a few percent of ours for $\Omega_M=0.3$).
The first is a direct empirical technique of using the measured redshifts
to calculate the masses of the individual galaxy groups with a virial analysis.
The second is a hybrid empirical-theoretical matching of the observed
luminosity function of groups to the mass function expected from $\Lambda$CDM
(we are not aware of any purely theoretical prediction, owing to the
highly uncertain effects of baryonic physics in structure formation).
As displayed in Fig.~\ref{fig:ML}, both approaches yield very similar results,
with $\Upsilon$ increasing from a value of $\sim$~60~$\Upsilon_{\odot,b_J}$ for group luminosities
of $L^*$, to a maximum value of $\sim$~300~$\Upsilon_{\odot,b_J}$ for clusters\footnote{\citet{2007arXiv0712.3255M}
used a different analysis on the same data and found $\Upsilon$ higher by a factor of $\sim$5.},
which is similar to the average $\Upsilon$ for the whole universe.

We plot the results for the NGC~1407 group, increasing the $B$-band luminosity
by 10\% to map to the photographic $b_J$-band (see \citealt{2004MNRAS.355..769E}).
Here we see that at the optical luminosity of the NGC~1407 group,
a typical $\Upsilon$ ought to be $\sim$~130--190~$\Upsilon_{\odot, b_J}$.
Our consensus solution is much higher than this, with 
$\Upsilon \sim$~800~$\Upsilon_{\odot, b_J}$, which is extreme even for a galaxy cluster.
However, the range of uncertainty for NGC~1407 is still large, and we note that
the shaded prediction region reflects the uncertainty in the mean value,
{\it not} the cosmological scatter which may be larger.
The NGC~1407 $\Upsilon$ values could be brought down if more luminosity were
included, e.g., in additional group member galaxies, extended stellar halos, or
intra-group light.
However, any additional members will be dwarfs and not contribute much to the
total luminosity, while the latter two issues should affect the 2dFGRS survey results 
similarly.
Note that in the $B$- and $K$-bands, the NGC~1407 group $\Upsilon$ is 
$850^{+1010}_{-340}$~$\Upsilon_{\odot,B}$ and
$190^{+300}_{-80}$~$\Upsilon_{\odot,K}$, 
respectively
(for solution GG, where no X-ray constraints or concentration priors are applied,
$\Upsilon$ is $\sim$~60\% higher).

The high $\Upsilon$ of the NGC~1407 group goes along with other unusual properties.
The low luminosity for its halo mass means that it is also deficient
in bright galaxies, at a level $\sim$~2--3 times
lower than found observationally for typical low-redshift groups of similar mass
\citep{2005MNRAS.358..217Y,2007arXiv0710.3780H}, although NGC~1407 appears to be
an outlier at only the $\sim$~1--2~$\sigma$ level.
The luminosity of the central galaxy NGC~1407 itself
is also low at the $\sim$~2~$\sigma$ level for groups of its mass 
(\citealt{2004ApJ...617..879L,2005MNRAS.358..217Y,2008MNRAS.385L.103B};
\citealt{2009MNRAS.392..801M};
but see \citealt{2008arXiv0805.3346C}).
The group's early-type galaxy fraction of $\sim$~80\%--90\% (B+06a; B+06b) is
higher than typical for a group of its mass or velocity dispersion
(\citealt{1998ApJ...496...39Z,2005MNRAS.358..217Y}; B+06a).
The only thing ``normal'' about the NGC~1407 group is the luminosity of the
central galaxy relative to the total luminosity
(OP04; \citealt{2004ApJ...617..879L}; B+06a; \citealt{2007arXiv0710.3780H}).

For comparison, we also plot in Fig.~\ref{fig:ML}
some literature results for the two nearest galaxy clusters, Fornax and Virgo.
These suggest that Fornax is more than twice as luminous as
the NGC~1407 group, yet $\sim$~30\% less massive, resulting in
a factor of $\sim$~3 difference in $\Upsilon$.
Virgo on the other hand may have an $\Upsilon$ consistent with that of the 
NGC~1407 group, suggesting that these two systems are physically similar despite their 
very different galaxy content.

One key uncertainty in these conclusions is the distance: 
although this is ideally the era of distance
measurements to 10\% accuracy, NGC~1407 is a somewhat problematic case.
As summarized by S+08, distance estimates to this galaxy using
various well-calibrated techniques range between 19 and 27 Mpc.  
We have adopted 21~Mpc since larger distances would seemingly
introduce problems for the GC luminosities and sizes \citep{2006MNRAS.366.1230F},
and incidentally would make it even harder to fit the X-ray data unless an 
implausibly low stellar $\Upsilon$ is adopted (see \S\ref{sec:exX}).
Most of our results in this paper have been independent of distance, e.g., the shape
of the $v_{\rm c}(r)$ profile, but changing from our default distance of 21 Mpc to 27 Mpc
would increase the halo mass and shift $\Upsilon$ downward
(though still leaving NGC~1407 as an outlier in Fig.~\ref{fig:ML}).

An alternative way to consider the NGC~1407 group is via its baryon fraction
$f_{\rm b} \equiv M_{\rm b}/M_{\rm vir}$, where $M_{\rm b}$ is the total
baryon mass including stars (in and between galaxies), gas, and any other
undefined baryonic form.
Assigning a conservative uniform value of $\Upsilon_{*,B}=4.45 \Upsilon_{\odot,B}$ to
all of the group galaxies, the total stellar mass is 
$\sim$~4~$\times10^{11} M_{\odot}$;
the hot gas mass is $\sim 10^{10} M_{\odot}$ (Z+07).
Our consensus mass result then implies that $f_{\rm b}\sim$~0.002--0.007.
Given the cosmological $f_{\rm b}=0.17$ \citep{2009ApJS..180..225H},
this means $\sim$96\%--99\% of the baryons in the system have been lost or else are invisible.
Again, incorporating higher $\Upsilon_*$ values and ``missed'' galaxy halo light
could boost $f_{\rm b}$, but only by a factor of $\sim$~2.
Such a result would contradict theoretical expectations for cluster-mass halos to 
remain essentially closed-boxes (e.g., \citealt{2004MNRAS.349.1101D};
\citealt{2006MNRAS.365.1021E,2007MNRAS.377...41C}),
and observational evidence for baryon ``disappearance'' to be relatively constant 
at $\sim$~25\% \citep{2007ApJ...666..147G}.

The under-representation of baryons may occur in many stellar systems
\citep{2008IAUS..244..136M},
which might be explained by expulsion from feedback mechanisms---even in
systems as massive as NGC~1407 (e.g., \citealt{2001ApJ...555..191F}).
Intra-group stellar light could also be considered a candidate for the missing baryons,
except that this would comprise $\sim$~40~times as much stellar light as the rest of
the group galaxies combined.
A third alternative is that most of the missing baryons reside in a relatively cool
and/or diffuse gaseous medium
(e.g., \citealt{2006ApJ...650..560C,2008Sci...319...55N}),
which upon collapse of the supergroup into a cluster
might heat up to emission levels detectable by X-ray telescopes.

\subsection{Implications}\label{sec:impl}

Bringing together all of the strands from the preceding subsections, we consider
their possible implications for the evolutionary history of NGC~1407 and its
group, and where this system fits in a cosmological context.
Revisiting first the membership status of NGC~1400 with its large peculiar velocity,
we note that there is no evidence from
photometry and kinematics that it is associated with a second group of galaxies
superimposed along the line-of-sight.
Also, the high inferred group mass does not hinge upon this single velocity,
and exceeds the minimum value
($\gsim 3\times10^{13} M_\odot$) necessary to keep NGC~1400 bound.
Interestingly, if NGC~1400 is now at pericenter on a near-radial orbit, 
then our best-fit solution ``GT'' implies that its apocenter is at $r_{\rm vir}$, 
and thus that this galaxy has only recently joined the group (as also suggested by S+08).

Without this new arrival, NGC~1407 would technically meet the definition for
a fossil group, i.e. a system that consists optically of a single
bright galaxy but has a group-mass halo
(e.g., \citealt{2003MNRAS.343..627J}).
The implication is that the GCG swallowed up any other bright primordial members
at early times,
resulting in a transient fossil phase as experienced by many groups
\citep{2005ApJ...630L.109D,2007MNRAS.382..433D,2008MNRAS.386.2345V}.

The GC kinematics in NGC~1407 and other GCGs (i.e., the moderate amount
of rotation and flat velocity dispersion profiles) appear broadly consistent with 
the formation of these galaxies in multiple major mergers of gas-free galaxies
with pre-existing GCSs (\S\ref{sec:gckin}).
However, the involvement of the group environment is unclear,
and more observational and theoretical work is needed to determine, e.g.,
when the GCGs formed relative to their surrounding groups,
and whether the GCSs are more closely associated with the galaxies or the groups
(cf. \citealt{1997AJ....114..482B,1999AJ....117.2398M,2008MNRAS.385..361S}).

The detailed mass profiles derived from the GC kinematics generally show
an onset of the group- or cluster-scale halo via an increasing $v_{\rm c}(r)$.
A comparison of the DM densities around NGC~1407, M87, and NGC~1399 implies
that the cores of the Virgo and Fornax clusters collapsed at earlier times
than the NGC~1407 group did.
Some commonalities in the GC kinematics and mass profiles of NGC~1407
and M87 suggest that the formational pathways of these two systems were similar,
albeit at different redshifts.
The properties of the DM itself could in principle be further probed via the central
slopes of the mass profiles and from deriving a mass-concentration relation---but
any such conclusions will require greater care in dealing with selection effects,
and with systematic issues such as orbital anisotropy.

Comparing the NGC~1407 group's global properties to empirical results for large
samples of groups, we find that its high $\Upsilon$ and correspondingly low
star formation make it a ``dark cluster'' as proposed by \citet{1993ApJ...403...37G}.
The overall impression is of a poor galaxy group trapped
in a high-mass halo that was somehow unable to form more galaxies, and
which evolved its galaxies quickly to the red sequence owing to interactions
or to the lack of fresh gas for star formation.
The system could represent the tail end of galaxy cluster formation in the universe
(with its peak at $z \sim 0.7$ for masses of $\sim 10^{14} M_{\odot}$; 
\citealt{2006MNRAS.367.1039H,2008arXiv0805.3346C}).
The future collapse of the Eridanus ``supergroup'' into a cluster
should bring the system's $\Upsilon$ closer to typical values (see Fig.~\ref{fig:ML}),
while the missing baryons might emerge as the hypothetical
unseen gas is heated into X-ray bright gas.
The only incongruity would be the unusually low luminosity of NGC~1407 as a
cluster-central galaxy, unless it experiences another major merger.

One could posit that high-$\Upsilon$ clusters such as Virgo were built up from 
progenitor groups like NGC~1407 with high $\Upsilon$ (but earlier collapse times),
akin to the notion that many brightest cluster galaxies
originated as fossil groups \citep{2006MNRAS.372L..68K}.
This scenario has intriguing relevance for the observational puzzle that
the stellar mass fraction $f_*$ in low-redshift galaxy groups and clusters
decreases with mass---contradicting
the expectation from CDM hierarchical merging that more massive systems
are formed from less massive systems with similar $f_*$
\citep{2008MNRAS.385.1003B}.
The NGC~1407 group on the other hand has a very low $f_* \sim 0.004$, which
would go in the direction needed by theory.
An obvious interpretation is that this group is a relic from an earlier epoch,
when the progenitor groups of clusters had systematically lower $f_*$.
However, as discussed by \citet{2008MNRAS.385.1003B}, the other present-day
groups should also have
low $f_*$ unless they had problematically recent bouts of star formation.
One way out is to suppose that NGC~1407 represents the tip of an iceberg, of groups
that were either missed observationally, or had their masses 
systematically underestimated.
A less mundane alternative is that structure formation on these scales is
{\it anti}-hierarchical (e.g., because of warm DM).

The notion that clusters are built up from a biased subset of groups would also
point to a pre-processing scenario for cluster galaxy evolution.
However, the trend of recent findings is that galaxy clusters
have fairly universal properties related to their total mass rather than
to their merger histories, i.e. the galaxy content is somehow systematically
altered as the clusters are assembled, rather than the clusters being
assembled from systematically distinct groups 
(e.g., \citealt{2006ApJ...650L..99L,2007ApJ...663..150M,2007arXiv0710.3780H}).
Additional nuances are being added to this picture
(e.g., \citealt{2006MNRAS.366....2W,2006ApJ...642..188P}),
and it is possible that a mass-correlated parameter such as {\it density}
could be the main driving factor.
In this context, the NGC~1407 group could be a telling system, since it is an outlier
in the usual optical correlations, and its mass is below the proposed boundary
for galaxy quenching to occur \citep{2006ApJ...642..188P}.

To better understand the implications of the NGC~1407 group, its uniqueness
must be clarified.  There are observational
problems with detecting such systems, as discussed by \citet{1993ApJ...403...37G}:
the ``finger of God'' effect of a high-mass poor group tends to be
lost in the crowd of neighboring redshifts (see also \citealt{2008ApJ...674L..13C}).
It would thus not be surprising if many such systems have been missed in
optical surveys.  On the other hand, the group {\it is} easily detected in
X-ray surveys, with a rough $T_{\rm X}$-based guess for $\Upsilon$
that is only a factor of $\sim 2$
lower than our detailed estimate (see \S\ref{sec:xray}).
Furthermore, it is an ordinary system when compared to X-ray correlations involving
$L_{\rm X}$, $T_{\rm X}$, $\sigma_{\rm p}$, $\beta_{\rm spec}$, $\beta_{\rm X}$,
$M_{\rm vir}$, $L_B$, etc.
(\citealt{2001MNRAS.328..461O}; OP04; B+06a).
Its only unusual aspect seems to be a $T_{\rm X}$ somewhat higher than
typical for its $L_{\rm X}$, although the $T_{\rm X}(r)$ profile in detail
may not be unusual 
for a group (\citealt{2007MNRAS.380.1554R}; cf. \citealt{2008ApJ...687..986D}).
This contrasts with its optical properties, almost all of which appear to be unusual 
(see \S\ref{sec:bary}).
There may be something of a contradiction between optical-
and X-ray-based mass functions which we cannot resolve in the present paper
(cf. \citealt{2008MNRAS.389..749M}).

There is anecdotal evidence for systems similar to the NGC~1407 group.
The fairly small nearby group catalogs of \citet{2005ApJ...618..214T} and B+06a 
include a few more candidates for ultra-dark groups---suggesting that the $\Upsilon$ of 
NGC~1407 may be well within the cosmological scatter.
Two lower-mass groups at intermediate redshift in the survey of 
\citet{2006ApJ...642..188P} were found to have unusually high early-type galaxy fractions,
prompting the authors to dub them ``bare'' massive-cluster cores.
The ``dark core'' in the cluster Abell 520 \citep{2007ApJ...668..806M}
could be a preserved descendant of a group like NGC~1407, 
although this core appears to be missing its GCG.
High-$\Upsilon$ systems could also be detected as ``dark clumps'' in weak lensing surveys
(cf. \citealt{2000A&A...355...23E,2006A&A...454...37V}).

Another fairly nearby group associated with NGC~1600 may also have 
$\Upsilon_B \sim$~1000~$\Upsilon_{B,\odot}$,
as found by \citet{2008ApJ...679..420S}, who
suggest that this system represents a new class of fossil group
with relatively low $L_{\rm X}$.
Alternatively, NGC~1407 and NGC~1600 might be heralds of a
lurking population of massive ultra-dark groups, which were recognized in these 
cases only because of their bright GCGs and X-ray emission.
A careful mining of modern observational surveys, and of cosmological simulations,
may reveal the frequency of such systems, 
and their place within the wider context of structure formation.

\section{Conclusions}
\label{sec:conc}

We have used Keck/DEIMOS to study the kinematics of GCs around NGC~1407,
arriving at a final sample (including literature data)
of 172 GC velocities out to galactocentric radii
of $\sim$~60~kpc.  
Outside the central regions, the GC system has weak rotation ($v_{\rm rot}/\sigma_p \sim 0.2$)
around the major axis, with the metal-poor and metal-rich GC subsystems showing
rotational misalignments.
We also detect a moving group of GCs at large radii.

The GC velocity dispersion profile declines rapidly to $\sim$~2$'$, whereafter it stays 
constant or rises gradually.
The rapid decline may be due to a contribution of centrally-located
unresolved dwarf-globular transition objects on high-eccentricity orbits;
we find two clear photometric examples of such objects.
We also identify one probable intra-group GC, and find that the ``blue tilt''
(a color-magnitude relation for metal-poor GCs) discovered with {\it HST} imaging
persists with ground-based photometry of a spectroscopically-confirmed GC sample.

Dynamical modeling of the GCS indicates a circular velocity profile that continues to
rise out to at least 60~kpc, 
consistent with a massive $\Lambda$CDM halo
with $M_{\rm vir}\sim 7\times10^{13} M_{\odot}$
and a group mass-to-light ratio of $\Upsilon_B \sim$~800~$\Upsilon_{\odot,B}$.
This agrees with results from group galaxy kinematics,
but our preliminary X-ray based analysis of the mass profile shows some discrepancies.
In order to reach agreement, the GCs would have to follow an implausible
orbital distribution, with highly radial anisotropy in the center rapidly
transitioning to tangential anisotropy in the outer parts.
We survey X-ray results in other systems, finding
some suggestions of discrepancies with kinematics-based results. 
Also, the common
phenomenon of X-ray ``kinks'' implies problems either with the assumption
of hydrostatic equilibrium,
with standard stellar $\Upsilon$ values, or with the CDM paradigm.
Resolving this issue is of obvious importance to cosmological inferences
from X-ray-based mass functions.

We compare NGC~1407 and other nearby group- and cluster-central
galaxies, finding similarities in their GC kinematics and mass profiles.
We suggest that the NGC~1407 group is similar to, but collapsed later than,
the ``seed'' progenitor group that became the core of the Virgo cluster.
Comparing its properties to large surveys of galaxy groups and clusters,
this group appears to be a strong outlier.
It is one of the ``darkest'' systems in the universe, whose $\Upsilon$ is similar
to those of the faintest dwarf galaxies \citep{2007ApJ...670..313S}, and
we are left with a puzzle as to why star formation was so attenuated in this
system, and where the initial cosmological baryon complement is hiding.
Finding more such high-$\Upsilon$ groups is part of the difficult but crucial inventory of 
mass and light in the universe, and 
could help disentangle the roles of halo density and mass in driving galaxy evolution.

Our new observations compound the existing evidence that GC systems
have isotropic or tangentially-biased velocity distributions,
which seems to dramatically contradict most theoretical expectations for halo
particles to reside on radially-based orbits.
However, there are open questions about the orbital dynamics of GC acquisition and
destruction
in a realistic theoretical context, and about possible
observational links between GC luminosities, sizes, and orbits.

NGC~1407 and its surrounding galaxy group are an intriguing system that
warrants further investigation.  More complete kinematics observations
and more detailed dynamical models are necessary to unravel its orbital and
mass-to-light ratio mysteries.  Our efforts continue along these lines.

\acknowledgements
We thank J\"urg Diemand, Sandy Faber, Caroline Foster, Oleg Gnedin, Brijesh Kumar, 
Bill Mathews, Nicola Napolitano, Julio Navarro, Joel Primack, Simon White, 
and Marcel Zemp for helpful discussions,
and Mark Wilkinson for a constructive review.
We thank Eugene Churazov, Vince Eke, Yasushi Fukazawa, Fill Humphrey, 
Kyoko Matsushita, and Haiguang Xu for providing their results in tabular form,
and Javier Cenarro, George Hau, Rob Proctor,
and Max Spolaor for supplying their data to us prior to publication.
Some of the data presented herein were obtained at the W. M. Keck Observatory, 
which is operated as a scientific partnership among the California Institute of Technology, 
the University of California, and the National Aeronautics and Space Administration. 
The Observatory was made possible by the generous financial support of the W. M. Keck Foundation.
The analysis pipeline used to reduce the DEIMOS data was developed at UC Berkeley with support from NSF grant AST-0071048.
Based in part on data collected at Subaru Telescope
(which is operated by the National Astronomical Observatory of Japan),
via a Gemini Observatory time exchange (Program ID GC-2006B-C-18).
We used methods and code provided in Numerical Recipes
\citep{1992nrca.book.....P}.
This material is based upon work supported
by the National Science Foundation under Grants AST-0507729 and AST-0808099,
and by NASA Grants HST-GO-09766.01-A and HST-GO-10402.06-A.
AJR was further supported by the FONDAP Center for Astrophysics CONICYT 15010003.
JS was supported by NASA through a Hubble Fellowship, administered by
the Space Telescope Science Institute, which is operated by the
Association of Universities for Research in Astronomy, Incorporated, under
NASA contract NAS5-26555.
LS and DF thank the ARC for financial support. 
\\

{\it Facilities:}
\facility{Keck:II, Keck:I, Subaru, HST}

\bibliography{ajr.bib}

\end{document}